\documentclass[12pt]{article} 
\pdfoutput=1
\usepackage[utf8]{inputenc}	
\usepackage[T1]{fontenc}
\usepackage[english]{babel}	
\usepackage{amsmath}		
\usepackage{amssymb}		
\usepackage{upgreek}
\usepackage{lipsum}		
\usepackage[pdftex]{graphicx}	
\usepackage{caption}		
\usepackage[skip=0pt]{subcaption}		%
\usepackage{epsfig}		
\usepackage{epstopdf}
\usepackage{slashed}		
\usepackage{float}		
\usepackage{placeins}		
\usepackage{multirow}		
\usepackage[toc,page]{appendix} 
\usepackage{braket}		
\usepackage{arydshln}		
\usepackage[headsepline]{scrpage2}	
\usepackage[bottom]{footmisc}	
\usepackage{footnote}		
\usepackage[pdftex]{hyperref}	
\usepackage[numbers,sort&compress]{natbib}
\usepackage{blindtext}		
\usepackage{dsfont}			
\usepackage{graphicx}		
\usepackage{float}
\usepackage{color}
\usepackage{nicefrac}
\usepackage{verbatim}

\usepackage[letterpaper, margin=2.0cm]{geometry}

\newcommand{\Deltaz}{\tilde{z}_{\rm c}}

\newcommand{\A}{\mathcal{A}}

\renewcommand{\H}{\mathcal{H}}

\newcommand{\G}{\mathcal{G}}
\newcommand{\N}{\mathcal{N}}
\newcommand{\dd}{\mathrm{d}}
\renewcommand{\k}{\textbf{k}}
\renewcommand{\Re}{{\rm Re}}
\renewcommand{\Im}{{\rm Im}}

\newcommand\ee{\mathrm{e}}	
\newcommand\ii{\mathrm{i}}	
\newcommand\sign{\mathrm{sign}}	
\newcommand{\vect}[1]{\mathbf{#1}}
\renewcommand{\S}{\mathcal{S}}

\newcommand\fodd{f^{\rm odd}}
\newcommand\feven{f^{\rm even}}

\newcommand\nodd{n^{\rm odd}}
\newcommand\neven{n^{\rm even}}

\newcommand{\GeV}{\mathrm{\,GeV}}

\newcommand{\bse}{\begin{subequations}}
\newcommand{\ese}{\end{subequations}}

\begin{document}
\date{}
\thispagestyle{empty}

\begin{flushright}
{
\small
TUM-HEP-1050/16
}
\end{flushright}

\vspace{0.4cm}
\begin{center}
\Large\bf\boldmath
Leptogenesis from Oscillations of Heavy
 Neutrinos\\
with Large Mixing Angles
\unboldmath
\end{center}

\vspace{0.4cm}

\begin{center}
{Marco Drewes$^1$, Bj\"orn~Garbrecht$^1$, Dario Gueter$^{1,2,3}$ and Juraj Klari\'{c}$^1$}\\
\vskip0.2cm
{$^1$\it Physik-Department T70, James-Franck-Stra{\ss}e,\\
Technische Universit\"at M\"unchen, 85748 Garching, Germany}\\
\vskip0.2cm
{$^2$\it Max-Planck-Institut f\"ur Physik (Werner-Heisenberg-Institut), F\"ohringer Ring 6,
80805 M\"unchen, Germany}\\
\vskip0.2cm
{$^3$\it Excellence Cluster Universe, Boltzmannstra{\ss}e 2,\\
Technische Universit\"at M\"unchen, 85748 Garching, Germany}\\
\vskip1.4cm
\end{center}

\vspace{-0.5cm}
\begin{abstract}
  \noindent 

\end{abstract}
The extension of the Standard Model by heavy right-handed neutrinos can simultaneously explain the observed neutrino masses via the seesaw mechanism and the baryon asymmetry of the Universe via leptogenesis. If the mass of the heavy neutrinos is below the electroweak scale, they may be found at the LHC, BELLE II, NA62, the proposed SHiP experiment or a future high-energy collider. In this mass range, the baryon asymmetry is generated via $CP$-violating oscillations of the heavy neutrinos during their production. We study the generation of the baryon asymmetry of the Universe in this scenario from first principles of non-equilibrium quantum field theory, including spectator processes and feedback effects. 
We eliminate several uncertainties from previous calculations and find that the baryon asymmetry of the Universe can be explained with larger heavy neutrino mixing angles, increasing the chance for an experimental discovery.
For the limiting cases of fast and strongly overdamped oscillations of right-handed neutrinos, the generation of the baryon asymmetry can be calculated analytically up to corrections of order one.
\newpage
\begin{footnotesize}
\tableofcontents
\end{footnotesize}

\section{Introduction}

\subsection{Motivation}
Over the past decades the Standard Model of particle physics (SM) has been
established as a powerful theory explaining almost all phenomena that are observed in particle physics. Its full particle content has been discovered eventually, and its predictions to this end pass all precision tests \cite{Agashe:2014kda}. Nevertheless, it is clear that the SM cannot be a complete theory of Nature.
Any attempt to explain the observed neutrino flavour oscillations with the SM field content relies on non-renormalizable interactions mediated by operators of mass dimension larger than four, which are generally associated with the existence of new heavy degrees of freedom that have been integrated out. 
Moreover, the SM fails to explain several problems in cosmology.
These include the origin of the matter-antimatter asymmetry in the Universe that
can be quantified by the baryon-to-photon ratio~\cite{Hinshaw:2012aka,Ade:2015xua,Cyburt:2015mya}
\begin{equation}
\eta_B = 6.1 \times 10^{-10}\,.
\end{equation}

The addition of $n_s\geq 2$ right-handed (RH) gauge-singlet (sterile) neutrinos $N_i$ ($i=1\ldots n_s$) can simultaneously explain the observed light neutrino masses via the seesaw mechanism \cite{Minkowski:1977sc,GellMann:1980vs,Mohapatra:1979ia,Yanagida:1980xy,Schechter:1980gr,Schechter:1981cv} and the baryon asymmetry of the Universe (BAU) via leptogenesis \cite{Fukugita:1986hr}.\footnote{The possibility that sterile neutrinos compose dark matter is discussed in detail in the review \cite{Adhikari:2016bei}.}.
The sterile neutrinos are connected with the SM solely through their Yukawa interactions $Y$ with the SM lepton doublets $\ell_a$ ($a=e,\mu,\tau$)
and the Higgs field $\phi$.
The Lagrangian of this model is given by
\begin{align}
\label{eq:Lagrangian}
{\cal L}=\mathcal{L}_{\rm SM}+\frac{1}{2}\bar{N}_i({\rm i} \partial\!\!\!/-M)_{ij} N_j
-Y_{ia}^*\bar{\ell}_a \varepsilon\phi P_{\rm R} N_i
-Y_{ia}\bar{N}_iP_{\rm L}\phi^\dagger \varepsilon^\dagger \ell_a\,,
\end{align}
where $\mathcal{L}_{\rm SM}$ is the SM Lagrangian. The spinors $N_i$ observe the Majorana condition $N_i^c=N_i$, where the superscript $c$ denotes charge conjugation. Besides, $\varepsilon$ is the antisymmetric ${\rm SU}(2)$-invariant tensor with $\varepsilon^{12}=1$.\footnote{Note that ${\rm SU}(2)$ group indices remain suppressed throughout this paper.} 
The eigenvalues $M_i$ of $M$, which in good approximation equal the physical masses of the $N_i$ particles, introduce new mass scales in nature. 
The requirement to explain neutrino oscillation data does not fix the magnitudes of the masses $M_i$, as oscillation experiments only constrain the combination\begin{equation}\label{mnudef}
m_\nu=v^2 Y^\dagger M^{-1}Y^*. 
\end{equation}
An overview of the implications of different choices of $M_i$ for particle physics and cosmology is {\it e.g.} provided by Ref.~\cite{Drewes:2013gca}. The relation between the parameters in the Lagrangian~(\ref{eq:Lagrangian}) and neutrino oscillation data is given in Appendix~\ref{App:Seesaw}.

The magnitude of the $M_i$ is often assumed to be much larger than the electroweak scale. However, values below the electroweak scale are phenomenologically very interesting because they may allow for an experimental discovery of the $N_i$ particles and to potentially unveil the mechanism of neutrino mass generation. Various experimental constraints on this low-scale seesaw scenario are summarised in Ref.~\cite{Drewes:2015iva} and references therein. 
In the present work, we focus on masses $M_i$ in the GeV range.
Apart from some theoretical arguments \cite{Shaposhnikov:2007nj,Araki:2011zg,Khoze:2013oga,Drewes:2014vaa}, the study of this mass range is motivated by the possibility to test it experimentally. 
Heavy neutrinos with $M_i<5$ GeV can be searched for in meson decays at B and K factories \cite{Kobach:2014hea,Canetti:2014dka,Shuve:2016muy,Abada:2015zea,Antusch:2015mia,Milanes:2016rzr,Cvetic:2016fbv,Asaka:2016rwd} or fixed target experiments \cite{Gorbunov:2007ak}, including NA62 \cite{Asaka:2012bb}, the SHiP experiment proposed at CERN \cite{Anelli:2015pba,Alekhin:2015byh,Graverini:2015dka} or a similar setup at the DUNE beam at FNAL \cite{Akiri:2011dv,Adams:2013qkq}.
Larger masses are accessible at the LHC \cite{Das:2012ze,Helo:2013esa,Dev:2013wba,Das:2014jxa,Hessler:2014ssa,Ng:2015hba,Izaguirre:2015pga,Hung:2015hra,Peng:2015haa,Dev:2015kca,Gluza:2015goa,Gago:2015vma,Das:2015toa,Kang:2015uoc,Dib:2015oka,Das:2016hof,Degrande:2016aje}, either via vector boson fusion ($M_i> 500$ GeV), $s$-channel exchange of W bosons ($500 {\rm GeV} > M_i > 80$ GeV) or in real gauge boson decays ($M_i<80$ GeV), but the perspectives would be best at a high energy lepton collider ILC \cite{Das:2012ze,Asaka:2015oia,Antusch:2015mia,Banerjee:2015gca,Hung:2015hra}, FCC-ee \cite{Blondel:2014bra,Antusch:2015mia,Antusch:2015gjw,Abada:2015zea} or the CEPC \cite{Antusch:2015mia,Antusch:2015rma}.

Since the $N_i$ are gauge singlets, they can interact with ordinary matter only via their quantum mechanical mixing with left-handed (LH) neutrinos that arises as a result of the Higgs mechanism and is the reason why the SM neutrinos become massive.
This mixing can be quantified by the elements of the matrix
\begin{equation}\label{thetadef}
\theta=v Y^\dagger M^{-1}.
\end{equation}
Event rates in experiments are proportional to combinations of the parameters
\begin{eqnarray}
U_{a i}^2=|(\theta U_N)_{a i}|^2,
\end{eqnarray}
which determine the interaction strength of the heavy neutrino $N_i$ with leptons of flavour $a$.
Here $U_N$ is a unitary matrix that diagonalises the heavy neutrino mass matrix.
For convenience, we also introduce the parameter
\begin{eqnarray}
U_i^2=\sum_a U_{a i}^2
\end{eqnarray}
that quantifies the total mixing of a given heavy neutrino of flavour $i$ as well as the quantity
\begin{equation}\label{totalU}
U^2 = \sum_i U_i^2 = {\rm tr}(\theta^\dagger\theta).
\end{equation}
It is of interest to determine for which range of values of $U_{ai}^2$ heavy neutrinos can simultaneously explain neutrino oscillation data and the BAU. In the present work, we improve on the network of equations that describes the generation of the BAU from ${\rm GeV}$-scale sterile neutrinos and develop analytic approximations to the solutions for phenomenologically relevant limiting cases.

\subsection{Leptogenesis Scenarios}\label{sec:benchmarkscenarios}
Any mechanism that generates a non-zero BAU has to fulfil the three Sakharov conditions \cite{Sakharov:1967dj} of i) baryon number violation, ii) $C$ and $CP$ violation and iii) a deviation from thermal equilibrium.\footnote{Leptogenesis is based on the idea that a matter-antimatter asymmetry $L$ is generated in the leptonic sector and partly converted into a baryon asymmetry $B$ by weak sphalerons, which violate $B+L$ and conserve $B-L$. This of course in addition requires a violation of $B-L$, which is provided by the Majorana mass $M$.}
Parity and baryon number are already sufficiently violated in the SM, the latter by weak sphalerons \cite{Kuzmin:1985mm} at temperatures larger than $T_{\rm ws}\simeq 130$ GeV \cite{D'Onofrio:2014kta}.
In the Lagrangian (\ref{eq:Lagrangian}), $CP$ is (in addition to the $CP$ violation in the SM) violated by complex phases in the Yukawa coupling matrix $Y$ and the mass matrix $M$. 
The non-equilibrium condition can be realised by the heavy neutrinos $N_i$ in different ways. These can qualitatively be distinguished by the relative magnitude of different time scales, which we express through the variable $z=T_{\rm ref}/T$. Here $T$ is the temperature of the primordial plasma and $T_{\rm ref}$ some arbitrarily chosen reference temperature, which we set to $T_{\rm ref}=T_{\rm ws}$ for convenience, such that sphalerons freeze out at $z=1$. We assume that the radiation dominated era starts with a vanishing abundance of $N_i$, which appears reasonable due to the smallness of their couplings $Y$ \cite{Bezrukov:2008ut}.
The heavy neutrinos are produced in a flavour state that corresponds to an eigenvector of $YY^\dagger$ (\emph{interaction basis}). Since $Y$ and $M$ are in general not diagonal in the same flavour basis, they start to undergo flavour oscillations at $z=z_{\rm osc}$. Their abundance reaches thermal equilibrium at $z=z_{\rm eq}$. They decouple (\emph{freeze out}) from the plasma and subsequently decay at $z=z_{\rm dec}$. While this picture qualitatively holds for all parameter choices in the Lagrangian, the values of $z_{\rm osc}$, $z_{\rm eq}$ and $z_{\rm dec}$ greatly vary.

In the original thermal leptogenesis scenario \cite{Fukugita:1986hr}, the $N_i$ are superheavy ($M_i\gg T_{\rm ws}$). In this case, their production, freezeout and decay all happen long before sphaleron freezeout ($z_{\rm osc}<z_{\rm eq}<z_{\rm dec}\ll 1$). The final lepton asymmetry is produced in the $CP$-violating decay of $N_i$ particles and partly converted into a BAU by the sphalerons. The non-equilibrium condition is satisfied by the deviation of the $N_i$ distribution functions from their equilibrium values at $z>z_{\rm dec}$.  
Oscillations amongst the $N_i$ in principle occur, but at $z\sim z_{\rm dec}$ they are so rapid that they can be averaged over, so that it is not necessary to track the non-diagonal correlations of the right-handed neutrinos states. Exceptions from this behaviour require accidental parametric cancellations that appear unlikely in phenomenological scenarios~\cite{Garbrecht:2011aw,Garbrecht:2014aga}. 
This scenario and various modifications have been studied in the literature in great detail and are reviewed in Refs.~\cite{Buchmuller:2005eh,Davidson:2008bu,Blanchet:2012bk}.

For $M_i$ in the $\GeV$ range under consideration here, however, the smallness of the light neutrino masses (\ref{mnudef}) implies that the Yukawa couplings $Y_{ia}$ must be very small. In this case the $N_i$ production proceeds much more slowly, and the non-equilibrium condition is satisfied by the initial approach of their distribution functions to equilibrium prior to sphaleron freezeout at $z=1$.
This scenario is often referred to as \emph{leptogenesis from neutrino oscillations} \cite{Akhmedov:1998qx} because coherent 
oscillations of the heavy neutrinos during their production lead to $CP$-violating correlations between their mass eigenstates at $z\sim z_{\rm osc}$. 
These are then transferred into matter-antimatter  asymmetries $\Delta_a=B-L_a/3$ in the individual SM flavours $a=e,\mu,\tau$ when scatterings convert some of the $N_i$ back into SM leptons. Here $L_a$ are flavoured lepton asymmetries, which are kept in equilibrium with the baryon asymmetry $B$ by sphaleron processes.
Since the violation of total lepton number due to the Majorana masses is suppressed at $T>T_{\rm ws}\gg M_i$, the total lepton number remains small initially: $|\Delta_a|\gg |\sum_a\Delta_a|\simeq0$.
A total asymmetry $\sum_a\Delta_a\neq0$ is, however, generated 
because part of the asymmetries $\Delta_a$ are converted into helicity asymmetries in the Majorana fields $N_i$ by \emph{washout} processes  with an efficiency that depends on the different flavours $a$.
If the washout is completed before sphaleron freezeout, all asymmetries are erased. If the washout is incomplete at $z=1$, then a baryon asymmetry $B$ survives, as $B$ is conserved for $z>1$.

Based on the relation among the time scales $z_{\rm osc}$ and $z_{\rm eq}$, which is controlled by the Yukawa couplings of the sterile neutrinos and their Majorana masses, we can distinguish between two regimes:
\begin{itemize}
\item
In the \emph{oscillatory} regime oscillations occur much earlier than the equilibration ($z_{\rm osc}\ll z_{\rm eq}$) such that the charges $\Delta_a$ are mainly generated at early times during the first few oscillations. This requires weak damping rates and hence small Yukawa couplings in order to prevent the charges from being washed out too early. In turn, this setup allows for a perturbative analysis in the Yukawa couplings.
\item
In the \emph{overdamped} regime the equilibration of at least one heavy neutrino happens before any full oscillation among the heavy neutrinos can be completed ($z_{\rm osc}\gg z_{\rm eq}$). This requires either some degree of mass degeneracy amongst the $M_i$ because the mass differences govern the oscillation time or anomalously large Yukawa couplings $Y$. Yet, for a successful generation of the BAU, we must have at least one sterile neutrino that does not fully equilibrate. This setup allows for an analytic approximation in terms of quasi-static solutions that are driven by the slow approach
of one of the sterile flavour eigenstates toward equilibrium.
\end{itemize}
A simple power counting argument suggests that the flavoured asymmetries $L_a$ are of order $\mathcal{O}[Y^4]$, cf.\
Eq.~(\ref{sol:act_av}), while the total $L$ (and hence $B$) is of order $\mathcal{O}[Y^6]$. This counting is however, valid only for times $z\ll (\mathcal{O}[Y^2] T)^{-1}$, and cannot be used in the overdamped regime defined below (see {\it e.g.} Eq.~\ref{BAU:maxflvasymm}), or to describe the late time washout.

\begin{table}
	\centering
	\begin{tabular}{cc|c|c}
		& & overdamped & oscillatory\\
		\hline
		$M=1\,{\rm GeV}$ & $\Re \omega = 3 \pi/4$
		&$\Delta M^2=10^{-6} M^2$ &$\Delta M^2=2\times 10^{-5} M^2$\\
		$\delta = 3 \pi/2$ & $\alpha_1=0$ & $\Im \omega=4.71$ & $\Im \omega=2.16$\\
		$\alpha_2=- 2 \pi$& & $U^2 = 3.6\times 10^{-7}$& $U^2 = 2.2\times 10^{-9}$
	\end{tabular}
	\caption{The parameters used for the examples presented in this work. For the light neutrino masses, a normal hierarchy is assumed.}
	\label{tab:params}
\end{table}

We shall introduce two theoretical benchmark scenarios that roughly correspond to the two regimes. 
The {\it naive seesaw} corresponds to a situation in which the Yukawa couplings are of the order
\begin{equation}
|Y_{ia}|^2\sim \sqrt{m_{\rm atm}^2 + m_{\rm lightest}^2}M_i/v^2, \label{F0}
\end{equation}
where $m_{\rm atm}^2$ is the larger of the two observed light neutrino mass splittings and $m_{\rm lightest}$ is the unknown mass of the lightest neutrino. In this scenario, there are no cancellations in the seesaw relation~(\ref{mnudef}). This leads to rather small mixing angles $U_{ai}^2$ and makes it very difficult to find the heavy neutrinos in experimental searches.
Larger mixing angles can be made consistent with the observed neutrino masses if there are cancellations in the seesaw relation~(\ref{mnudef}). One way to motivate this is to promote $B-L$, which is accidentally conserved in the SM, to a fundamental symmetry that is slightly broken. This possibility is usually referred to as {\it approximate lepton number conservation}, as it implies that the violation of the total $L$ at low energies is suppressed compared to the violation of individual lepton numbers $L_a$.  
In this limit one finds that heavy neutrinos with Yukawa couplings much larger than suggested by the relation~(\ref{F0}) must be organised in pairs of mass eigenstates $N_i$ and $N_j$ which in the limit of exact $B-L$ conservation form a Dirac-spinor $\Psi_N=(N_i+\ii N_j)/\sqrt{2}$. This implies
\begin{eqnarray}
M_i=M_j \ , \ U_{a i}^2=U_{a j}^2 \ {\rm } \ {\rm for} \ a = e,\mu,\tau\,.\label{equalcouplings}
\end{eqnarray}
Moreover, if the $B-L$ symmetry is slightly broken, the heavy neutrino mass basis (where $M$ is diagonal) and interaction basis (where $YY^\dagger$ is diagonal) are maximally misaligned in the flavours $i$ and $j$. 
One of the interaction eigenstates does not couple to the SM at all, corresponding to a zero eigenvalue in $YY^\dagger$, while the other one can have arbitrarily large Yukawa couplings without generating large neutrino masses or a rate of neutrinoless double $\beta$ decay that is in conflict with present
observational bounds.
Within this work, we illustrate our analytic and numerical results for both scenarios through two parametric example points that are specified in Table~\ref{tab:params}.

\subsection{Goals of this Work}
The seesaw Lagrangian~(\ref{eq:Lagrangian}) contains $7 n_s-3$ new parameters, where $n_s$ is the number of sterile neutrinos. For five of these (two mass splittings and three light neutrino mixing angles) best fit values can be obtained from neutrino oscillation data~\cite{Gonzalez-Garcia:2014bfa}, see Appendix~\ref{App:Seesaw}.
In view of upcoming experimental searches, it is highly desirable to identify the range of the remaining parameters that allow to explain the BAU via
leptogenesis from neutrino oscillations.
This question has been addressed by a number of authors in the past \cite{Akhmedov:1998qx,Asaka:2005pn,Shaposhnikov:2008pf,Canetti:2010aw,Canetti:2012vf,Canetti:2012kh,Drewes:2012ma,Canetti:2014dka,Khoze:2013oga,Frigerio:2014ifa,Shuve:2014zua,Garbrecht:2014bfa,Abada:2015rta,Hernandez:2015wna,Kartavtsev:2015vto,Drewes:2016lqo,Hernandez:2016kel,Drewes:2016jae}.

The viable parameter space in the minimal model with $n_s=2$ has first been mapped in Refs.~\cite{Canetti:2010aw,Canetti:2012vf,Canetti:2012kh}.\footnote{There have to be at least two  RH neutrinos for two reasons. First, for every non-zero SM neutrino mass the type-I seesaw mechanism requires one sterile neutrino (except for models with extended scalar sectors), and two non-zero mass differences of active neutrinos have been confirmed experimentally.
Second, leptogenesis is only possible with two or more sterile neutrinos, as the $CP$-violation arises from a quantum interference involving $N_i$ that couple with different phases.}
The results of this analysis have been used to examine the physics case for the SHiP experiment~\cite{Alekhin:2015byh} and the discovery potential of a future lepton collider~\cite{Blondel:2014bra}. 
More recent studies~\cite{Abada:2015rta,Hernandez:2015wna} suggest that the viable parameter region is smaller. 
In particular, the maximal values of $U_i^2$ that are  for given $M_i$ compatible with successful leptogenesis are smaller than claimed in Refs.~\cite{Canetti:2012vf,Canetti:2012kh}, making an experimental discovery more difficult.
With the present paper, we aim to clarify this question. 
For this purpose, we derive approximate analytic solutions for the time evolution of the asymmetries in the oscillatory and overdamped regimes. This is in contrast to the initial study in Refs.~\cite{Canetti:2012vf,Canetti:2012kh}, which was entirely numerical.
Analytic solutions for the \emph{oscillatory} regime have previously been found in Refs.~\cite{Asaka:2005pn,Drewes:2012ma,Abada:2015rta,Hernandez:2015wna}, but cannot be used to identify the maximal $U_i^2$ compatible with leptogenesis because the $N_i$ oscillations tend to be overdamped when some of the $U_{ai}^2$ are comparably large.  
We confirm numerically that our analytic solutions are accurate up to factors of order one in the regimes where they are applicable. We make use of the analytic understanding to identify the parameter region that leads to the largest possible $U_{ai}^2$ that is consistent with successful leptogenesis. Within this region, we search for the maximal value of $U^2$ numerically. Compared to the previous numerical scan in Refs.~\cite{Canetti:2012vf,Canetti:2012kh}, we apply the results of improved calculations of the thermal production and washout rates in the plasma \cite{Anisimov:2010gy,Besak:2012qm,Garbrecht:2013bia,Ghisoiu:2014ena}, include spectator processes, 
and use an updated result for the value of $T_{\rm ws}$.

The parameter space in the model with $n_s=3$ is considerably larger and has been studied only partially in the context of leptogenesis from neutrino oscillations to date \cite{Drewes:2012ma,Canetti:2014dka,Hernandez:2015wna,Shuve:2014zua,Garbrecht:2014bfa}.
In Ref.~\cite{Drewes:2012ma} it has been pointed out that in this scenario the generation of the BAU does not necessarily rely on a mass degeneracy amongst the $M_i$, which is required in the case with $n_s=2$ \cite{Asaka:2005pn} as well as for resonant leptogenesis from $N_i$ decays~\cite{Covi:1996wh,Flanz:1996fb,Pilaftsis:1997dr,Pilaftsis:1997jf,Pilaftsis:2003gt}. 
This results have been confirmed in Refs.~\cite{Canetti:2014dka,Hernandez:2015wna,Shuve:2014zua,Garbrecht:2014bfa}. It has also been pointed out that leptogenesis can be achieved for larger values of $U_i^2$ for $n_s>2$ \cite{Canetti:2014dka,Hernandez:2015wna}. A complete parameter scan for the model with $n_s=3$  would be highly desirable, but is numerically challenging. Our analytic understanding in specific corners of the parameter space will be helpful in this context, as it allows to identify the relevant physical effects and time scales.

This paper is structured as follows: In Section~\ref{sec:evo_eq} we present the evolution equations for both the sterile neutrinos and the SM asymmetries, and we discuss the qualitative behaviour of the solutions. In Sections~\ref{sec:weak} and~\ref{sec:str} we derive analytic approximations to the solutions in the oscillatory and the overdamped regimes, respectively. Constraints on the active-sterile mixing are derived in Section~\ref{sec:mixing}. We discuss the implications of our results and conclude in Section~\ref{sec:disc_and_concl}. Technical details can be found in a number of appendices. In Appendix~\ref{App:Seesaw}, we summarise the parametrisation of the
masses and couplings in the seesaw Lagrangian~(\ref{eq:Lagrangian}) that is employed in this paper. We also explain the phenomenological interesting case of scenarios with an approximate lepton number conservation that can lead to a large active-sterile mixing. Appendix~\ref{AppendixKineticEquations} contains an extensive derivation of the kinetic equations for the sterile neutrinos based on first principles of
non-equilibrium field theory, while in Appendix~\ref{AppendixSMCharges} the kinetic equations for the
SM particles, that also include spectator effects, are reviewed more briefly. Finally, Appendix~\ref{App:Oscillatory} contains some details on the oscillations of the sterile neutrinos
that are omitted in the main text.

\section{Evolution Equations}
\label{sec:evo_eq}

We need to describe the real-time evolution of the fields appearing in the seesaw Lagrangian~(\ref{eq:Lagrangian})
as well as of the spectator fields these couple to in the early Universe from the hot big bang down to $T=T_{\rm ws}$ (or $z=1$). 
Since quantum correlations of the different mass eigenstates
of the heavy neutrinos are of crucial importance, there is an immediate need to go beyond a formulation in terms of Boltzmann equations for classical distribution functions.
The evolution of sterile neutrinos in the early Universe is often described by density matrix equations~\cite{Akhmedov:1998qx,Asaka:2005pn,Shaposhnikov:2008pf,Canetti:2010aw,Canetti:2012vf,Canetti:2012kh,Shuve:2014zua,Abada:2015rta,Hernandez:2015wna}
that can be motivated in
analogy to the more detailed derivation for systems of SM neutrinos~\cite{Sigl:1992fn}.

An alternative way to derive quantum kinetic equations and systematically include all quantum and thermodynamic effects
from first principles is offered by the closed-time-path (CTP) formalism of non-equilibrium quantum field theory~\cite{Schwinger:1960qe,Keldysh:1964ud,Calzetta:1986cq}. We describe this approach in Appendix~\ref{App:Seesaw}.
The main advantage is that it allows to derive effective kinetic equations that hold at the desired level of accuracy from first principles in a series of controlled approximations. 
More specifically, overcounting issues as well as ambiguities related to the definition of asymptotic states in a dense plasma can be avoided, and necessary resummations of infrared enhanced rates at finite temperature
are straightforward.

\paragraph{Charge and Number Densities}
We can safely assume that the charged fields
are maintained in kinetic equilibrium by gauge interactions such that we can describe these by chemical potentials, which are in linear approximation proportional to the comoving charge densities,
\begin{align}
\label{ChargeDefs}
q_X=\left\{\begin{array}{l}\frac{a_{\rm R}^2}{3}\mu_X\;\textnormal{for massless bosons}\\\frac{a_{\rm R}^2}{6}\mu_X\;\textnormal{for (massless) chiral fermions}\end{array}\right.
\,.
\end{align}
We use a parametrisation where
\begin{equation}
a_{\rm R}=m_{\rm Pl}\sqrt{\frac{45}{4\pi^3 g_\star}}=T^2/H
\end{equation}
corresponds to a comoving temperature in an expanding Universe with Hubble parameter $H$. Here, $m_{\rm Pl}=1.22\times 10^{19} \GeV$ is the Planck mass and $g_\star=106.75$ the effective number of relativistic degrees of freedom. The physical temperature is given by $T=a_{\rm R}/a$, where $a$ is the scale factor.

The main quantity of interest is the baryon asymmetry of the Universe or, more precisely,
the comoving density $B$ of baryon number as a function of time.
It is violated by sphaleron processes that are fast compared to the expansion rate for $z<1$ and connect $B$ to the comoving lepton number density $L=\sum_{a=e,\mu,\tau}L_a$. 
The slowly evolving quantities relevant for leptogenesis are 
\begin{eqnarray}
\label{Delta:a}
\Delta_a=B/3 - L_a\,,
\end{eqnarray}
which are conserved by all SM interactions (including weak sphalerons). Here
\begin{eqnarray}
L_a = g_w q_{\ell a} + q_{{\rm R} a}\,,
\end{eqnarray}
where $q_{\ell a}$ and $q_{{\rm R} a}$ are the comoving lepton charge densities of flavour $a$ stored within left and right chiral SM leptons, respectively, and $g_w=2$ accounts for the ${\rm SU}(2)$ doublet multiplicity.

Among the SM degrees of freedom, only $\ell$ and $\phi$ directly interact with the sterile neutrinos. Nonetheless, the remaining degrees of freedom can also carry asymmetries and participate in
chemical equilibration. They are referred to as spectator fields~\cite{Barbieri:1999ma,Buchmuller:2001sr,Garbrecht:2014kda}. 
The main effect of the spectators is to hide a fraction of the asymmetries from the washout, which only acts on the $L_a$.
Taking account of these, one arrives at relations 
\begin{eqnarray}
q_{\ell a}=\sum_b A_{ab}\Delta_b \ {\rm and} \ q_\phi=\sum_a C_a \Delta_a\,,
\end{eqnarray}
where the coefficients 
\begin{align}
\label{spectator}
A=
\frac{1}{711}
\left(
\begin{array}{ccc}
-221 & 16 & 16\\
16 & -221  & 16\\
16 & 16 & -221
\end{array}
\right)
\,,
\quad
C=
-\frac{8}{79}
\left(
\begin{array}{ccc}
1 & 1 & 1
\end{array}
\right)
\end{align}
are derived in Appendix~\ref{spectatorSec}.

The Majorana fields $N_i$ strictly speaking cannot carry any lepton charges. However, at temperatures $T\gg M_i$, their helicity states effectively act as particles and antiparticles.
We describe the $N_i$ by the deviation $\delta n_h$ of
their number density from equilibrium, that is formally defined
in Eq.~(\ref{nDeviationDef}). Here, $h=\pm$ denotes the sign of the helicity $\pm\frac12$,
and $\delta n_h$ is matrix valued.
In the flavour basis where $M$ is diagonal, the diagonal elements are the number densities and the off-diagonal entries correspond to quantum correlations. 
This allows to define \emph{sterile charges}
\begin{align}
\label{def:sterile_av}
q_{Ni}\equiv2\delta\nodd_{ii}\,,
\end{align}
in terms of the helicity-odd deviations of the occupation numbers from their equilibrium values, which is introduced more precisely in Appendix~\ref{AppendixKineticEquations}.
The Yukawa interactions $Y$ violate individual lepton flavour numbers $L_a$ at order $\mathcal{O}[Y^2]$ ({\it e.g.} by light neutrino oscillations). 
The Majorana mass $M$ also violates the total lepton number\begin{equation}
L=\sum_a L_a.
\end{equation}
However, at temperatures $T\gg M_i$ most particles are relativistic and spin flips are suppressed, such that the quantity 
\begin{equation}\label{Lgeneral1}
\tilde{L}=L+\sum_i q_{N i}
\end{equation}
is approximately conserved (up to terms of order $M_i^2/T^2$).
Since the $N_i$ start from initial conditions that are far from equilibrium, the assumption of kinetic equilibrium is not justified for them in principle. We briefly discuss the error introduced by the use of momentum averaged equations in Appendix~\ref{app:act_charg}, see also Ref.~\cite{Asaka:2011wq}.

In terms of these charge densities, we next write down the set of quantum kinetic equations used in our analysis.
A detailed derivation for the evolution of the sterile neutrinos within the CTP framework is given in Appendix~\ref{AppendixKineticEquations}, while a sketch of the derivation for the equations of SM charges is presented in Appendix~\ref{AppendixSMCharges}.

\paragraph{Evolution of Sterile Neutrino Densities}

In terms of the variable $z$ the time evolution of the number densities and
flavour correlations of the sterile neutrinos is governed by the equation
\begin{align}
\label{Diff:Sterile}
\frac{\dd}{\dd z}\delta n_{h} = -\frac{\ii}{2}[H_N^{\rm th}+z^2 H_N^{\rm vac},\delta n_{h}]-\frac{1}{2}\{\Gamma_N,\delta n_{h}\}+\sum_{a,b=e,\mu,\tau}\tilde{\Gamma}_N^a (A_{ab} + C_b/2)\Delta_b\,.
\end{align}
The flavour matrix $H_N^{\rm vac}$ can be interpreted as an effective Hamiltonian in vacuum, and $H_N^{\rm th}$ is the Hermitian part of the finite temperature correction. 
The contributions involving the matrix $\Gamma_N$ and the vector $\tilde{\Gamma}_N$ are collision terms.
Explicit expressions for these are derived in Appendix~\ref{AppendixKineticEquations}, 
\bse
\label{RHN:rates}
\begin{align}
H^{\rm vac}_N &=
\frac{\pi^2 }{18 \zeta(3)}\frac{a_{\rm R}}{T_{\rm ref}^3}
\left(\Re[M^\dagger M] + \ii h  \Im[M^\dagger M]\right)\,,
\label{avg:hamiltonian}\\
H^{\rm th}_N&=\mathfrak{h}_{\rm th}\frac{a_{\rm R}}{{T_{\rm ref}}}
\left(\Re[Y^* Y^t]-\ii h\Im [Y^* Y^t]\right)\,,\label{avg:hamiltonianth}\\
\label{avg:decayrate}
\Gamma_N &=\gamma_{\rm av} \frac{a_{\rm R}}{{T_{\rm ref}}}
\left(\Re[Y^* Y^t]-\ii h \Im [Y^*Y^t]\right)\,,\\
\label{avg:backreaction}
(\tilde{\Gamma}^a_N)_{ij}&= \frac{h}{2}\gamma_{\rm av} \frac{a_{\rm R}}{T_{\rm ref}}
\left(
\Re [Y^*_{ia}Y^t_{aj}] - \ii h  \Im [Y^*_{ia}Y^t_{aj}]
\right)\,.
\end{align}
\ese
with $\gamma_{\rm av}=0.012$ and $\mathfrak{h}_{\rm th}\approx 0.23$, {\it cf.} Eqs.~(\ref{col:av},\ref{eq:higgs_thermal}) and the discussion of these.
As pointed out in the previous section, we make use of the freedom of choice of the reference temperature scale $T_{\rm ref}$ to fix it as the temperature $T_{\rm ws}$ of weak sphaleron freezeout. However, for the sake of generality we keep the notation $T_{\rm ref}$ throughout this paper. 

It is worthwhile to emphasise that the above equations only hold in the regime where the $N_i$ are relativistic. We have essentially neglected their masses everywhere except in $H^{\rm vac}_N$, where they are absolutely crucial because they are responsible for the flavour oscillations.
The relativistic approximation has two different consequences.
One one hand, the derivation following Eq.~(\ref{eq:quasiparticle_app}) assumes that the right-handed neutrino masses $M_i$ are kinematically negligible in scatterings and decays. Putting both $N_i$ on the same mass shell is certainly a reasonable assumption, as $|M_i^2 - M_j^2| \ll M_{i}^2$ in the entire parameter space under consideration here. Entirely neglecting the masses leads to errors $\sim M_i^2/T^2$ to the rates $\Gamma_N$ and $\tilde{\Gamma}^a_N$, which should, however, not have a huge effect on our results.\footnote{If the asymmetry is generated near the electroweak scale, in principle also the masses of the top quark and gauge bosons and the contribution to the heavy neutrino masses generated by the Higgs mechanism should be taken into account. On one hand, this affects the particle kinematics. On the other hand, one should replace (\ref{avg:hamiltonianth}) by $H^{\rm th}_N =\frac{a_{\rm R}}{{T_{\rm ref}}} \left(\Re[Y^* Y^t][\mathfrak{h}_{\rm th} + \mathfrak{h}_{\rm EV}(z)] - \ii h\Im [Y^* Y^t]\mathfrak{h}_{\rm th}\right)$, where $\mathfrak{h}_{\rm EV}(z)=\frac{2\pi^2}{18\zeta(3)}\frac{v^2(z)}{T_{\rm ref}^2}z^2$ and $v(z)$ is the temperature dependent Higgs field expectation value. We neglect these effects here.
}  This is in contrast to standard thermal leptogenesis scenarios of out-of-equilibrium decay, where the dynamics is dependent on the relation between the absolute right-handed neutrino mass and the Hubble rate.
On the other hand, we neglect lepton number violating scatterings, which are suppressed by $M_i^2/T^2$. This assumption is clearly justified for $M_i$ of a few GeV and in the oscillatory regime, but is in principle questionable for $M_i$ near the electroweak scale in the overdamped regime, where the BAU is generated shortly before sphaleron freezeout. We believe that our equations can still be used in this regime because the rates for lepton number violating processes are suppressed by the small parameters $\epsilon_i$ and $\mu_i$ introduced in Eq.~(\ref{DefEpsMu}), but this statement should be checked quantitatively in the future to identify their precise range of validity.

Before explicitly solving Eq.~(\ref{Diff:Sterile}), we discuss the basic properties of the solutions. 
For this purpose we neglect the {\it backreaction} term with $\tilde{\Gamma}_N$.
The qualitative behaviour of the system is governed by the eigenvalues of $H_N^{\rm vac}$ and $\Gamma_N$, which determine the time scales on which the sterile neutrinos oscillate and come into equilibrium. 
While $H_N^{\rm vac}$ is diagonal in the flavour basis where $M$ is diagonal (mass basis), $\Gamma_N$ is diagonal in the same basis as $Y Y^\dagger$ (interaction basis). The misalignment between the two leads to sterile neutrino oscillations. That means, particles are produced in the interaction basis and then oscillate due to the commutator involving $H_N^{\rm vac}$. 
At sufficiently high temperatures the correction $H_N^{\rm th}$ due to thermal masses is larger than $H_N^{\rm vac}$, but by itself cannot initiate oscillations because it is diagonal in the same basis as $\Gamma_N$. 
For $n_s$ flavours of heavy neutrinos, there are of course $n_s$ 
relaxation times $z_\text{eq}$ and $n_s(n_s-1)/2$ oscillation times $z_\text{osc}$, all of which in general can be different.
For a qualitative classification of the oscillatory and overdamped regimes it is useful to consider the largest eigenvalues of the matrices 
$H_N^{\rm vac}$ and $\Gamma_N$.
We use the norm $||\cdot||$ of a
Hermitian matrix as the modulus of its largest eigenvalue. In case of $Y^* Y^t$ it is, for instance, associated with the interaction eigenstate with the strongest coupling to the primordial plasma.
The first oscillation involves the sterile neutrino mass states $N_i$ and $N_j$ with the largest mass splitting and occurs at a time
\begin{align}
\label{time_osc}
z_\text{osc} \approx \left(a_{\rm R} |M_i^2 - M_j^2|\right)^{-1/3}T_{\rm ref}\,,
\end{align}
such that $z_{\rm osc}^3 ||H_N^{\rm vac}||={\cal O}(1)$. The relaxation time scale
at which a sterile neutrino interaction state comes into thermal equilibrium is given by
\begin{align}
\label{time_eq}
z_{\rm eq}\simeq T_{\rm ref}/(\gamma_{\rm av} a_{\rm R}||Y^*Y^t||)\,,
\end{align}
such that $z_{\rm eq} \Gamma={\cal O}(1)$ with $\gamma_{\rm av}$ being the averaged relaxation rate (over temperature). 
 
If the slowest oscillation time scale is shorter than the fastest relaxation time scale, then leptogenesis occurs in the oscillatory regime.
In this case the heavy neutrinos undergo a large number of coherent oscillations before coming into equilibrium, which in terms of the variable $z$ become increasingly rapid.
The baryon asymmetry is most efficiently generated during the first few oscillations, before the oscillations become fast (compared to the rate of Hubble expansion), {\it cf.} Figure~\ref{fig:weak_washout_example}. 
There is a clear separation between the time $z_{\rm osc}$ when the asymmetry gets generated and the time $z_{\rm eq}$ when the $N_i$ come into equilibrium and the washout becomes efficient. This allows to treat these two processes independently.
We discuss this regime in Section~\ref{sec:weak}.

\begin{figure}
	\centering
	\includegraphics[scale=0.5]{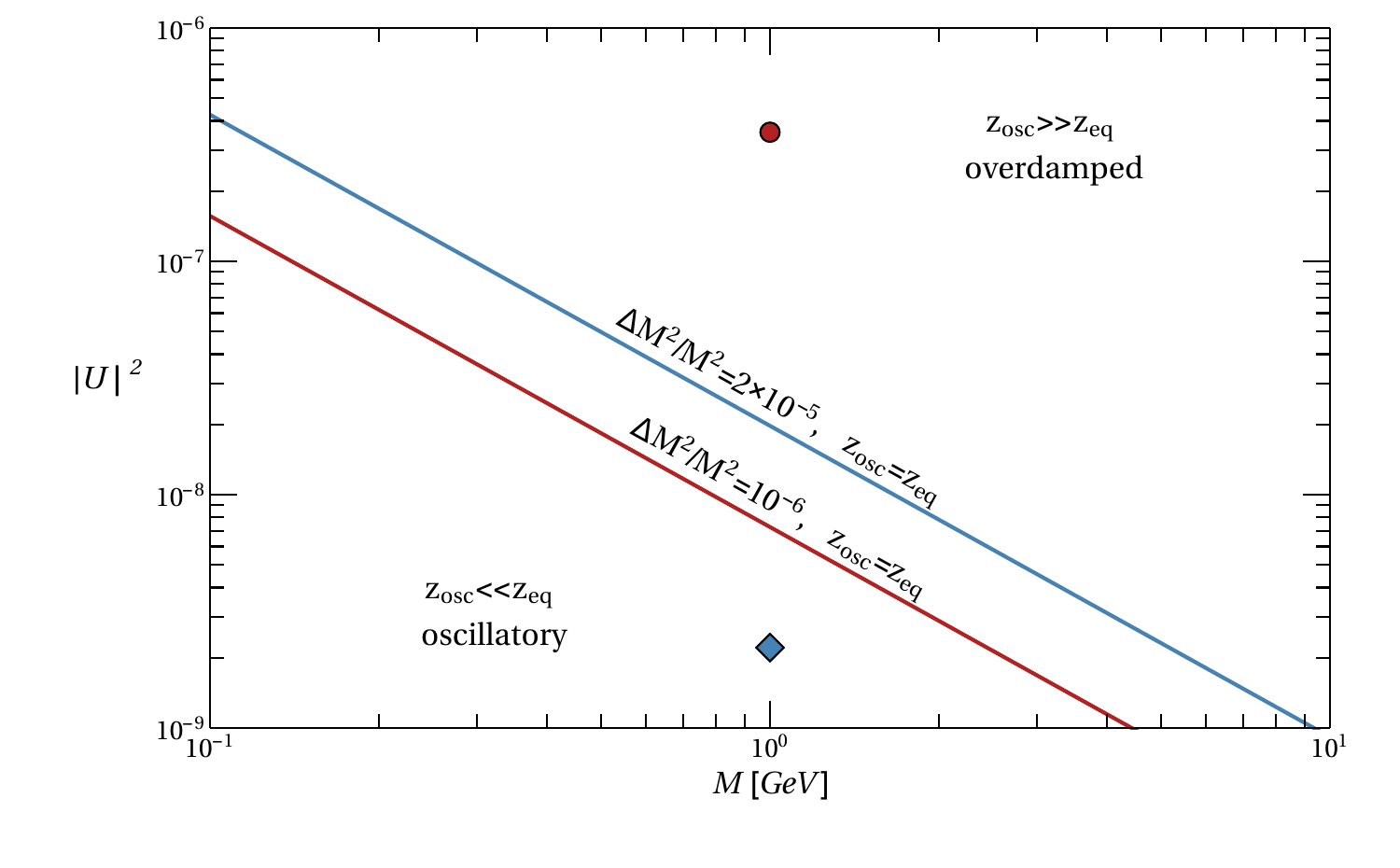}
	\caption{
Parameter regions for the effective mixing angle $\sum_a U_a^2$ [using the estimate~(\ref{cond:regime})] in case of two sterile flavours with corresponding average mass $M$ and a squared-mass splitting
$\Delta M^2=M_1^2-M_2^2$. The regions above/below the blue/red lines correspond to the overdamped/oscillatory regimes for the mass splittings indicated in the plot. The blue and red dots correspond to the two example parameter sets specified in Table~\ref{tab:params}. We can see that the blue point lies in the oscillatory and the red point in the overdamped regime.
}
	\label{fig:mixangles}
\end{figure}

If, on the other hand, at least one heavy neutrino flavour eigenstate comes into equilibrium before a neutrino that is produced in this state has performed a complete flavour oscillation,
then the oscillations are overdamped,
{\it cf.} Figure~\ref{fig:strong_washout_example}. As we illustrate in Section~\ref{sec:str}, this allows for baryogenesis with larger Yukawa couplings and consequently also larger active-sterile mixing
angles $U_{ai}^2$.
In the scenario with $n_s=2$, the largest possible values of $U_{ai}^2$ can be realised when the first oscillation happens rather late ($z_{\rm osc}\sim1$), as otherwise the washout tends to erase all asymmetries before sphaleron freezeout. As a result of the integration over a long time, the power counting in $Y$ that allows to estimate the magnitude of the asymmetries in the oscillatory regime may not be applied, and the \emph{backreaction} term involving $\tilde{\Gamma}_N$ may not be neglected.
Eqs.~(\ref{time_osc},\ref{time_eq}) allow to relate the mass difference to the Yukawa couplings in order to determine  which regime a given parameter choice corresponds to:
\begin{align}
\label{cond:regime}
\frac{||Y^* Y^t||\gamma_{\rm av}a_{\rm R}^{2/3}}{|M_i^2-M_j^2|^{1/3}}
\begin{cases}
&\ll 1 \quad \text{oscillatory}\\
&\gg 1 \quad \text{overdamped}
\end{cases}
\,.
\end{align}
Figure.~\ref{fig:mixangles} schematically illustrates where the oscillatory and the overdamped regime are
located in the $M_i-U^2$ plane for various mass splittings. We also indicate the points
from Table~\ref{tab:params} that we use in our examples in order to illustrate the two parametric
regimes.
For $n_s>2$ the situation becomes more complicated because there are more oscillation and equilibration time scales, which can be ordered in various different ways.
Moreover, the constraints on the relative size of the individual $U_{ai}^2$ from neutrino oscillation data are much weaker and allow for a flavour asymmetric washout (while for $n_s=2$ there is not enough freedom in the unconstrained parameters in Eq.~(\ref{CasasIbarraDef}) to realise vastly different values of individual $U_{ai}^2$ \cite{Ruchayskiy:2011aa,Asaka:2011pb}).

\paragraph{Evolution of SM Charge Densities} 
The time evolution of the asymmetries $\Delta_a$ is governed by the equation\footnote{
Let us recall that we work in the heavy neutrino mass basis here, and Eq.~(\ref{Diff:Active}), and similarly Eq.~(\ref{eq:washout}), are not manifestly flavour covariant. 
One reason for this is that we, following the common convention, do not include the diagonal charge $q_{N_i}$ on the RHS of Eq.~(\ref{eq:washout}) in the definition of the source term $S_a$.
This implies that the separation into "source" and "backreaction" terms in Section~\ref{sec:str} is different from the one presented here, as we again define the source as coming from the off-diagonal correlations alone, and this definition is not flavour covariant.
}
\begin{align}
\label{Diff:Active}
\frac{\dd \Delta_a}{\dd z}=
&\frac{\gamma_{\rm av}}{g_w} \frac{a_{\rm R}}{T_{\rm ref}}\sum_{i} Y_{ia}Y_{ai}^\dagger
\,\left(\sum_b (A_{ab} + C_b/2)\Delta_b
-q_{Ni}\right)-\frac{S_a}{T_{\rm ref}}\,.
\end{align}
A sketch of its derivation is presented in Appendix~\ref{AppendixSMCharges}.
Note that we neglect the correlations of the different active charges here, which are deleted by the lepton-flavour violating interactions
mediated by the SM Yukawa-interactions,
thereby breaking the flavour covariance of the evolution equations.
The first term on the right-hand side is the washout that is complementary to
the damping rate for the sterile charges, while the second term is referred to as the source term
\begin{align}
\label{Source}
S_a=2\frac{\gamma_{\rm av}}{g_w} a_{\rm R}\sum\limits_{\overset{i,j}{i\not=j}}Y^*_{ia}Y_{ja}
\left[\ii{\rm Im}(\delta n^{\rm even}_{ij})+{\rm Re}(\delta n^{\rm odd}_{ij})\right]\,.
\end{align}
It describes the generation of SM asymmetries in the presence of off-diagonal correlations of
sterile neutrinos.

\paragraph{Numerical solution}
In order to compare our analytic approximations we explicitly solve the system of differential equations~(\ref{Diff:Sterile},\ref{Diff:Active}) in the
basis where the mass matrix is diagonal, without any further approximations from $z=0$ to the electroweak phase transition at $z=1$.
Note that we assume zero initial abundance for the active charges $\Delta_a(z=0)=0$, as well as zero initial abundance for the right-handed neutrinos, meaning that their deviation from equilibrium is $\delta n_{h,i,j}(z=0) = - \delta_{i,j} n^\mathrm{eq}$.

\section{Oscillatory Regime}
\label{sec:weak}

\begin{figure}
	\centering
	\includegraphics[scale=0.6]{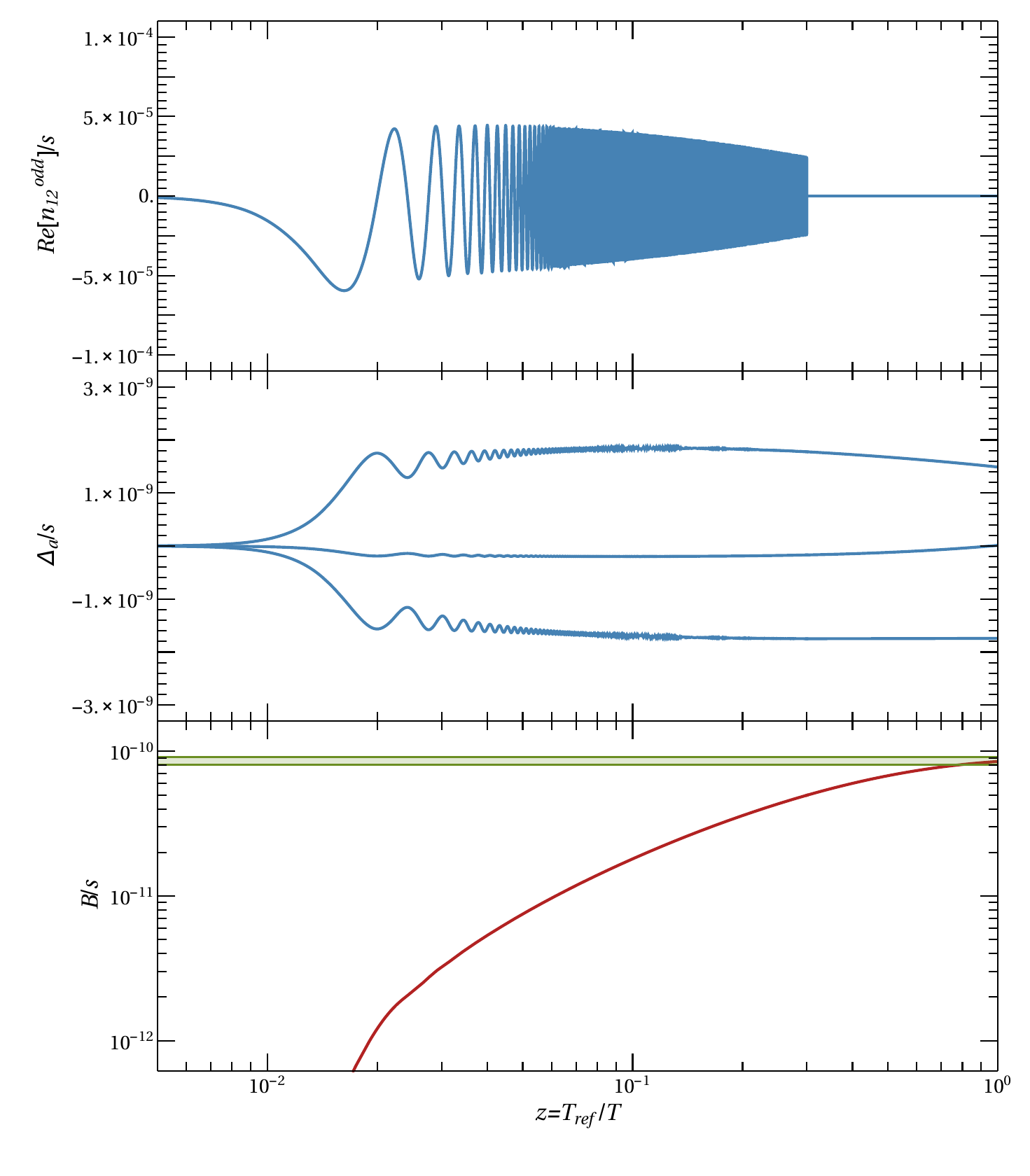}
	\caption{
The upper panel illustrates the $CP$-violating oscillations of heavy neutrinos, as characterised by the helicity odd off-diagonal flavour correlations in their mass basis. These act as a source for the generation of flavoured lepton asymmetries.
We cut off the oscillations at the point when they become too rapid to make a significant contribution to the source term, as indicated in the plot.
The middle panel shows the individual asymmetries generated in the three SM flavours. It is clearly visible that the total lepton asymmetry is only generated when the washout begins, and that its modulus remains smaller than that of the asymmetries in individual flavours at all times.
The lowest panel shows the generated baryon asymmetry, where the green band indicates the error bars of the observed value.   
}
	\label{fig:weak_washout_example}
\end{figure}

We now study the \emph{oscillatory regime}, where the
first oscillations of the
off-diagonal correlations of the sterile neutrinos
happen much earlier than their relaxation toward equilibrium, {\it i.e.}
$z_{\rm osc}\ll z_{\rm eq}$.
 The separation of scales $z_{\rm osc}\ll z_{\rm eq}$ allows to treat the generation of flavoured asymmetries from $N_i$ oscillations and their washout (which leads to $B\neq 0$) independently.
 At early times when $z\sim z_{\rm osc}$, we can expand the solution to the coupled system of Eqs.~(\ref{Diff:Sterile},\ref{Diff:Active})
in the Yukawa couplings $|Y^*Y^t|$, as we specify within Section~\ref{subsection:early:oscillations}
in detail. 
At late times,
when $z\sim z_{\rm eq}$, the off-diagonal correlations have either decayed or their effect
averages out due to the rapidity of their oscillations. Therefore, we can neglect the
commutator term in Eq.~(\ref{Diff:Sterile}) as well as the source term in Eq.~(\ref{Diff:Active}) (\textit{i.e.}
the contributions explicitly depending on $\delta n_{ij}$ for $i\not=j$).
This is done in Section~\ref{sec:latetimewashout}. 
Our solutions hold for arbitrary $n_s$ as long as the slowest oscillation time scale is faster than the fastest equilibration time scale. Throughout this section, we work in the mass basis (where $M$ is diagonal).
In Figure~\ref{fig:weak_washout_example}, we present a characteristic example for the evolution of the particular charge densities for $n_s=2$.

\subsection{Early Time Oscillations}
\label{subsection:early:oscillations}

We now identify in more detail the truncations that may be applied to Eqs.~(\ref{Diff:Sterile}) and~(\ref{Diff:Active}) when $z\sim z_{\rm osc}$ and solve the problem thus simplified analytically.

\paragraph{Oscillations of Sterile Neutrinos}
First, consider the thermal correction to
the oscillation frequency of the sterile neutrinos due to thermal masses. While in the parametrisation of Eq.~(\ref{Diff:Sterile}),
the oscillation frequency induced by the vacuum term $H_N^{\rm vac}$ is growing with $z^2$, the thermal
contributions given by $\mathfrak{h}_{\rm th}$ remain constant. As a result, at very early times,
the thermal effects exceed the contributions from
the vacuum masses. However, because $H_N^{\rm th}$ is generated by forward scatterings mediated by the Yukawa interactions, it is diagonal in the same flavour basis as $\Gamma_N$, {\it i.e.} the interaction basis in which heavy neutrinos are produced.  $H_N^{\rm th}$ therefore commutes with $\delta n_h$ at early times (before $H_N^{\rm vac}$ becomes sizeable) and does not lead to oscillations.\footnote{One may wonder whether the large thermal masses can lead to a big enhancement at $z\ll z_{\rm osc}$ by somehow amplifying a small population of the helicity-odd occupation numbers generated during the first fraction of an oscillation.
However, it turns out that the main part of the  charges $\Delta_a$ in the oscillatory regime is produced well
during the first full oscillation. This is confirmed by our numerical solutions, which take full account of the thermal masses.
}
For this reason, the thermal masses  only lead to subdominant corrections in the oscillatory regime, and we neglect these in the following.
A more detailed discussion about these time scales is presented in Appendix~\ref{app:time_scales}.
The relation $z_{\rm osc}\ll z_{\rm eq}$ also leaves the backreaction mediated through $\tilde{\Gamma}$ in Eq.~(\ref{Diff:Sterile}) as
a higher order effect at early times $z\sim z_{\rm osc}$,
such that it only becomes important later, when the  charges $\Delta_a$ have already been generated by the source term. 
In summary, for $z\sim z_{\rm osc}$, and given the relation $z_{\rm osc}\ll z_{\rm eq}$, Eq.~(\ref{Diff:Sterile}) can be simplified to
\begin{align}
\label{osc:weakwashoutav}
\frac{\dd}{\dd z}\delta n_{h} +\frac{\ii}{2}z^2[H_N^{\rm vac},\delta n_{h}]=-\frac{1}{2}\{\Gamma_N,\delta n_{h}\}\,.
\end{align}

In order to compute $q_{\ell a}$ as well as $q_{Ni}=2n_{ii}^{\rm odd}$ we have to solve Eq.~(\ref{osc:weakwashoutav}) both for helicity-even and helicity-odd distributions. 
The relation $z_{\rm osc}\ll z_{\rm eq}$ allows for a perturbative expansion in the coupling term $|Y^* Y^t|$. Solutions to order $\mathcal{O}(|Y^* Y^t|^0)$ are obtained when neglecting the right hand side of Eq.~(\ref{osc:weakwashoutav}), what results in the diagonal terms
\begin{align}
\label{eq:initial_abundance}
\delta n_{ii}^{\rm even}&=-n^{\rm eq}+\mathcal{O}(|Y^* Y^t|)\,,\quad \quad \delta n_{ii}^{\rm odd}=0+\mathcal{O}(|Y^* Y^t|)\,,
\end{align} 
with the equilibrium solution~(\ref{n_eq}), whereas the off-diagonal entries vanish. Note that the first term of Eq.~(\ref{eq:initial_abundance}) corresponds to vanishing initial abundances of the right-handed neutrinos. The first non-vanishing contribution to the charges $\Delta_a$ is  $\mathcal{O}(|Y^* Y^t|^2)$, and it
arises from the off-diagonal components of $\delta n^{\rm odd}$. These can be obtained by solving Eq.~(\ref{osc:weakwashoutav}) with the replacement
\begin{align}
\delta n_{hij} \rightarrow -n^{\rm eq}\delta_{ij}\,,
\end{align}
on the right hand side, such that we are left with solving 
\bse
\begin{align}
\label{off_diagonal_de}
\frac{\dd}{\dd z}\nodd_{ij}+{\rm i} \Omega_{ij} z^2 \nodd_{ij}&=-\ii {\rm Im}[Y^*Y^t]_{ij}G\,,\\
\frac{\dd}{\dd z}\neven_{ij}+{\rm i} \Omega_{ij} z^2 \neven_{ij}&={\rm Re}[Y^*Y^t]_{ij}G\,,
\end{align}
\ese
with 
\begin{align}
\Omega_{ij}=\frac{a_{\rm R}}{T_{\rm ref}^3}\frac{\pi^2}{36 \zeta(3)}(M_{ii}^2-M_{jj}^2)\,, \quad \quad
G=\gamma_{\rm av}\frac{a_{\rm R}}{T_{\rm ref}}
n^{\rm eq}\,.
\end{align}
The general solutions to these equations are
\begin{subequations}
\label{sol:f0hav}
\begin{align}
\nodd_{ij}&=-\ii{\rm Im}[Y^*Y^t]_{ij}G\mathcal{F}_{ij}\,,\quad \quad 
\neven_{ij}={\rm Re}[Y^*Y^t]_{ij}G\mathcal{F}_{ij}\,,\\
\label{F_tilde}
\mathcal{F}_{ij}&=\left[C_{ij}-\frac{z}{3}E_{2/3}\left(-\frac{\ii}{3} \Omega_{ij} z^3\right)\right]\exp \left(-\frac{\ii}{3} \Omega_{ij}z^3\right)\,,
\end{align}
\end{subequations}
where $C_{ij}$ is an integration constant that in case of zero initial charge is determined to be
\begin{align}
\label{Aij}
C_{ij}&=\lim_{z\rightarrow 0}\left[\frac{z}{3} E_{2/3}\left(-\frac{\ii}{3} \Omega_{ij} z^3\right)\right]=\frac{\Gamma\left(\frac{1}{3}\right)}{3^{\frac 23}(-\ii \Omega_{ij})^{\frac 13}}\,,
\end{align}
and
\begin{align}
E_n(x)=\int\limits_1^\infty \!\dd t\,\frac{{\rm e}^{-xt}}{t^n}\,.
\end{align}

\paragraph{Sterile charges} 
The helicity-odd off-diagonal elements $\delta\nodd_{ij}$ are crucial for the generation of flavoured asymmetries $q_{\ell a}$. The diagonal elements (in the mass basis), on the other hand, can be interpreted as sterile charges $q_N$, {\it cf.} Eq.~(\ref{def:sterile_av}).
Within the present approximations, they vanish at $z_{\rm osc}$, when the flavoured asymmetries are generated. 
To show this, we solve Eq.~(\ref{osc:weakwashoutav}) for diagonal, helicity-odd charge densities,
\begin{align}
\label{relaxation_force}
\frac{\dd}{\dd z}\delta \nodd_{ii}
=-(\Gamma_N)_{ii} \delta \nodd_{ii}+F_i(z)\,,
\end{align}
where
\begin{subequations}
\begin{align}
(\Gamma_N)_{ii}&=\gamma_{\rm av}\frac{a_{\rm R}}{T_{\rm ref}} {\rm Re[Y^*Y^t]}_{ii}\,,\\
F_i(z)&=-\gamma_{\rm av}\frac{a_{\rm R}}{T_{\rm ref}} 
\sum\limits_{\overset{j}{j\not=i}}
\left(
{\rm Re}[Y^*Y^t]_{ij}{\rm Re}[\delta \nodd_{ij}]
+{\rm Im}[Y^*Y^t]_{ij}{\rm Im}[\delta \neven_{ij}]
\right)\,.
\end{align}
\end{subequations}
The solutions~(\ref{sol:f0hav}) lead to
\begin{subequations}
\begin{align}
{\rm Re}[\delta \nodd_{ji}]&=-{\rm Im}[Y^*Y^t]_{ij}{\rm Im}[\mathcal{F}_{ji}]G\,, \\
{\rm Im}[\delta \neven_{ji}]&={\rm Re}[Y^*Y^t]_{ij}{\rm Im}[\mathcal{F}_{ji}]G\,,
\end{align}
\end{subequations}
such that, when using the symmetry properties of
the various tensors, $F_i(z)$ vanishes and consequently
so does $\delta \nodd_{ij}$ since we assume zero sterile charge as an initial condition.
In total this results in 
\begin{align}
\label{sol:sterile_av}
q_{Ni}=2\delta \nodd_{ii}=0\,,
\end{align}
which is valid at $\mathcal{O}(|Y^* Y^t|^2)$. In Appendix~\ref{app:2_3_flavours} we show that for $n_s=2$ sterile neutrino flavours this even holds to all orders. However, in case of $n_s\geq 3$ flavours, already at $\mathcal{O}(|Y^* Y^t|^3)$ there appears a non-vanishing contribution that is however negligible
in the oscillatory regime.

\paragraph{Asymmetries in Doublet Leptons and Sterile Neutrinos}
Likewise, in order to calculate the charge densities $\Delta_a$
in the oscillatory regime, we can neglect the washout term in Eq.~(\ref{Diff:Active}) during the initial production process around $z\sim z_{\rm osc}$. 
Since the generalised lepton number $\sum_a q_{\ell a} + \sum_i q_{Ni}$ is conserved when $T\gg M_i$ and we have previously shown that $q_{Ni}\simeq 0$ at $z\sim z_{\rm osc}$ in the oscillatory regime, we can conclude that $B\simeq0$ and $\Delta_a\simeq - q_{\ell a}$ at $z\sim z_{\rm osc}$.
This immediately leads to the solution
\begin{align}
\label{q:source}
\Delta_a(z)= -\int_0^z  \!\frac{\dd z'}{T_{\rm ref}}S_{a}\,.
\end{align}
Now, when neglecting the washout that only becomes important at later times, we can obtain
the flavoured lepton charge densities by substituting the source~(\ref{Source}) into Eq.~(\ref{q:source}).
To evaluate the resulting expression, we make use of the solutions~(\ref{sol:f0hav}) and integrate
\begin{align}
\label{eq:hypergeo}
\int\limits_0^z  \!\dd z'\, \Im\left[\mathcal{F}_{ij}(z')\right]=\frac{z^2}{2}\Im\,{}_2 F_2\left(\left\{\frac23,1\right\};\left\{\frac43,\frac53\right\};-\frac{\ii}{3} |\Omega_{ij}| z^3\right)\,\sign(M_{ii}^2-M_{jj}^2)\,,
\end{align}
with the generalised hypergeometric function 
\begin{align}
{}_pF_q(\{a_1,\dots,a_p\};\{b_1,\dots,b_q\};w)=\sum_{k=0}^\infty\prod_{i=1}^p\frac{\Gamma(k+a_i)}{\Gamma(a_i)}\prod_{j=1}^q\frac{\Gamma(b_j)}{\Gamma(k+b_j)}\frac{w^k}{k!}\,,
\end{align}
for $p,q\in \mathbb{N}_0$ and $w\in \mathbb{C}$, where $\Gamma(x)$  is the Gamma function. 
Because soon after the first few oscillations the  charges $\Delta_a$  saturate close to their maximal values $\Delta_a^{\rm sat}$, {\it cf.} also Figure~\ref{fig:weak_washout_example}, we can use
\begin{align}
\Delta_a(z)= -\int_0^z  \!\frac{\dd z'}{T_{\rm ref}}S_{a}\approx -\int_0^\infty \!\frac{\dd z'}{T_{\rm ref}}S_{a}\equiv \Delta_{a}^{\rm sat}\,,
\end{align}
where the approximation holds for $z$ moderately larger than $z_{\rm osc}$. 
On the other hand, as we have shown,
the diagonal  sterile charges $q_{Ni}$ are negligible at early times [{\it cf.} Eq.~(\ref{sol:sterile_av})], so that the only asymmetries present in the plasma are flavoured asymmetries in the SM fields.
To obtain these, we need the limit $z\to \infty$ of Eq.~(\ref{eq:hypergeo})
\begin{align}
\int\limits_0^\infty  \!\dd z\, \Im\left[\mathcal{F}_{ij}(z)\right]=-\frac{\pi^{\frac 12}\Gamma(\frac 16)}{2^{\frac23} 3^{\frac{4}{3}} |\Omega_{ij}|^{\frac 23}}\,\sign(M_{ii}^2-M_{jj}^2)\,.
\end{align}
Putting these elements together and dividing by the comoving entropy density
$s=2\pi^2 g_\star a_{\rm R}^3/45$, we obtain
\begin{align}
\label{sol:act_av}
\nonumber
\frac{\Delta_a^{\rm sat}}{s}&=\frac{\rm i}{g_\star^{\frac 53}}
\frac{3^{\frac{13}{3}} 5^{\frac53}\Gamma(\frac16)\zeta(3)^{\frac53}}{2^{\frac{8}{3}}\pi^{\frac{41}{6}}}
\sum\limits_{\overset{i,j,c}{i\not=j}}
\frac{
Y_{ai}^\dagger Y_{ic} Y_{cj}^\dagger Y_{ja}
}{{\rm sign}(M_{ii}^2-M_{jj}^2)}\left(\frac{m_{\rm Pl}^2}{|M_{ii}^2-M_{jj}^2|}\right)^{\frac23}\frac{\gamma_{\rm av}^2}{g_w}\\
&\approx
-\sum\limits_{\overset{i,j,c}{i\not=j}}
\frac{\Im[Y_{ai}^\dagger Y_{ic} Y_{cj}^\dagger Y_{ja}]}{{\rm sign}(M_{ii}^2-M_{jj}^2)}
\left(\frac{m_{\rm Pl}^2}{|M_{ii}^2-M_{jj}^2|}\right)^{\frac23}
\times 3.4
\times 10^{-4}\frac{\gamma_{\rm av}^2}{g_w}\,.
\end{align}

\begin{figure}
	\centering
	\includegraphics[scale=0.45]{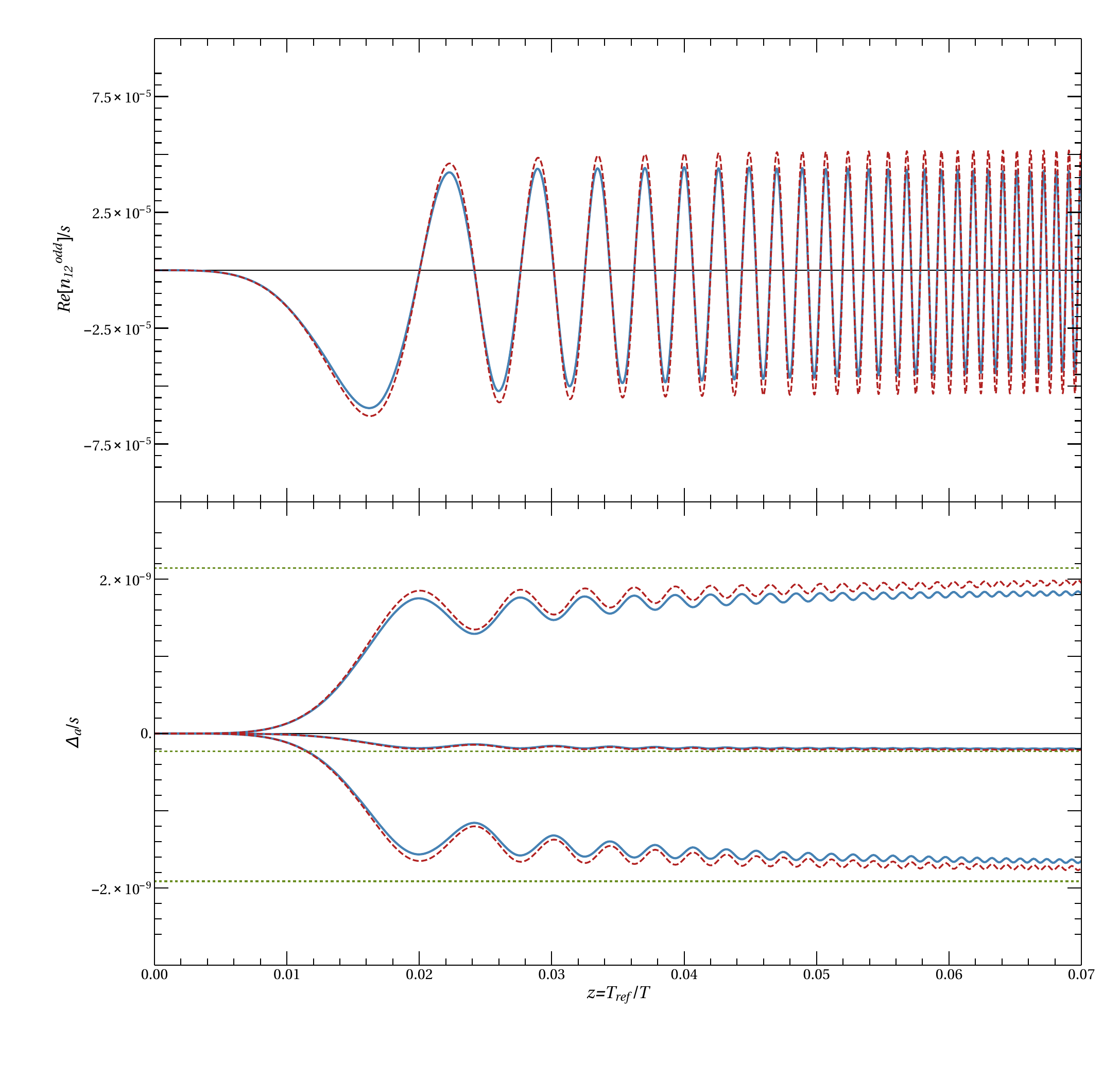}
\caption{Comparison of the numerical solution (blue, solid), to the approximate analytic result
(red, dashed) for the time evolution of the $CP$-violating correlation of the sterile neutrinos
$\Re[\delta n_{12}^\text{odd}]$ (upper panel), as well as the resulting time evolution of the three active
charges (blue, solid), compared to their saturation
limit given by Eq.~(\ref{q:source}) (red, dashed).
The approximation~(\ref{sol:act_av}) does not include washout effects since 
the washout time scale is assumed to be much later than the time scale of the
oscillations. 
Furthermore, backreaction of the produced asymmetries on the $N_i$ evolution, as well as effects due to thermal masses 
are neglected.
Note that the sum of the three charges $\Delta_a$ vanishes since lepton number
violation only occurs at order $|Y^*Y^t|^3$ when washout effects are included.}
	\label{fig:weakwashoutapprox}
\end{figure}

In Figure~\ref{fig:weakwashoutapprox} we compare the analytic results
for $\delta n^{\rm odd}_{12}$ as well as for the late-time asymmetries~(\ref{sol:act_av})
with the numerical solution. The discrepancies can be attributed to the fact that
backreaction and washout effects are neglected so far.
In a similar way as Figure~\ref{fig:weak_washout_example}, Figure~\ref{fig:weakwashoutapprox} also illustrates the validity of the
approximation in Eq.~(\ref{q:source}), where $z$ is taken to infinity, because
$\Delta_{a}$ indeed saturates after the first few oscillations.

\subsection{Late Time Washout}\label{sec:latetimewashout}

\begin{figure}
	\centering
	\includegraphics[scale=0.5]{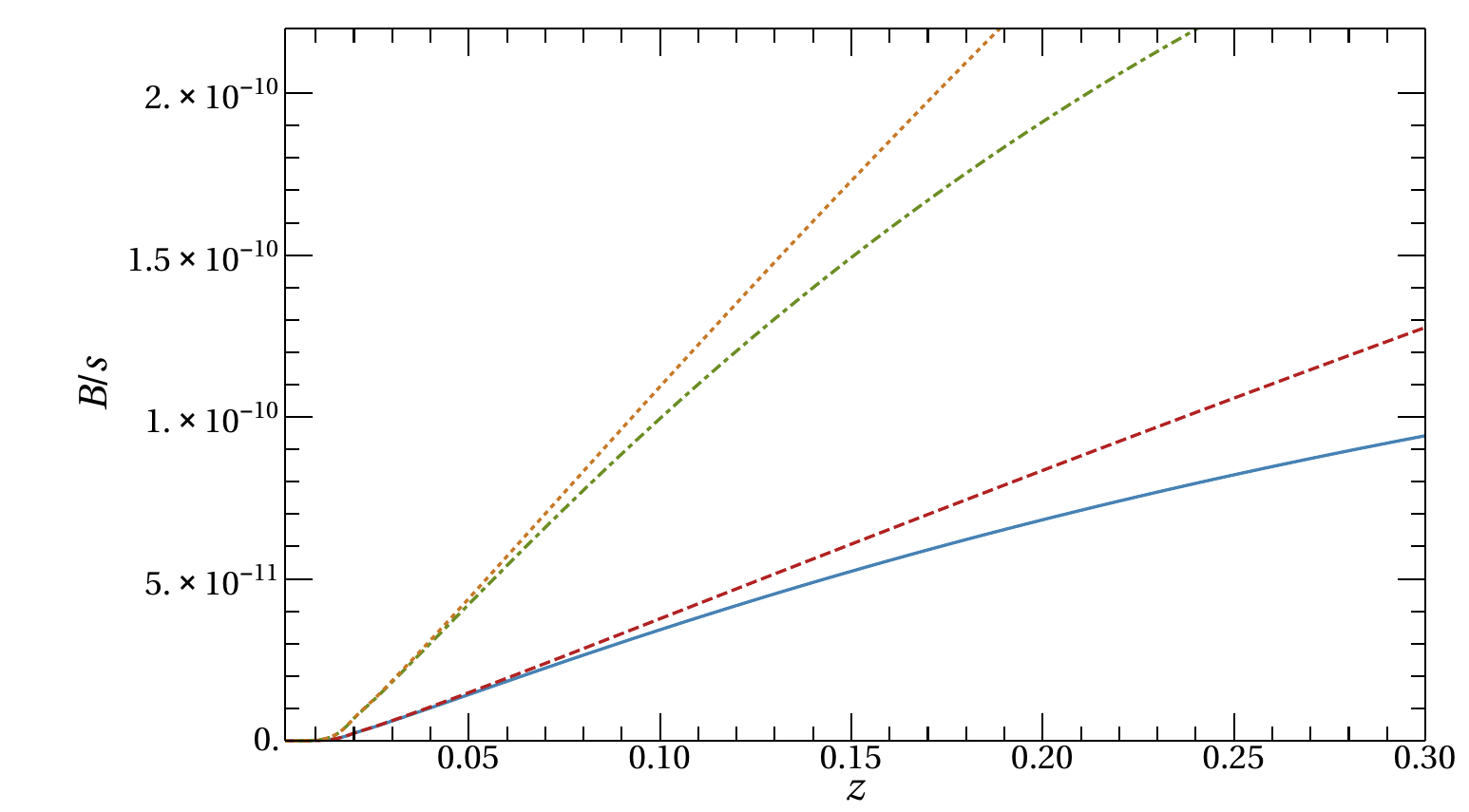}
	\caption{The numerical solution for the asymmetry $B/s$ in the 
oscillatory regime, with spectator and backreaction effects included (blue, solid) compared with the solutions without spectator effects (green, dot-dashed), without backreaction (red, dashed) and without spectator or backreaction effects (orange, dotted).}
	\label{fig:spectators_comparison}
\end{figure}

At late times, when $z\sim z_{\rm eq}$, we can neglect the oscillations of the sterile neutrinos
because they have already decayed or they are so rapid that their effect
averages out. In particular, there is no sizeable source for the flavoured asymmetries any more
and also no other effects from off-diagonal correlations of the sterile neutrinos.
This implies that the network of kinetic equations can be reduced to the following form
\begin{subequations}
\label{eq:washout}
\begin{align}
\label{eq:washout_1}
\frac{\dd \Delta_a}{\dd z}&=
\frac{\gamma_{\rm av}}{g_w} \frac{a_{\rm R}}{T_{\rm ref}}\sum_{i} Y_{ia}Y_{ai}^\dagger
\,\left( \sum_{b}(A_{ab} + C_b/2)\Delta_b-q_{Ni}\right)\,,
\\
\label{eq:washout_2}
\frac{\dd q_{Ni}}{\dd z}&=-\frac{a_{\rm R}}{T_{\rm ref}}\gamma_{\rm av}
\sum\limits_a Y_{ia}Y^\dagger_{ai}
\left(q_{Ni}-\sum_b (A_{ab} + C_b/2)\Delta_b\right)\,,
\end{align}
\end{subequations}
where we use $\Delta_{a}^{\rm sat}$  and $q_N=0$ as initial conditions.
Equation~(\ref{eq:washout_1}) is easily obtained from Eq.~(\ref{Diff:Active}) when dropping the source term. In order arrive at Eq.~(\ref{eq:washout_2}), we
keep the decay term $\Gamma_N$ as well as the backreaction term $\tilde{\Gamma}_N$,
while dropping the commutator in Eq.~(\ref{Diff:Sterile}) and solve it for the helicity-odd, diagonal charges.
This procedure is justified since the oscillations of the sterile charges around $z=z_{\rm eq}$ are fast enough
for their effect to average out.
Note that the backreaction terms can be identified with the contributions involving $q_{Ni}$ in Eq.~(\ref{eq:washout_1}) as well as $\Delta_{ b}$ and $q_\phi$ in Eq.~(\ref{eq:washout_2}).
In Figure~\ref{fig:spectators_comparison} the effect of the backreaction and spectator effects is presented
where in particular the latter can have a substantial impact on the final result.
The matrix $A$ and the vector $C$ appearing here specify the way how the spectator processes redistribute charges in the SM. Spectator processes have been neglected in most studies to date (except \cite{Shuve:2014zua}), which corresponds to setting $C=0$ and $A=-1$. 
The importance of including spectator effects is more pronounced than for conventional leptogenesis without flavour effects~\cite{Garbrecht:2014kda}
because in the present scenario, the asymmetries are purely flavoured and the net result
is due to an incomplete cancellation in the relation~(\ref{baryon_charge_Delta}) that is rather
sensitive to corrections in the individual terms.

\begin{figure}
	\centering
	\includegraphics[scale=0.5]{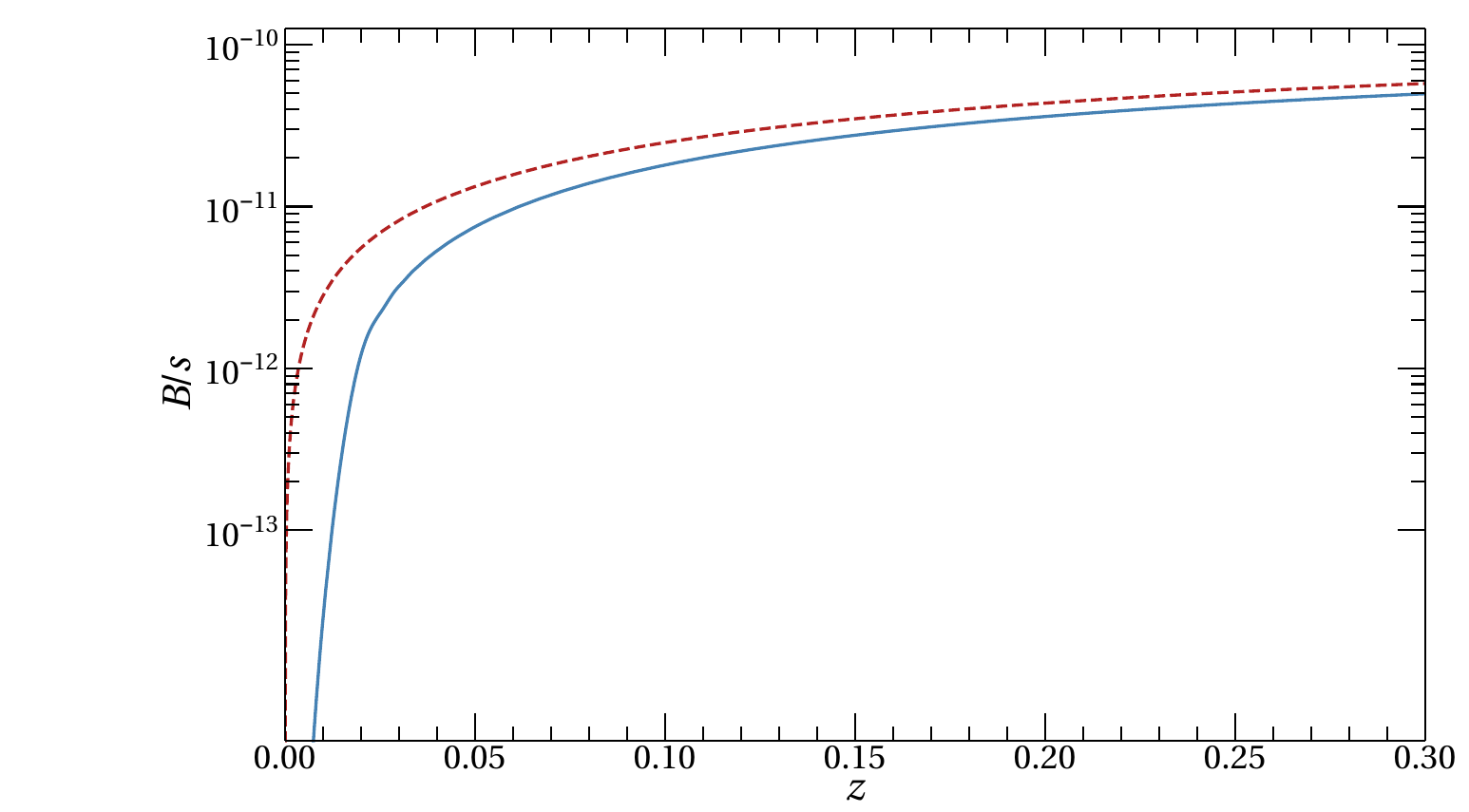}
	\caption{Comparison of the analytic treatment of the baryon asymmetry $B/s$ (red, dashed) in the oscillatory regime to the numerical solution (blues, solid).}
	\label{fig:analytic_numeric_washouts}
\end{figure}

Due to the hierarchy $z_{\rm osc}\ll z_{\rm eq}$, we can use the charge densities generated through sterile oscillations
around $z\sim z_{\rm osc}$, {\it cf.} Eqs.~(\ref{sol:act_av}) and~(\ref{sol:sterile_av}), as initial conditions for solving the equations governing the washout process. For $n_s$ sterile flavours we can reduce  Eqs.~(\ref{eq:washout}) to a linear first-order differential equation for $(3+n_s)$-dimensional vectors $V_{\Delta N}=(\Delta^t, q_N^t)^t$,
\begin{align}
\frac{\dd}{\dd z}
V_{\Delta N}=\frac{a_{\rm R}}{T_{\rm ref}}\gamma_{\rm av}
K V_{\Delta N}
\,, \quad \quad K=\left(
\begin{array}{ccc}
K^{\Delta \Delta} & K^{\Delta N} \\
K^{N \Delta} & K^{N N} 
\end{array}
\right)\,,
\end{align}
where the components of the matrices $K^{\Delta \Delta}, K^{\Delta N}, K^{N \Delta}$ and  $K^{N N}$ read

\begin{align}
\nonumber
K^{\Delta \Delta}_{ab}&=\frac{1}{g_w}\sum_{k=1}^{n_s} Y_{ak}^\dagger Y_{ka}\left(A_{ab}+\frac{1}{2}\right)\,, \quad  \quad
K^{\Delta N}_{aj}=-\frac{1}{g_w} Y_{aj}^\dagger Y_{ja}\,,\\
K^{N \Delta}_{ib}&=\sum_{d=1}^3Y_{id}Y_{di}^\dagger \left(A_{db}+\frac{1}{2}C_b\right)\,,\quad \quad\quad\quad\,
K^{N N}_{ij}=-\sum_{d=1}^3 Y_{id}Y_{di}^\dagger \delta_{ij}\,,
\end{align}
with $i,j=1,2,\dots,n_s$ sterile and $a,b=1,2,3$ active flavours. Here $A$ and $C$ as defined in Eq.~(\ref{spectator}) account for the spectator processes.
After diagonalising the Matrix $K$
\begin{align}
K^{\rm diag}={\rm T}^{-1}K {\rm  T}\,,
\end{align}
where ${\rm T}$ is a transformation matrix with the eigenvectors of $K$ as column vectors, we are left with the solution
\begin{align}
\left(
\begin{array}{ccc}
\Delta(z) \\
q_N(z)
\end{array}
\right)={\rm T}\,\exp\left(\frac{a_{\rm R}}{T_{\rm ref}}\gamma_{\rm av}  K^{\rm diag} \,z\right){\rm T}^{-1}
\left(
\begin{array}{ccc}
\Delta^{\rm in} \\
q_N^{\rm in}
\end{array}
\right)\,,
\end{align}
with $\Delta^{\rm in}=\Delta^{\rm sat}$ and $q_{N}^{\rm in}=0$ the asymmetries generated during the oscillation process at early times
$z\sim z_{\rm osc}$, {\it cf.} Eqs.~(\ref{sol:act_av}) and~(\ref{sol:sterile_av}).
As the washout processes are suppressed during the initial creation of
the asymmetries and because of relation $z_{\rm osc}\ll z_{\rm eq}$, we can impose
these initial conditions at $z=0$. The baryon charge $B$ gets frozen in as soon as the weak sphalerons freeze out. Since we choose the reference temperature $T_{\rm ref}$ such that this occurs when $z=1$, it follows from Eq.~(\ref{baryon_charge_Delta})
\begin{align}\label{BapproxOscillating}
B=\frac{28}{79}
[\Delta_1(z)+\Delta_2(z)+\Delta_3(z)]_{z=1}\,.
\end{align}
A comparison of the evolution of the baryon asymmetry in the analytic treatment with the full numerical solution is shown in Figure~\ref{fig:analytic_numeric_washouts}.

\section{Overdamped Regime}
\label{sec:str}

\begin{figure}
	\centering
	\includegraphics[scale=0.6]{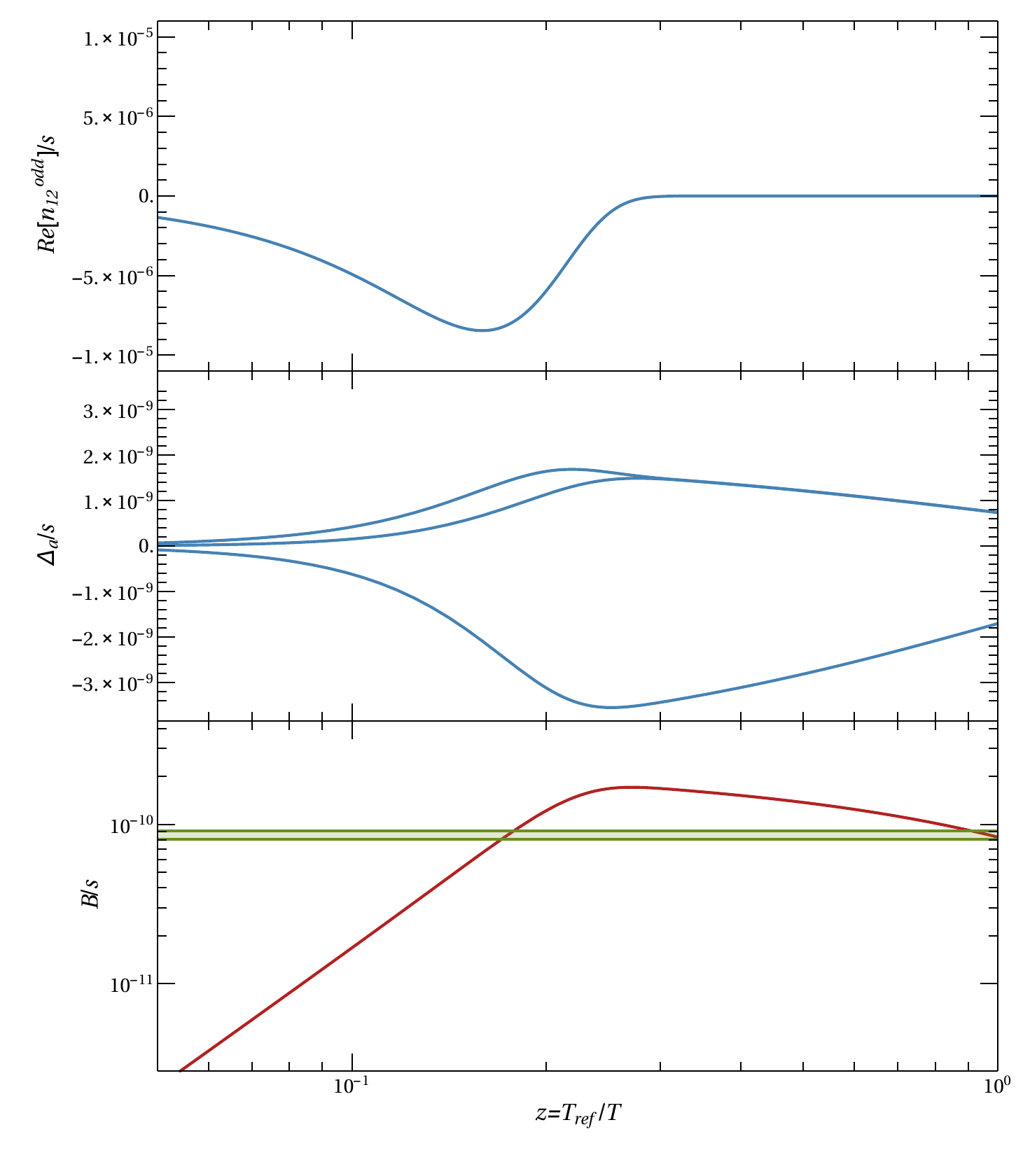}
	\caption{This example plot shows the production of the baryon asymmetry  $B/s$ (bottom panel) in the overdamped regime for two sterile flavours.
The top panel shows the helicity-odd part of the correlation $\delta n_{12}$. In comparison to the oscillatory regime, see Figure~\ref{fig:weak_washout_example}, this oscillation happens rather late and is overdamped. The generation of the SM charges $\Delta_a/s$ is shown in the middle panel. The bottom panel show the resulting baryon asymmetry, where the green band indicates the error bars of the observed value.
}
	\label{fig:strong_washout_example}
\end{figure}

There are phenomenologically interesting parameter choices where the equilibration of one of the heavy neutrino interaction eigenstates happens before the first oscillation is completed, leading to an overdamped behaviour of the oscillations.
This is particularly important in the case of mass-degenerate heavy
neutrinos, for which the first oscillation can happen at times as late as
sphaleron freezeout, and in scenarios in which the $Y_{ia}$ are much larger than the naive seesaw expectation (\ref{F0}).
Both of this can \emph{e.g.} be motivated in scenarios with an approximate $B-L$ conservation.
In these scenarios one eigenvalue of $Y Y^\dagger$ is always much smaller than the other, see Appendix \ref{App:Seesaw}, so that one interaction eigenstate couples only very feebly to the plasma. Instead of being produced through direct scatterings, the feebly coupled state gets populated through oscillations with a sterile neutrino that has already equilibrated. 
Using the same perturbative approximation as in the oscillatory regime is no
longer justified, because the larger decay rate cannot be treated as a small
perturbation to the vacuum oscillation any more.
Instead, we use a quasi-static approximation in a similar manner to
applications to resonant leptogenesis from $N_i$ decay
\cite{Garbrecht:2014aga, Iso:2014afa}.
In the following we derive analytic expressions to treat the overdamped regime for $n_s=2$. 
Throughout this computation, we work in the interaction basis of the sterile neutrinos.
An example plot for the generation of net baryon charge in the overdamped regime for two sterile flavours is shown in Figure~\ref{fig:strong_washout_example}.

\subsection{Source of the Asymmetry}
In the interaction basis, where $Y Y^\dagger$ is diagonal, the fact that one interaction state decouples in the $B-L$ conserving limit implies
that we can write the Yukawa couplings and the right-handed neutrino masses as:
\begin{align}\label{DefEpsMu}
	Y^\dagger&=
\begin{pmatrix}
Y_e & \epsilon_e \\
Y_\mu & \epsilon_\mu \\
Y_\tau & \epsilon_\tau
\end{pmatrix}
\,,
&&M=
\begin{pmatrix}
 \upmu_1 & \bar{M}\\
 \bar{M} & \upmu_2
\end{pmatrix}\,,
\end{align}
see Appendix~\ref{App:Seesaw}. In the interaction basis we have therefore
$\sum_a |Y_{a}|^2 \gg \sum_b |\epsilon_{b}|^2$, as well as $\sum_a Y_{a}^* \epsilon_a =0$, as the matrix $Y Y^\dagger$ is diagonal.
We can treat the smaller Yukawa coupling $|\epsilon_{2a}|$
as an expansion parameter throughout the following calculation.
We will solve the equations for the positive helicity distribution $\delta n_{+,ij}$, while
all remaining distributions can
be obtained through complex conjugation of the mass and Yukawa matrices.

The momentum averaged sterile neutrino decay matrix $\Gamma_N$
inherits the flavour structure of the
Yukawa matrices $YY^\dagger$. Therefore, in the  interaction basis the decay rate $\Gamma_N$ 
as well as the thermal mass matrix $H_N^{\rm th}$ are both diagonal:
\bse
\begin{align}
    \Gamma_N &=
	\gamma_{\rm av} \frac{a_{\rm R}}{T_{\rm ref}}
	\begin{pmatrix}
	\sum_a |Y_a|^2 & 0 
	\\ 0 & \sum_a |\epsilon_a|^2
	\end{pmatrix}
	\,,\\
	H_N^{\rm th} &=
	\mathfrak{h}_{\rm th} \frac{a_{\rm R}}{T_{\rm ref}}
	\begin{pmatrix}
	\sum_a |Y_a|^2 & 0 
	\\ 0 & \sum_a |\epsilon_a|^2
	\end{pmatrix}\,,
\end{align}
\ese
From now we neglect the smaller eigenvalue, \emph{i.e.} all terms of 
$\mathcal{O}\left( |\epsilon_a|^2 \right)$.
The contribution to the effective Hamiltonian from the
vacuum mass matrix $H_N^{\rm vac}$ is not necessarily diagonal in the interaction basis, {\it i.e.}
it takes the form
\begin{align}
    H_N^{\rm vac}=
	\frac{\pi^2}{18 \zeta(3)}\frac{a_{\rm R}}{T_{\rm ref}^3}
    \begin{pmatrix}
		\bar{M}^2 + |\mu_1|^2 & \bar{M}(\mu_1 + \mu_2^*)
		\\ \bar{M}(\mu_1^* + \mu_2) & \bar{M}^2 + |\mu_2|^2
    \end{pmatrix}\,.
\end{align}
Note that we have not yet expanded in $\mu_{1,2}$ in order to keep equations valid in a more general case as well.
We consider the regime where the equilibration of $N_1$ happens before the oscillations between the sterile
flavours begin, which means that the rate at which $\delta n_{11}$ reaches it's quasi-static value is much faster than the rate of the oscillations,
\begin{align}
	\frac{z_{\rm eq}}{z_{\rm osc}} =
	\frac{\sqrt[3]{|M^2_{1}-M^2_{2}|/a_{\rm R}^2}}
	{\gamma_{\rm av} \sum_a |Y_a|^2}
	\ll 1\,.
\end{align}

We separate the evolution equations into the directly damped equations, containing
$[ YY^\dagger]_{11}$,
\bse
\label{osc2x2:equilibrated}
\begin{align}
\frac{\dd \delta n_{11}}{\dd z} &= - (\Gamma_N)_{11} \delta n_{11}
	-\frac{\ii}{2} z^2
	\left[ (H_N^{\rm vac})_{12} \delta n_{21} - (H_N^{\rm vac})^*_{12} \delta n_{12}\right]\,,\\
	\frac{\dd \delta n_{12}}{\dd z} &= - \frac{(\Gamma_N)_{11}}{2} \delta n_{12}
	- \ii \frac{(H_N^{\rm th})_{11}}{2} \delta n_{12}
    - \frac{\ii}{2} z^2 \sum_k \left[ (H_N^{\rm vac})_{1k} \delta n_{k2} -
	\delta n_{1k} (H_N^{\rm vac})_{k 2} \right]\,,
\end{align}
\ese
and the ones that are damped indirectly, through mixing with other sterile flavours,
\begin{align}
\label{osc2x2:noneq}
 \frac{\dd \delta n_{22}}{\dd z} &= - \frac{\ii}{2} z^2 \left[ (H_N^{\rm vac})^*_{12} \delta n_{12} -
	 (H_N^{\rm vac})_{12} \delta n_{21}\right]\,.
\end{align}
At this point we make the quasi-static approximation~\cite{Garbrecht:2014aga, Iso:2014afa}
to the solutions of Eqs.~(\ref{osc2x2:equilibrated}) by assuming that the interactions of
the highly damped neutrino $N_1$ and its flavour correlations
instantaneously reach values that are determined by the deviation of the
feebly coupled state $N_2$ from equilibrium, {\it i.e.}
\begin{align}
	\label{osc2x2:quasistaticcond}
\dd \delta n_{11}/\dd z\,=\,\dd \delta n_{12}/\dd z\,=\,\dd \delta n_{21}/\dd z\,\approx\, 0\,,
\end{align}
which allows us to express $\delta n_{11}$, $\delta n_{12}$,
and $\delta n_{21}=\delta n_{12}^*$ in terms of $\delta n_{22}$,
\bse
\begin{align}
	\label{osc2x2:n11n12}
\delta n_{11} &= \frac{z^4 |(H_N^{\rm vac})_{12}|^2}
{(\Gamma_N)_{11}^2+(H_N^{\rm th})_{11}^2+ z^2 2(H_N^{\rm th})_{11}
\left[ (H_N^{\rm vac})_{11}- (H_N^{\rm vac})_{22} \right]
+ z^4 (\tilde{H}_N^{\rm vac})^2} \delta n_{22}\,,
\\
\delta n_{12} &=-\frac{z^2 (H_N^{\rm vac})_{12} \left\{\ii (\Gamma_N)_{11}+(H_N^{\rm th})_{11}+z^2
\left[ (H_N^{\rm vac})_{11}- (H_N^{\rm vac})_{22} \right]
\right\}}
{(\Gamma_N)_{11}^2+(H_N^{\rm th})_{11}^2+ z^2 2(H_N^{\rm th})_{11}
\left[ (H_N^{\rm vac})_{11}- (H_N^{\rm vac})_{22} \right]
+ z^4 (\tilde{H}_N^{\rm vac})^2} \delta n_{22}\,,
\end{align}
\ese
where we have introduced
\begin{align*}
	(\tilde{H}_N^{\rm vac})^2 \equiv|(H_N^{\rm vac})_{12}|^2 +
	\left[ (H_N^{\rm vac})_{11}- (H_N^{\rm vac})_{22} \right]^2\,.
\end{align*}

Inserting these results into the equation for the weakly washed-out sterile neutrino $N_2$ yields the
differential equation
\begin{align}
\label{osc2x2:n22}
\notag
\frac{\dd \delta n_{22}}{\dd z} &= - \frac{z^4 |(H_N^{\rm vac})_{12}|^2 (\Gamma_N)_{11}}
{(\Gamma_N)_{11}^2+(H_N^{\rm th})_{11}+ z^2 2(H_N^{\rm th})_{11}
\left[ (H_N^{\rm vac})_{11}- (H_N^{\rm vac})_{22} \right]
+ z^4 (\tilde{H}_N^{\rm vac})^2} \delta n_{22}
\\ 
&=- (\Gamma_N)_{11} \frac{|(H_N^{\rm vac})_{12}|^2}
{(\tilde{H}_N^{\rm vac})^2}
\frac{z^4}{(z^2+\Deltaz^2)(z^2+\Deltaz^{*2})}
\delta n_{22}\,,
\end{align}
with the parameter
\begin{align}
	\Deltaz = \sqrt{\frac{(H_N^{\rm th})_{11}}{\tilde{H}_N^{\rm vac}}
	\left[ \frac{(H_N^{\rm vac})_{11}-(H_N^{\rm vac})_{22}}{ \tilde{H}_N^{\rm vac}}+
		\ii \sqrt{\frac{|(H_N^{\rm vac})_{12}|^2}{(\tilde{H}_N^{\rm vac})^2}+
\frac{\gamma_{\rm av}^2}{\mathfrak{h}_{\rm th}^2}}\right]}\,.
\end{align}
Its absolute value introduces a new time scale
\begin{align}
	|\Deltaz| = \sqrt{\frac{(H_N^{\rm th})_{11}}{\tilde{H}_N^{\rm vac}}}\sqrt[4]{1+\frac{\gamma_{\rm av}^2}{
		\mathfrak{h}_{\rm th}^2}}
		\sim z_{\rm osc} \sqrt{\frac{z_{\rm osc}}{z_{\rm eq}}
		\frac{\mathfrak{h}_{\rm th}}{\gamma_{\rm av}}} \gg z_{\rm osc}
		\,.
\end{align}
The time scale $|\Deltaz|$ indicates the instance when the vacuum part of the
Hamiltonian $z^2 H^{\rm vac}_{N}$ becomes comparable to the thermal contribution $H^{\rm th}_{N}$.
The general solution to Eq.~(\ref{osc2x2:n22}) is given by
\begin{align}
	\delta n_{22}= \delta n_{22}(0) \exp\left\{-(\Gamma_N)_{11} \frac{|(H_N^{\rm vac})_{12}|^2}
	{(\tilde{H}_N^{\rm vac})^2}
    \left[z-\frac{\Im \left(\Deltaz^3 \arctan \frac{z}{\Deltaz}\right)}{\Im\Deltaz^2}\right]\right\}\,.
\end{align}
For times $z\ll |\Deltaz|$, we can approximate this solution by
\begin{align}
	\delta n_{22} \approx \delta n_{22}(0) \exp\left( -(\Gamma_N)_{11} \frac{|(H_N^{\rm vac})_{12}|^2}
	{(\tilde{H}_N^{\rm vac})^2} \frac{z^5}{5 |\Deltaz|^4}\right)\,,
	\label{osc2x2:n22simplsol}
\end{align}
which results in the equilibration time-scale for $N_2$
\begin{align}
	z_{N_2}^{\mathrm{eq}} = |\Deltaz| \sqrt[5]{
		\frac{5}
	{(\Gamma_N)_{11}|\Deltaz|}
	\frac{(\tilde{H}_N^{\rm vac})^2}{|(H_N^{\rm vac})_{12}|^2}}\,.
\end{align}
Therefore, unless $|(H_N^{\rm vac})_{12}|^2 \ll (\tilde{H}_N^{\rm vac})^2$, $N_2$ will reach equilibrium before
$|\Deltaz|$, justifying the usage of Eq.~(\ref{osc2x2:n22simplsol}). Note that
this situation naturally occurs in the pseudo-Dirac scenario, where the flavour and mass bases are maximally
misaligned, such that $(H_N^{\rm vac})_{11}=(H_N^{\rm vac})_{22}$.
Furthermore, in the pseudo-Dirac scenario one can also expand in
$\mu_{1,2}\ll \bar{M}$, leading to a simplified expression for the equilibration time-scale:
\begin{align}
	z_{N_2}^{\mathrm{eq}} =
	\sqrt[5]{\frac{405 \zeta^2(3) \mathfrak{h}_{\rm th}^2}{\pi^2 \gamma_{\rm av}} \frac{T_{\rm ref}^5 \sum_a |Y_a|^2}{a_{\rm R} \bar{M}^2 \mu^2}}
\end{align}
with $\mu=|\mu_1+\mu_2^*|/2=|M_1^2-M_2^2|/(4 \bar{M})$, and $\bar{M}^2= (M_1^2+M_2^2)/2$.

\begin{figure}
	\centering
	\includegraphics[scale=0.6]{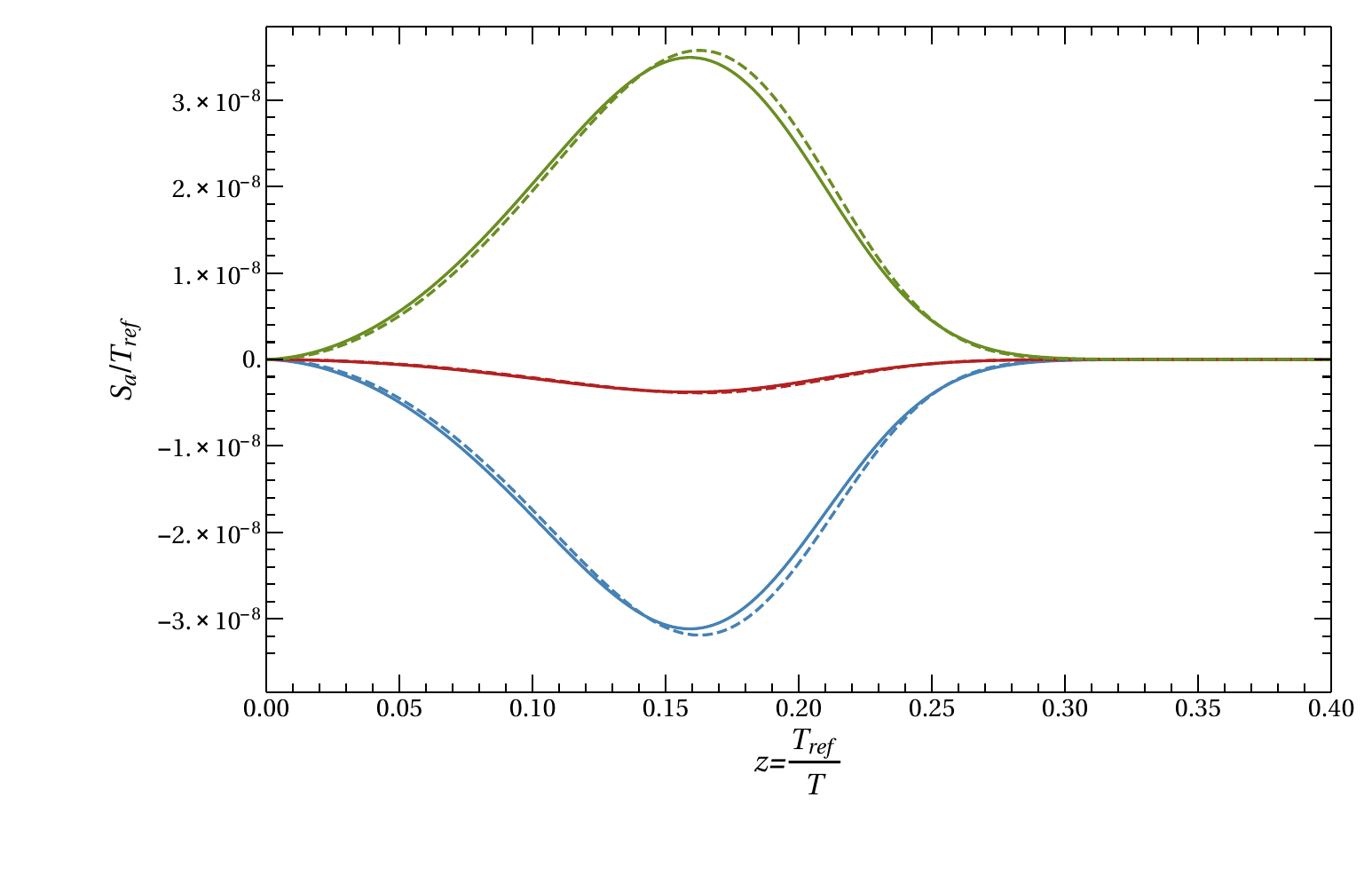}
	\caption{Source of the lepton asymmetries for the three SM flavours calculated numerically (solid) and analytically (dashed).}
	\label{fig:sources}
\end{figure}

The source of the lepton asymmetry is caused by the $CP$-odd correlation
\begin{align}
	\delta n_{+\,12} - \delta n_{-\,12}^* = - \frac{2 z^2 \ii (H_N^{\rm vac})_{12} (\Gamma_N)_{11}}
	{(\tilde{H}_N^{\rm vac})^2 (z^2+\Deltaz^2)(z^2+\Deltaz^{*2})} \delta n_{22}(z)\,,
\label{osc2x2:n12cpo}
\end{align}
which yields the source term
\begin{align}
\label{eq:source_sw}
\nonumber
	S_{a} &= a_{\rm R} \frac{\gamma_{\rm av}}{g_w}
	\sum\limits_{\overset{i,j}{i\not=j}} Y^*_{ia} Y_{ja} \left( \delta n_{+\,ij} - \delta n^*_{-\,ij} \right)\\ 
	&=4 
	\frac{\gamma_{\rm av}^2 a_{\rm R}^2}{g_w T_{\rm ref}}
	\frac{\sum_b |Y_{b}|^2}{(\tilde{H}_N^{\rm vac})^2}\frac{z^2}{|z^2+\Deltaz^2|^2}
	{\rm Im}\left[Y^*_{a} (H_N^{\rm vac})_{12} \epsilon_{a} \right]\delta n_{22}(z)\end{align}
that is non-vanishing only at first order in the smaller Yukawa $|\epsilon_{a}|$. The $z$ dependence of the source term divided by $T_{\rm ref}$ and the
entropy density $s$ is shown in Figure~\ref{fig:sources}. Note that the trace of the source $\sum_a S_a$ vanishes as we have
$\sum_a Y_a \epsilon^*_a=0$ in the interaction basis.
In the limit $\mu \ll \bar{M}$, the source term further simplifies to:
\begin{align}
	\frac{S_{a}}{s\,T_{\rm ref}}&\approx
	-\frac{45 \sqrt{5}}{g_\star^{3/2} g_w 4 \pi^{7/2}} \frac{\gamma_{\rm av}^2}{\mathfrak{h}_{\rm th}^2}
	\frac{m_{\rm Pl} \bar{M} \mu}{T_{\rm ref}^3} \frac{\Im[ Y^*_{a} \epsilon_{a}]}{\sum_b |Y_b|^2}
	z^2 \exp\left( -\frac{z^5}{{z_{N_2}^{\mathrm{eq}}}^5} \right)\\
	&= -5.65 \times 10^{-7} \times \frac{m_{\rm Pl} \bar{M} \mu}{T_{\rm ref}^3} \frac{\Im[ Y^*_{a} \epsilon_{a}]}{\sum_b |Y_b|^2}
	z^2 \exp\left( -\frac{z^5}{{z_{N_2}^{\mathrm{eq}}}^5} \right)
	\,.
	\label{eq:source_swpd}
\end{align}

\paragraph{Validity of the Approximations}
For times $(\Gamma_N)_{11}^{-1}\ll z\ll|\Deltaz|$
\footnote{Note however, that in presence of another non-vanishing charge that contributes to
the size of $\delta n$, \textit{e.g.}
	$\Delta_a$, its derivatives
	will be proportional to the derivatives of $\Delta_a$, which may further
	extend the validity of the overdamped approximation, as it is the case
	for $\delta n^{\rm odd}$ once we include the backreaction of the
	active charges.},
Eq.~(\ref{osc2x2:n22simplsol}) implies that  $\dd \delta n_{22}/\dd z$ is small. Furthermore, we can approximate
\bse
\begin{align}
\delta n_{11} &= \frac{|(H_N^{\rm vac})_{12}|^2}{(\tilde{H}_N^{\rm vac})^2} \frac{z^4}{|\Deltaz|^4}\delta n_{22}\,,
\\
\delta n_{12} &=-\frac{(H_N^{\rm vac})_{12}}{(\tilde{H}_N^{\rm vac})^2} \frac{z^2}{|\Deltaz|^4}\left[ (H_N^{\rm th})_{11}+\ii (\Gamma_N)_{11} \right]
\delta n_{22}\,.
\end{align}
\ese
Hence, it is straightforward to see that the assumption made in Eq.~(\ref{osc2x2:quasistaticcond}) is justified in this regime, as the derivatives of $\delta n_{11}$ and $\delta n_{12}$
are much smaller than any of the individual terms on the right hand sides of Eq.~(\ref{osc2x2:equilibrated}),
\bse
\begin{align}
	\frac{\dd \delta n_{11}}{\dd z} &= \frac{4}{z}\delta n_{11} + \frac{\dd \delta n_{22}}{\dd z}
	\frac{\delta n_{11}}{\delta n_{22}} \ll (\Gamma_N)_{11} \delta n_{11}\,,\\
	\frac{\dd \delta n_{12}}{\dd z} &=
	\frac{2}{z}\delta n_{12} + \frac{\dd \delta n_{22}}{\dd z}
	\frac{\delta n_{12}}{\delta n_{22}} \ll (\Gamma_N)_{11} \delta n_{12}\,.
\end{align}
\ese

\subsection{Time Evolution of the SM Charges in the Overdamped Regime}
At least one of the damping rates for the charges $\Delta_a$
 is of the same order in $|Y_{1a}|^2$
as the larger of the sterile neutrino production rates. This implies that the
washout of the active leptons typically happens at the same time as the overdamped
oscillation of the sterile neutrinos.
Neglecting the backreaction of the active flavours onto the sterile sector,
as suitable for the oscillatory regime during the initial production
of the asymmetries, is no longer an applicable
approximation here.
However, as all  charges $\Delta_a$ are of first order in the smaller Yukawa
coupling $|Y_{2a}|$, see Eq.~(\ref{eq:source_sw}), the calculation of the sterile charges at
\emph{zeroth} order in $|Y_{2a}|$ remains unchanged.
To correctly describe the evolution of the  charge $\Delta_a$, one has to solve the whole set of coupled differential
equations at first order in $|Y_{2a}|$.

\begin{figure}
	\centering
	\includegraphics[scale=0.6]{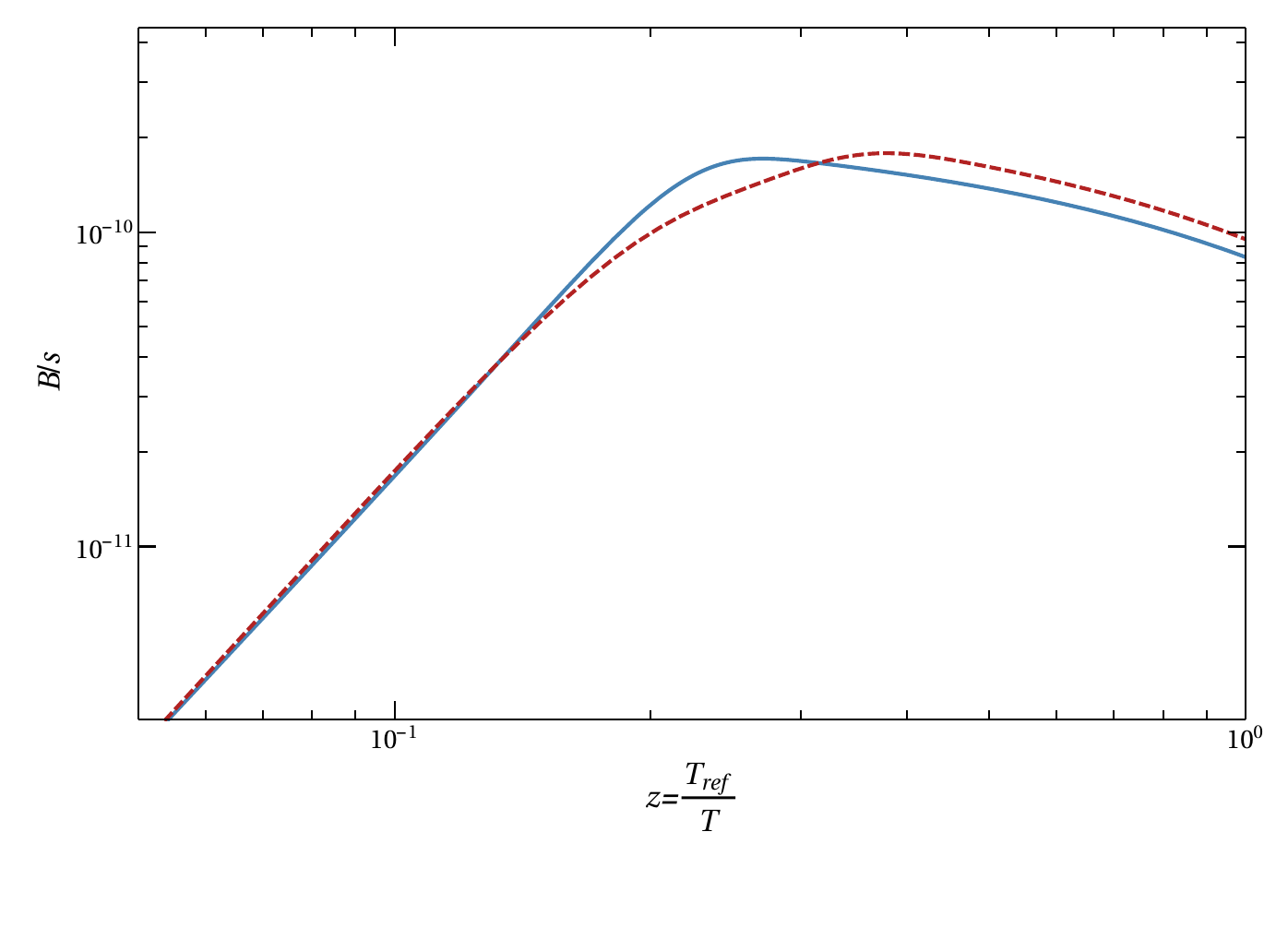}
	\caption{Total baryon asymmetry calculated numerically (blue,full)
	and analytically (red,dashed).}
	\label{fig:lept_asym} 
\end{figure}

\begin{figure}
	\centering
	\includegraphics[scale=0.6]{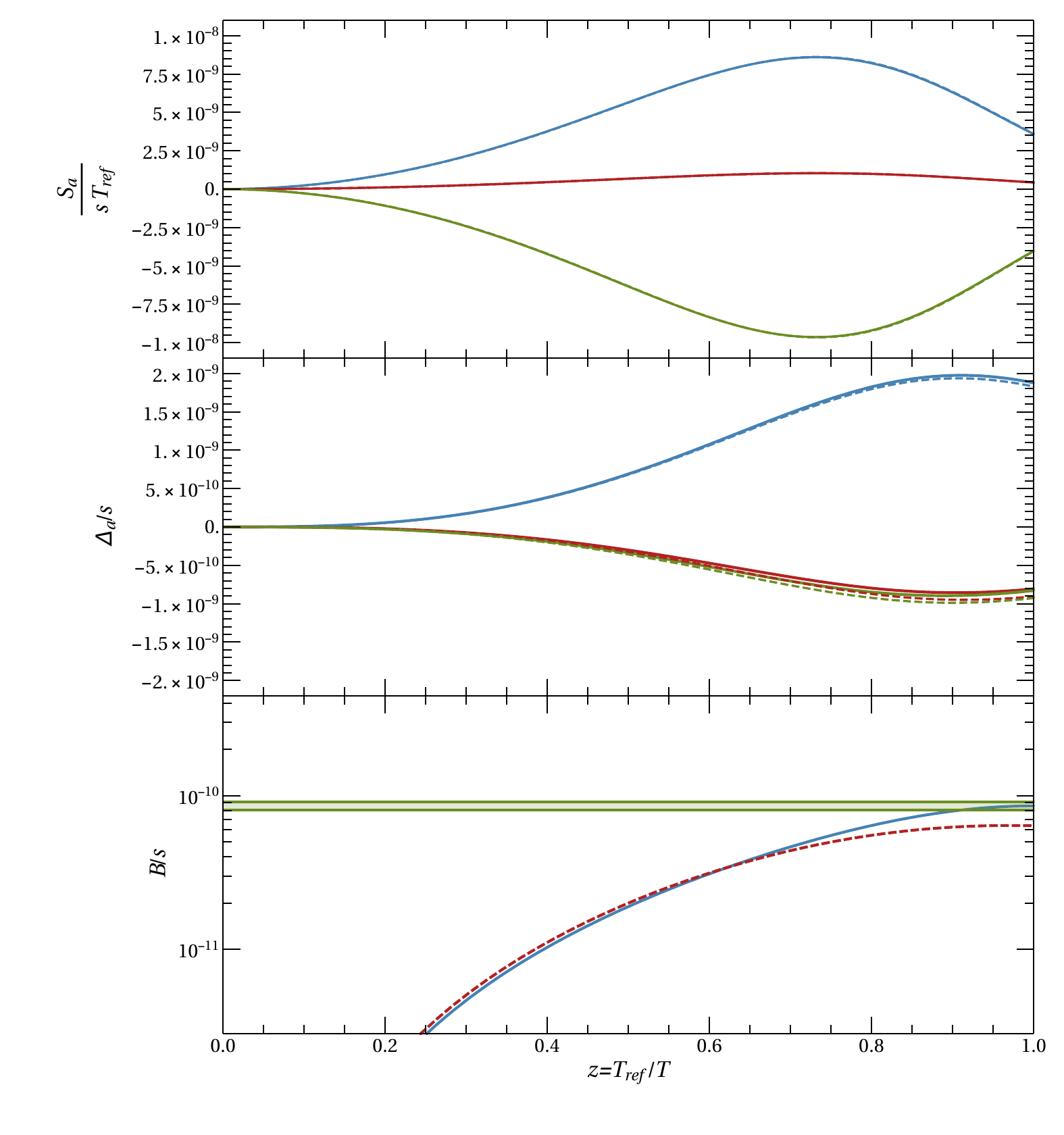}
	\caption{The comparison between numerical and analytic solutions for the source term, individual lepton charges and the baryon asymmetry
		for parameter choices that lead to maximal mixing angles for right-handed neutrino masses of $\bar{M}=1\GeV$ in the case of normal hierarchy.
	The analytical approximations are always presented with a dashed line, for the source term they are indistinguishable from
	the numerical result. The parameters used for this plot are
	$\Delta M^2=4.002 \times 10^{-8} \bar{M}^2$, $\omega=\frac{5 \pi}{4}+5.26\ii$, $\alpha_1=0$, $\alpha_2=0$,$\delta=\pi/2$, and the discrete
	parameter $\xi=1$. The small $CP$-violating parameters are $\mu=1.001 \times 10^{-8} \bar {M}$ and
	$\sum_a |\epsilon_a|^2= 3.65 \times 10^{-10} \sum_a |Y_a|^2$.}
	\label{fig:maxmixingNO_test} 
\end{figure}

\begin{figure}
	\centering
	\includegraphics[scale=0.6]{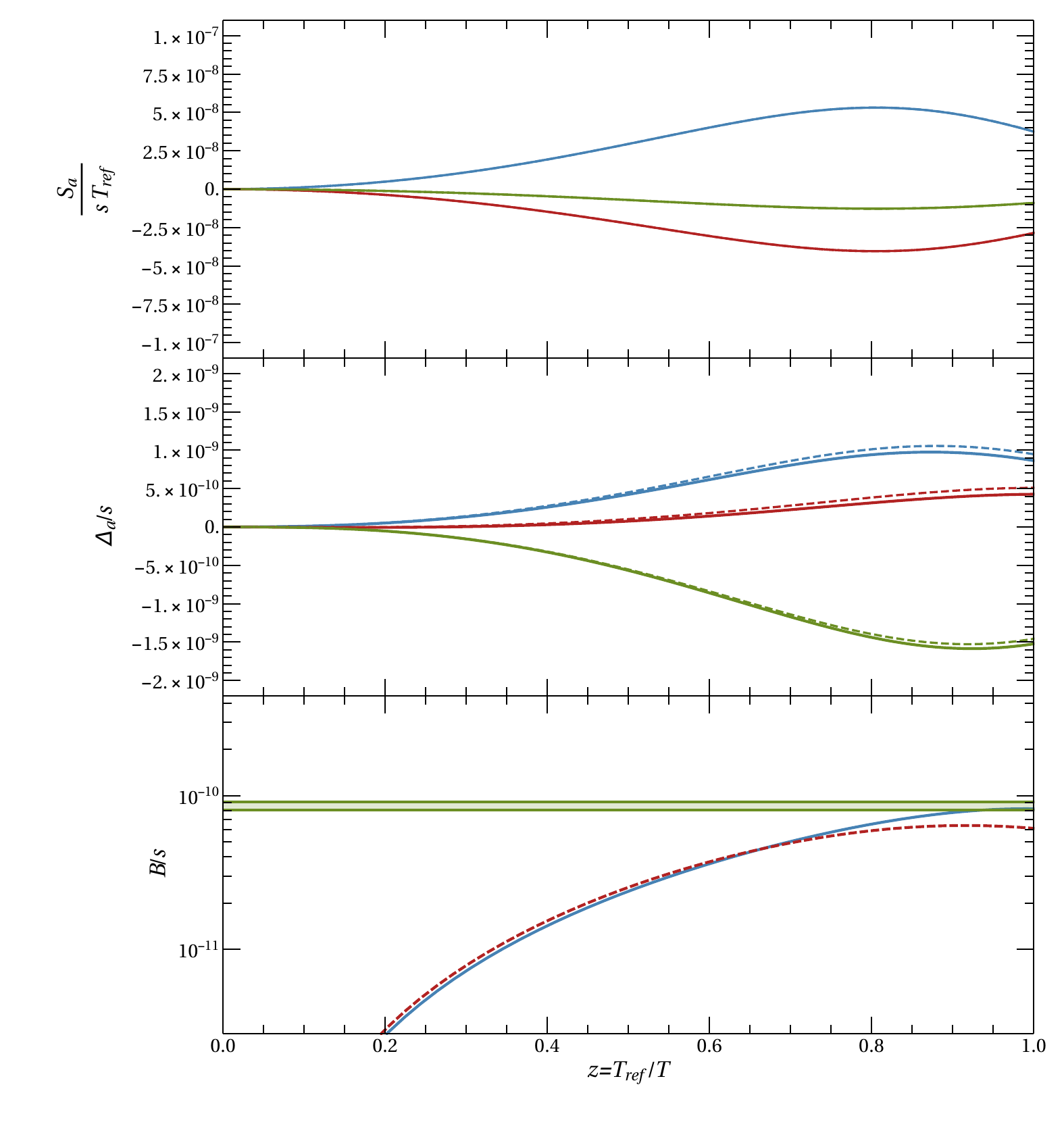}
	\caption{The comparison between numerical and analytic solutions for the source term, individual lepton charges and the baryon asymmetry
	for parameter choices that lead to maximal mixing angles for right-handed neutrino masses of $\bar{M}=1\GeV$ in the case of inverted hierarchy.
	The analytical approximations are always presented with a dashed line, for the source term they are indistinguishable from
	the numerical result. The parameters used for this plot are
	$\Delta M^2=5.306 \times 10^{-8} \bar{M}^2$, $\omega=\frac{\pi}{4}+5.55\ii$, $\alpha_1=0$, $\alpha_2=\pi$,$\delta=\pi$, and the discrete
	parameter $\xi=1$. The small $CP$-violating parameters are $\mu=1.36 \times 10^{-8} \bar {M}$ and
	$\sum_a |\epsilon_a|^2= 2.22\times 10^{-10} \sum_a |Y_a|^2$.}
	\label{fig:maxmixingIO_test} 
\end{figure}

\paragraph{Suppression due to Backreaction}
To include effects coming from the backreaction of the active flavours onto the sterile sector,
we consider once more the system of Eqs.~(\ref{Diff:Sterile}, \ref{Diff:Active}).
Among the $CP$-odd
sterile distributions, the entry $\delta n_{11}^{\rm odd}$ receives
the biggest correction  due to backreaction.
When neglecting the smaller Yukawa coupling $|\epsilon_{a}|$, the
 matrices $\tilde{\Gamma}_N$
take the form
\begin{align}
	\tilde{\Gamma}_N^a =\frac12 \gamma_{\rm av}
	\frac{a_{\rm R}}{T_{\rm ref}}
	\begin{pmatrix}
		|Y_{a}|^2 && 0 \\ 0 && 0
	\end{pmatrix}
	\,.
\end{align}
By applying the quasi-static approximation to the sterile neutrinos as in the previous section,
we obtain the approximate densities of $\delta n_{11}^{\rm odd}=2 q_{N_1}$,
\begin{align}
	\label{chg:sterile-balanced}
	\nonumber
	\delta n_{11}^{\rm odd} &\approx
	\sum_{b,c} \frac{|Y_{b}|^2}{2\sum_d |Y_d|^2} (A_{bc}+C_c/2)\Delta_c
	\left( 1 - \frac{|(H_N^{\rm vac})_{12}|^2}{(\tilde{H}_N^{\rm vac})^2}\frac{z^4}{|z^2+\Deltaz^2|^2} \right)
	\\
	&+\frac{|(H_N^{\rm vac})_{12}|^2}{(\tilde{H}_N^{\rm vac})^2}\frac{z^4}{|z^2+\Deltaz^2|^2} \delta n_{22}^{\rm odd}
	\,,
\end{align}
as well as the off-diagonal correlations $\delta n_{12}$.
Inserting the quasi-static solutions back into the evolution equations of the SM leptons and the
indirectly damped neutrino $\delta n_{22}$ gives
\begin{align}
	\label{chg:effeq}
	\frac{\dd \Delta_a}{\dd z} &= \tilde{W}_{ab} \Delta_b 
	- \frac{S_{a}(z)}{T_{\rm ref}}
	\\&\notag
	+\frac{a_{\rm R}}{T_{\rm ref}}\frac{\gamma_{\rm av}}{g_w}|Y_{a}|^2
	\frac{|(H_N^{\rm vac})_{12}|^2}{(\tilde{H}_N^{\rm vac})^2}\frac{z^4}{|z^2+\Deltaz^2|^2}
	\left( 2 \delta n_{22}^{\rm odd} - \sum_{b,c} \frac{|Y_{b}|^2}{\sum_d |Y_d|^2}
	(A_{bc} + C_c/2)\Delta_c\right)
	\\
	\label{chg:deltan22odd}
	\frac{\dd \delta n_{22}^{\rm odd}}{\dd z} &= -(\Gamma_N)_{11}
	\frac{|(H_N^{\rm vac})_{12}|^2}{(\tilde{H}_N^{\rm vac})^2}\frac{z^4}{|z^2+\Deltaz^2|^2}
	\frac12 \left( 2\delta n_{22}^{\rm odd} - \sum_{b,c} \frac{|Y_{b}|^2}{\sum_d |Y_d|^2} (A_{bc}+C_c/2)\Delta_c\right)\,,
\end{align}
with the effective washout matrix
\begin{align}
	\tilde{W}_{ab}&=\frac{a_{\rm R}}{T_{\rm ref}}\frac{\gamma_{\rm av}}{g_w}
	|Y_{a}|^2
	\sum_c \left( \delta_{ac} - \frac{|Y_{c}|^2}{\sum_{d} |Y_d|^2} \right)A_{cb}\,.
\end{align}
When we express the $\delta n_{22}^{\rm odd}$ dependence in
Eq.~(\ref{chg:effeq}) through the derivative $\dd \delta n_{22}/\dd z$,
the expression simplifies to
\begin{align}
	\frac{\dd \Delta_a}{\dd z} &= \sum_b \tilde{W}_{ab} \Delta_b 
	- \frac{S_{a}(z)}{T_{\rm ref}} - \frac{2}{g_w}\frac{|Y_{a}|^2}{\sum_d |Y_d|^2} \frac{\dd \delta n_{22}^{\rm odd}}{\dd z}\,.
\end{align}
To calculate the individual charges $\Delta_a$, we can neglect the derivative $\dd \delta n_{22}^{\rm odd}/\dd z$, as it is small for times
$z\ll \Deltaz$.
The solution for $\Delta_a(z)$ can now be computed by integrating

\begin{align}
	\label{chg:a_approx}
	\Delta_a(z)\approx
	\sum_{b,c=1,2} v_{ab}^T \ee^{w_b z}   \!\int_0^z \dd z'\, \ee^{-w_b z'} v_{bc}
	\frac{S_{c}(z')}{T_{\rm ref}}\,,
\end{align}
where $w_{1,2}$ are the two non-vanishing eigenvalues of the matrix
$\tilde{W}_{ab}$, and $v_{bc}$ the set of the corresponding eigenvectors.
As a result of the conservation of the generalised lepton number (\ref{Lgeneral1}), there is a vanishing eigenvalue. The  lepton number $L$ remains conserved when neglecting the derivatives of both sterile charges $\dd \delta n_{ii}/\dd z$.
The sterile charge density $\delta n_{22}^{\rm odd}$ can formally be obtained
by integrating Eq.~(\ref{chg:deltan22odd}) with the approximate form for
the SM charges from Eq.~(\ref{chg:a_approx}). For practical purposes
it is sufficient to completely neglect it for times before the equilibration
of $N_2$, $z\ll z_{N_2}^{\mathrm{eq}}$, and to replace it by its quasi-static
value for later times.
By including corrections to $\delta n^{\rm odd}_{11}$ of order
$\dd \Delta_{a}/\dd z$, and partially integrating the rate of change of the baryon
asymmetry $\dd B/\dd z$, we can obtain the baryon asymmetry of the Universe
\begin{align}\label{BapproxOverdamped}
	B(z)\approx \frac{28}{79}
	\left[
		\sum_{ab} \Delta_{a}(z) (A_{ab}+C_b/2)
		\frac{|Y_{b}|^2}{g_w \sum_d |Y_d|^2}
		+\frac{2}{g_w}\delta n_{22}^{\rm odd}(z)
	\right]\,,
\end{align}
up to an $\mathcal{O}(50\%)$ error for $z \geq z_{N_2}^{\mathrm{eq}}$.
For the parametric example from Table~\ref{tab:params}, a comparison between this analytic approximation
and the numerical result is shown in Figure~\ref{fig:lept_asym}.
A comparison of the numerical and analytic solutions for the source, active lepton charges and the final baryon asymmetry
for the points that lead to the maximal mixing angles for right-handed neutrino masses of $\bar{M}= 1\GeV$ are shown on
Figures~\ref{fig:maxmixingNO_test} and~\ref{fig:maxmixingIO_test}.

\paragraph{The baryon asymmetry in the case of highly flavour asymmetric washout}
In the special case of a highly flavour asymmetric washout, where the washout rate of one of the active neutrino flavours is much smaller
than the $N_2$ equilibration rate $\gamma_{\rm av} |Y_a|^2 a_{\rm R}/T_{\rm ref} \ll (z_{N_2}^{\rm eq})^{-1}$, while the other flavours
have a strong washout compared to it $\gamma_{\rm av} |Y_b|^2 a_{\rm R}/T_{\rm ref} \gg (z_{N_2}^{\rm eq})^{-1}$, the formal solution of the
evolution equations~(\ref{chg:a_approx}) can be further simplified.

Since we assumed that the washouts of the other two flavours are large, they have reached quasi-static equilibrium early, and the relation between the
three charges is approximately given by $\Delta_{b}=-\Delta_{a}/2$,\footnote{This can be found by looking at the eigensystem of the washout matrix $\tilde{W}_{ab}$. There are two approximately vanishing eigenvalues, corresponding to the conserved generalised lepton number and the negligible washout in one flavour.} 
where the flavour with the smallest washout, $\Delta_{a}$, is formally given by:
\begin{align}
	\frac{\Delta_{a}(z)}{s} = - \exp\left( -\frac{\gamma_{\rm av} a_{\rm R}}{2 g_w T_{\rm ref}} |Y_a|^2 z\right)
	\int_0^z \mathrm{d} z^\prime \frac{S_a(z^\prime)}{s T_{\rm ref}}
	\exp\left( \frac{\gamma_{\rm av} a_{\rm R}}{2 g_w T_{\rm ref}} |Y_a|^2 z^\prime \right)
	\label{chg:flvasymmwashout}
\end{align}
By neglecting the washout of the flavour $\Delta_a$ for $z\ll z_{N_2}^\mathrm{eq}$, the exponential within the integral can be approximated to be constant,
which leads us to the approximate flavour asymmetry:
\begin{align}
	\frac{\Delta_{a}(z)}{s} &= 
	-\frac{405 \zeta^{6/5}(3)}{60^{1/5}\,2 \pi^5}\frac{\gamma_{\rm av}^{7/5}}{g_w \mathfrak{h}_{\rm th}^{4/5} g_\star^{6/5}}
	\left[ \frac{m_{\rm Pl}^2}{\bar{M} \mu (\sum_b |Y_b|^2)^2} \right]^{1/5}\,
	\Im[ Y^*_{a} \epsilon_{a}]
	\gamma\left( \frac35, \frac{z^5}{{z_{N_2}^{\mathrm{eq}}}^5}\right)
	\exp\left( -\frac{\gamma_{\rm av} a_{\rm R}}{2 g_w T_{\rm ref}} |Y_a|^2 z\right)\\
	&\approx
	-4.44 \times 10^{-6}
	\left[ \frac{m_{\rm Pl}^2}{\bar{M} \mu (\sum_b |Y_b|^2)^2} \right]^{1/5}\,
	\Im[ Y^*_{a} \epsilon_{a}]
	\gamma\left( \frac35, \frac{z^5}{{z_{N_2}^{\mathrm{eq}}}^5}\right)
	\exp\left( -\frac{\gamma_{\rm av} a_{\rm R}}{2 g_w T_{\rm ref}} |Y_a|^2 z\right)\,,
\end{align}
where $\gamma(s,x)$ is the lower incomplete gamma function
\begin{align}
	\gamma(s,x)\equiv \int_0^x t^{s-1} \ee^{-t} \mathrm{d} t\,.
\end{align}
Inserting this expression back into Eq.~\ref{BapproxOverdamped}, and neglecting corrections of $\mathcal{O}(|Y_a|^2)$, gives us the total baryon asymmetry:
\begin{align}
	\frac{B(z)}{s}&\approx\frac{28}{79}\left( \frac{\Delta_a(z)}{6 g_w s} + \frac{2}{g_w s} \delta n^{\rm odd}_{22}(z) \right)
	\label{Bapprox1}\\
	&\approx
	-1.31 \times 10^{-7}
	\left[ \frac{m_{\rm Pl}^2}{\bar{M} \mu (\sum_b |Y_b|^2)^2} \right]^{1/5}\,
	\Im[ Y^*_{a} \epsilon_{a}]\nonumber\\
	&\times \ \gamma\left( \frac35, \frac{z^5}{{z_{N_2}^{\mathrm{eq}}}^5}\right)
	\exp\left( -\frac{\gamma_{\rm av} a_{\rm R}}{2 g_w T_{\rm ref}} |Y_a|^2 z\right)
	[1+\theta(z-z_{N_2}^\mathrm{eq})]
	\,,
	\label{BAU:maxflvasymm}
\end{align}
Equation (\ref{Bapprox1}) can be obtained from (\ref{BapproxOverdamped}) by setting $\Delta_{b}=-\Delta_{a}/2$.
To obtain (\ref{BAU:maxflvasymm}), we in addition used the fact that the derivative on the LHS of (\ref{chg:deltan22odd})
can be neglected for $z>z_{N_2}^\mathrm{eq}$, \textit{i.e.},
we assumed that $ \delta n^{\rm odd}_{22}$ reaches quasi-static equilibrium instantaneously at $z_{N_2}^\mathrm{eq}$, which 
is reflected by the Heaviside theta function.
The increase in $B$ after $z=z_{N_2}^\mathrm{eq}$ can be understood physically: For earlier times, $N_2$ is essentially decoupled from the system. Due to the approximate conservation of the generalised lepton number $\tilde{L}\simeq0$, the SM charges $L$ and $B$ are determined by the amount of $\tilde{L}$ that is stored in $N_1$.
After $N_1$ reaches equilibrium, the backreaction also stores a fraction of $\tilde{L}$  in $N_2$, leading to a larger deficit (and hence larger $|L|$, $|B|$) in the SM fields.

\section{Limits on the Heavy Neutrino Mixing}
\label{sec:mixing}
With Eqs. (\ref{BapproxOscillating}) and (\ref{BapproxOverdamped}), we have found approximate analytic expressions for the BAU in the limiting cases that the oscillations of the
sterile neutrinos occur either deeply in the oscillatory regime (the slowest oscillation time scale is much faster than the fastest equilibration time scale) or in the strongly overdamped regime (equilibration of one interaction eigenstate occurs long before the onset of the oscillations).
From an experimental viewpoint it is interesting to identify the maximal values of $U_i^2$ for which leptogenesis is possible. 
If one ignores constraints from direct searches for heavy neutrinos
(see {\it e.g.} Ref.~\cite{Drewes:2015iva} and references therein for a recent
summary), then these maximal values occur in the overdamped regime, which
is characterised by a strong washout.
There are two possibilities for preserving the baryon asymmetry at the
electroweak scale from this washout. Either there is a strong hierarchy
among the Yukawa couplings of heavy neutrinos to the different SM flavours
$e,\mu,\tau$, causing one of the charges $\Delta_a$ to be approximately
conserved, or the asymmetry is produced close to the electroweak scale, such that
there is no time for a complete washout before sphalerons freeze out. In the case of  $n_s=2$, a strong hierarchy among the doublet
Yukawa couplings is not possible while being consistent with neutrino
oscillation data. Therefore we need to resort to a strong mass degeneracy in order to
delay the generation of the  asymmetry until $z\simeq 1$. Yet, we are interested in
maximising the mixing angles while keeping the washout rate
of the SM flavours as small as possible, which constrains the parameters
$\delta$ and $\alpha_2$. Minimising the washout rate of the active flavours also introduces a difference between the normal and inverted hierarchies, as the minimal washout for the inverted hierarchy can be an order of magnitude smaller than the one for normal hierarchy given the same total mixing angle.
Furthermore, maximising the analytic expression
for the source also determines $\Re\, \omega$. Therefore it is only necessary
to scan over the remaining three parameters: $\Im\, \omega$, $M$ and $\Delta M$.

When solving Eq.~(\ref{Diff:Sterile}) for the helicity-even correlation
function, we can use the fact that a solution with a rescaled time dependence
$\delta n^{\rm even}(z \zeta)$ corresponds to a solution of the same equation
with the vacuum Hamiltonian replaced by
$H_{N}^{\rm vac}\rightarrow \zeta^{3} H_{N}^{\rm vac}$, the
thermal mass by $H_N^{\rm th}\rightarrow \zeta H_N^{\rm th}$, and the
rate $\Gamma_N\rightarrow \zeta \Gamma_N$.
For parameter choices with large mixing angles, one of the eigenvalues of
the decay rate of the sterile
neutrinos is typically much larger than the other,
$(\Gamma_N)_{11}\gg (\Gamma_N)_{22}$, and the misalignment between the
mass and flavour eigenstates is maximal,
which implies that the only parameters playing a role in the evolution of the
$\delta n^{\rm even}$ correlation are the average Majorana mass $M$,
the mass splitting $\Delta M^2 = M_1^2-M_2^2$, and the imaginary angle in the Casas-Ibarra parametrisation $\Im\, \omega$.
Any change of the mass scales $M\rightarrow \xi M$, or
$\Delta M^2 \rightarrow \eta \Delta M^2$, can therefore be compensated by
a shift in $\Im\, \omega \rightarrow \Im\, \omega + \log(\eta/\xi^3)/6$, as
well as replacing
$\delta n^{\rm even}(z) \rightarrow  \delta n^{\rm even}(\eta^{1/3}z)$.
Note that although the oscillation and equilibration time scales change,
their ratio remains the same.

\begin{figure}
	\centering
	\includegraphics[scale=1.6]{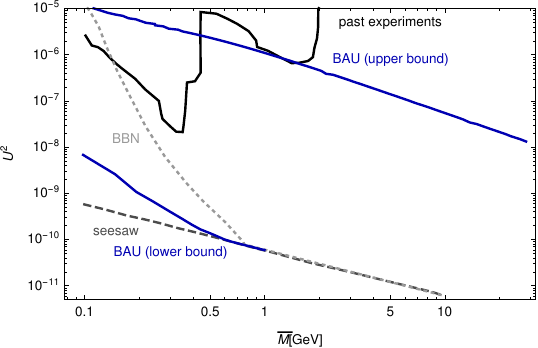}
	\includegraphics[scale=1.6]{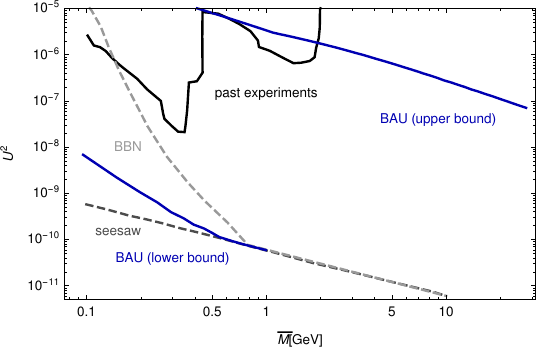}
	\caption{
The solid, dark blue lines show the largest and smallest value of $U^2$ we find to be consistent with neutrino oscillation data and the requirement to explain the observed BAU as a function of $\bar M=(M_1+M_2)/2$. They are compared to the upper bound from direct search experiments summarised in Ref.~\cite{Drewes:2015iva} (solid black line), the lower bound from neutrino oscillation data (grey dashed ``seesaw'' line) and the lower bound from the requirement that the $N_i$ have a lifetime of less than $0.1$s so that their decay does not modify primordial nucleosynthesis (dotted grey ``BBN'' line). The upper panel corresponds to normal neutrino mass hierarchy, the lower panel corresponds to inverted hierarchy.
	}
	\label{fig:max_mixing}
\end{figure}

To determine how the helicity-odd charges $\delta n^{\rm odd}$ and
$\Delta_a$ change under this parameter transformation, we first need to
determine the change in the source term.
In contrast to the decay rate where we can typically neglect the smaller
Yukawa coupling $|Y_{2a}|$ in the interaction basis, it is essential for the source term.
By correctly applying the scaling transformation, the source term and with 
it the baryon asymmetry are rescaled according to Table~\ref{table:rescale}.
\begin{table}
	\centering
	\begin{tabular}{r||cc|cc|c}
		At the original scale & $M$ & $\Delta M^2$ & $\Im\, \omega$
		& $S(z)$ & $B(z=1)$\\
		\hline
		Rescaled & $\xi M$ & $\eta \Delta M^2$
		& $\Im\, \omega + \log(\eta/\xi^3)/6$
		& $S(\eta^{1/3} z) \xi \eta^{-1/3}$
		& $B(\eta^{1/3}) \xi \eta^{-1/3}$
	\end{tabular}
	\caption{Rescaling of the asymmetry}
	\label{table:rescale}
\end{table}
As a result, even if we do not achieve the observed BAU for some choice of
parameters, by keeping the ratios $\Delta M^2:|Y_{1e}|^2:|Y_{1\mu}|^2:|Y_{1\tau}|^2$ constant, these transformation rules tell us how to find the
parameters that lead to the desired result for the BAU just by changing the absolute mass
and the mass splitting of the right-handed neutrinos.
Furthermore, by maximising $B(\eta^{1/3})/\eta^{1/3}$, we can find the
optimal mass splitting for producing the baryon asymmetry
and then find the corresponding mass by determining
$\xi = B_{\rm obs} \eta^{1/3}/B(\eta^{1/3})$. For that mass these parameters
give the maximal mixing consistent with leptogenesis.
By using the scaling of the baryon asymmetry from Table~\ref{table:rescale},
we find the maximal mixing angles consistent with baryogenesis for the mass
range between $0.1-10\GeV$ as presented in Figure~\ref{fig:max_mixing}.

\section{Discussion and Conclusion} 
\label{sec:disc_and_concl}

In this work we study the production of lepton and baryon asymmetries from the 
 oscillations of sterile neutrinos with $\GeV$-scale masses in the minimal seesaw model. The main goal is to obtain an 
 understanding of the maximal heavy neutrino mixing angles $U_{a i}^2$ consistent with the requirement to explain the observed BAU, while correctly accounting for backreaction and spectator effects. This is of crucial importance in order to assess the possibility of an experimental discovery of heavy neutrinos that may be responsible for the generation of light neutrino masses via the seesaw mechanism and for the BAU via low-scale leptogenesis.  
 
Baryogenesis via heavy neutrino oscillations can happen in different regimes, which can qualitatively be understood in terms of three time scales: the oscillation time $z_{\rm eq}$ at which the first heavy neutrino flavour oscillation occurs,
the equilibration time $z_{\rm eq}$ at which the first heavy neutrino eigenstate comes into thermal equilibrium with the primordial plasma 
 and the time  $z_{\rm ws}$ when weak sphalerons freeze out and baryon number becomes a conserved quantity, {\it i.e.}, the BAU is frozen in. 
The generation of a baryon asymmetry can be understood analytically in the two extreme cases $z_{\rm osc}\ll z_{\rm eq}<z_{\rm ws}$ (oscillatory regime) and $z_{\rm ws}>z_{\rm osc}\gg z_{\rm eq}$ (overdamped regime).
For heavy neutrino parameters that interpolate between these two regimes, we have to resort to solving the kinetic equations numerically.

In the oscillatory regime asymmetries in the individual lepton flavours are generated within the first few oscillations of the right-handed neutrinos at $z\simeq z_{\rm osc}$. At a much later time $z\simeq z_{\rm eq}$, the flavour asymmetric and lepton number violating washout generates a non-zero total lepton number from these, which is partly converted into a net baryon number by weak sphalerons. Once all heavy neutrinos come into equilibrium, all lepton asymmetries are washed out. However,
if the washout is incomplete at $z=z_{\rm ws}$, a non-zero baryon asymmetry remains. The latter requirement implies that the Yukawa interactions of the sterile neutrinos must be sufficiently weak, and the analytic treatment is based on a perturbative expansion in the Yukawa couplings. To this end, our results agree with those previously found in the literature~\cite{Asaka:2005pn,Drewes:2012ma,Abada:2015rta}.

In the overdamped regime at least some of the heavy neutrino flavour eigenstates have Yukawa couplings that are much larger than the naive seesaw relation would suggest in absence of cancellations in the neutrino mass matrix. 
This can be made consistent with the smallness of the light neutrino masses if an approximate conservation of $B-L$ is realised in Nature. This underlying symmetry implies that each strongly coupled heavy neutrino flavour eigenstate is accompanied by a feebly coupled eigenstate that completely decouples in the limit of exact $B-L$ conservation. The two corresponding mass eigenstates form a Dirac spinor in that limit.
We find an approximate analytic description in this regime by expanding in the tiny Yukawa coupling, and by employing a quasi-static approximation to the evolution of the strongly coupled flavour eigenstate, which comes into equilibrium before the flavour oscillations amongst the two can begin. 
In contrast to the oscillatory regime, the effect of the thermal masses and the backreaction of the produced lepton asymmetries on the heavy neutrino evolution
cannot be neglected in this regime. Both of these tend to suppress the generated asymmetry. 
A complete washout of all asymmetries due to the large Yukawa couplings can be prevented in two different ways:
Either one of the SM leptons couples to heavy neutrinos much more weakly than the two others (leading to a highly flavour asymmetric washout and a survival of the asymmetry stored in the weakly coupled SM flavour), or the heavy neutrinos have  degenerate masses (in which case the oscillations and asymmetry generation occur very late at $z\sim z_{\rm ws}$ and there is no time for a complete washout before sphalerons freeze out).
In the scenario with only two heavy neutrinos, a strong hierarchy amongst the couplings to different SM lepton flavours is ruled out by neutrino oscillation data, and leptogenesis can only be achieved with degenerate heavy neutrino masses.
If there are more than two heavy neutrinos, then the extended parameter space allows to make a highly flavour asymmetric washout compatible with neutrino oscillation data, and baryogenesis is possible without a mass degeneracy~\cite{Drewes:2012ma,Hernandez:2015wna}.

The main new results of the present work are:
\begin{itemize}
\item The equations of motion have been derived from first principles of quantum field theory in the CTP formalism. We have, for the first time, included the effects of thermal masses, backreaction from the generated asymmetries and spectator fields in this derivation.
\item We have derived analytic approximations to the baryon asymmetry in case of both the oscillatory and the overdamped regime. 
While analytic solutions in the oscillatory regime have previously been found by several authors~\cite{Asaka:2005pn,Drewes:2012ma,Abada:2015rta}, the solutions in the overdamped regime are, to the best of our knowledge, presented here for the first time.
Up to $\mathcal{O}(1)$ corrections they are consistent with numerical cross-checks.
\item Based on these results, we have identified the largest possible heavy neutrino mixings consistent with leptogenesis in the scenario with two heavy neutrinos. Spectator effects, which account for the redistribution of SM charges due to fast SM interactions, and thermal masses have been included in both the analytic and the numerical treatment. While they have been neglected in recent studies so far, we have shown that they have a non-negligible impact on the final baryon asymmetry. Quantitatively we find that leptogenesis is possible for larger mixing angles than previously thought, which increases the chances of an experimental discovery.
\end{itemize}
In spite of this significant progress, several technical issues remain to be clarified in the future:
\begin{itemize}
\item Our treatment relies on  momentum-averaged kinetic equations. Since the assumption of kinetic equilibrium is not justified for the heavy neutrinos, this introduces an error of order one.
\item Throughout this paper we have considered all SM Yukawa interaction to be in equilibrium, which is true for temperatures $T\lesssim 10^5 \GeV$, when the electron has finally equilibrated. However, the physical interesting regime, \emph{i.e.} the time of the first oscillation, may already occur at higher temperatures.
\item We assume the weak sphalerons to freeze out suddenly, which however is not completely true when electroweak symmetry is broken in a crossover, as it is the case for the SM. This could be phenomenologically important in the strongly overdamped regime, when the creation of
the baryon asymmetry continues throughout the electroweak crossover.
\item Our analytic solutions in the oscillatory regime are valid for an arbitrary number of heavy neutrinos. The treatment of the overdamped regime is, however, focused on the minimal realistic model, in which only two of these exist. A generalisation to the case with three or more heavy neutrinos, which includes a larger number of different oscillation and equilibration time scales, may be very helpful for efficient phenomenological studies.
\end{itemize}

To fully explore the discovery potential of present and future experiments, it would be highly desirable to perform a complete parameter scan of low-scale leptogenesis in the scenario with three heavy neutrinos. This should consistently include constraints from a wide range of past experiments that are sensitive to the existence of heavy neutrinos, in particular direct searches for these particles and indirect searches for lepton number or flavour violation.
The present analytic results, in particular the new description
of the overdamped regime, should also be applicable to assess possibilities of generating the BAU  in extensions of the minimal seesaw model~(\ref{eq:Lagrangian}).

\label{sec:concl}

\section*{Acknowledgements}
This research was supported by the DFG cluster of excellence 'Origin and Structure of the Universe' (www.universe-cluster.de).

\begin{appendix}
\numberwithin{equation}{section}

\section{Parametrisation of the Seesaw Model and Neutrino Oscillation Data}
\label{App:Seesaw}

The extension of the SM by $n_s$ sterile neutrinos introduces $7n_s-3$ new physical parameters, {\it i.e.} $11$ or $18$ for the cases $n_s=2$ or $n_s=3$ considered in this paper. Various experimental constraints on these parameters are discussed in detail in Ref.~\cite{Drewes:2015iva}.
The relation between the parameters in the Lagrangian (\ref{eq:Lagrangian})
and constraints on the (presently incompletely determined \cite{Gonzalez-Garcia:2015qrr}) light neutrino mixing matrix $U_\nu$, light neutrino mass matrix $m_\nu$ can be expressed in term of the Casas-Ibarra parametrisation \cite{Casas:2001sr}
\begin{align}\label{CasasIbarraDef}
Y^\dagger=\frac{1}{v}U_\nu\sqrt{m_\nu^{\rm diag}}\mathcal{R}\sqrt{M^{\rm diag}}\,.
\end{align}
The PMNS matrix can be factorised as
\begin{align}
\label{PMNS}
U_\nu=V^{(23)}U_\delta V^{(13)}U_{-\delta}V^{(12)}{\rm diag}(\ee^{\ii \alpha_1/2},1,\ee^{\ii \alpha_2 /2})\,,
\end{align}
with $U_{\pm \delta}={\rm diag}(\ee^{\mp\ii \delta_1/2},1,\ee^{\pm\ii \delta /2})$ and where the non-vanishing entries of the matrix\\ $V=V^{(23)}V^{(13)}V^{(12)}$ are given by:
\bse
\begin{align}
V^{(ij)}_{ii}&=V^{(ij)}_{jj}=\cos \uptheta_{ij}\,,\\
V^{(ij)}_{ij}&=-V^{(ij)}_{ji}=\sin \uptheta_{ij}\,,\\
V^{(ij)}_{kk}&=1 \quad \text{for $k\neq i,j$}\,.
\end{align}
\ese
The parameters $\uptheta_{ij}$ are the mixing angles, $\delta$ is referred to as the Dirac phase and $\alpha_{1,2}$
as Majorana phases.\footnote{In case of two sterile flavours $\alpha_{1,2}$ are redundant such that we are effectively just left with one Majorana phase. For normal hierarchy we have $m_1=0$ such that $Y$ only depends on $\alpha_2$ but not on $\alpha_1$, while for inverted hierarchy we have $m_3=0$ and it is the difference $\alpha_1-\alpha_2$ on which $Y$ depends.}

The misalignment between sterile mass and interaction eigenstates is given by the complex orthogonal matrices $\mathcal{R}$ that fulfil $\mathcal{R}\mathcal{R}^T=\mathds{1}$. In case of three flavours it can be written as
\begin{align}
\mathcal{R}=\mathcal{R}^{(23)}\mathcal{R}^{(13)}\mathcal{R}^{(12)}\,,
\end{align}
where the non-vanishing entries read
\bse
\begin{align}
\mathcal{R}^{(ij)}_{ii}&=\mathcal{R}^{(ij)}_{jj}=\cos \omega_{ij}\,,\\
\mathcal{R}^{(ij)}_{ij}&=-\mathcal{R}^{(ij)}_{ji}=\sin \omega_{ij}\,,\\
\mathcal{R}^{(ij)}_{kk}&=1 \quad \text{for $k\neq i,j$}\,,
\end{align}
\ese
with three complex angles $\omega_{ij}$, while for two flavours we have to deal with one complex angle $\omega$ and additionally a distinction between normal hierarchy (NO) and inverted hierarchy (IO) has to be applied:
\begin{align}
\mathcal{R}^{\rm NO}=
\begin{pmatrix}
0 && 0\\
\cos \omega && \sin \omega \\
-\xi \sin \omega && \xi \cos \omega
\end{pmatrix}\,,\quad \quad 
\mathcal{R}^{\rm IO}=
\begin{pmatrix}
\cos \omega && \sin \omega \\
-\xi \sin \omega && \xi \cos \omega \\
0 && 0
\end{pmatrix}
\,,
\end{align}
where $\xi=\pm 1$. In both cases $\Im(\omega)$ determines the absolute size of the largest eigenvalue of the combination $YY^\dagger$.
One can express the overall size of the mass eigenstates $N_1$ and $N_2$ defined in Eq.~(\ref{totalU}) as 
\bse
\begin{eqnarray}
U^2&=&\frac{M_2-M_1}{2M_1 M_2} (m_2-m_3)\cos(2 \Re\omega)+\frac{M_1+M_2}{2M_1 M_2}(m_2+m_3)\cosh(2 \Im\omega) \\
&& {\rm for} \ {\rm normal} \ {\rm hierarchy}\nonumber, \\
U^2&=&\frac{M_2-M_1}{2M_1 M_2} (m_1-m_2)\cos(2 \Re \omega)+\frac{M_1+M_2}{2M_1 M_2}(m_1+m_2)\cosh(2 \Im \omega) \\
&& {\rm for} \ {\rm inverted} \ {\rm hierarchy}.\nonumber
\end{eqnarray}
\ese

Finally, we shall make connection to the benchmark scenarios defined in Sec.~\ref{sec:benchmarkscenarios}.
The \emph{naive seesaw} is characterised by small values of ${\rm Im}\omega$ (or ${\rm Im}\omega_{ij}$).
In the approximately \emph{lepton number conserving scenario}  unitary transformations amongst the heavy neutrino fields can be used to bring $Y$ and $M$ into the form \cite{Abada:2007ux,Fernandez-Martinez:2015hxa}
\bse
\begin{align}
Y^\dagger&=\left(
\begin{tabular}{c c c}
$Y_e$ & $\epsilon_e$ & $\epsilon_e'$\\
$Y_\mu$ & $\epsilon_\mu$ & $\epsilon_\mu'$\\
$Y_\tau$ & $\epsilon_\tau$ & $\epsilon_\tau'$
\end{tabular}
\right)
\
,
\
&&M=\left(
\begin{tabular}{c c c}
$\upmu_1$ & $\bar{M}$ & $\upmu_3$\\
$\bar{M}$ & $\upmu_2$ & $\upmu_4$\\
$\upmu_3$ & $\upmu_4$ & $M_3$
\end{tabular}
\right)\label{BmLcons} \
{\rm for} \ n_s=3\\
Y^\dagger&=\left(
\begin{tabular}{c c }
$Y_e$ & $\epsilon_e$ \\
$Y_\mu$ & $\epsilon_\mu$ \\
$Y_\tau$ & $\epsilon_\tau$ 
\end{tabular}
\right)
\
,
\
&&M=\left(
\begin{tabular}{c c }
 $\upmu_1$ & $\bar{M}$\\
 $\bar{M}$ & $\upmu_2$
\end{tabular}
\right)\label{BmLcons2} \
{\rm for} \ n_s=2
\end{align}
\ese
Here $\epsilon_a,\epsilon_a'\ll Y_a$ and $\upmu_i\ll M_3, \bar{M}$ are lepton number violation (LNV) parameters, which must vanish if $B-L$ is exactly conserved. $\bar{M}$ is the common mass of the two heavy neutrino mass eigenstates $N_1$ and $N_2$ that have comparable large mixing angles, the $\mu_i$ quantify the mass splitting $M_1-M_2$.
The deviation from maximal misalignment between the heavy neutrino mass basis (where $M$ is diagonal) and interaction basis (where $YY^\dagger$ is diagonal) in the flavours is quantified by the $\epsilon_a$. 
It is straightforward to see that $U_{a1}^2=U_{a2}^2$ in the mass basis, {\it i.e.}, both mass eigenstates couple with the same strength to SM leptons.
The maximal misalignment implies that one interaction eigenstate has couplings of order $Y_a$ while the interactions of the other one are suppressed by the small parameters $\epsilon_a$, {\it i.e.}, $Y Y^\dagger$ has two eigenvalues of very different magnitude $\sim Y_a^2$ and $\sim\epsilon_a^2$. The analytic solution in Section~\ref{sec:str} is effectively obtained in an expansion in $\epsilon_a$. 
In the parametrisation (\ref{CasasIbarraDef}) the $B-L$ conserving limit corresponds to large values of $|{\rm Im}\omega|\gg 1$.
A third heavy neutrino (if it exists) must decouple in the $B-L$ conserving limit, all its interactions are suppressed by $\epsilon_a'$.

\section{Derivation of the Quantum Kinetic Equations}\label{AppendixKineticEquations}

In this Appendix we provide a brief derivation of the quantum kinetic equation~(\ref{Diff:Sterile})based on first principles of non-equilibrium quantum field theory using the Schwinger-Keldysh CTP approach. For a pedagogical review of this topic see \emph{e.g.} Refs.~\cite{Chou:1984es,Berges:2004yj}.

\subsection{General Considerations and Definitions}
We start our discussion assuming Minkowski background spacetime and generalise it to the radiation dominated Friedmann-Robertson-Walter metric in the subsequent Subsection.

\paragraph{Correlation Functions in a Medium} The use of $S$-matrix elements is not always suitable to describe non-equilibrium systems because there is no well-defined notion of asymptotic states, and the properties of quasiparticles in a medium may significantly differ from those of particles in vacuum.
In contrast, observables can always be expressed in terms of correlation functions of the quantum fields, without reference to asymptotic states or free particles.
There are two linearly independent two-point functions for each field. 
For a generic fermion $\Psi$ these can be expressed in terms of the Wightman functions
\begin{eqnarray}
\ii S^{>}_{\alpha\beta}(x_{1},x_{2})=\langle \Psi_{\alpha}(x_{1})\bar{\Psi}_{\beta}(x_{2})\rangle \ , \quad
\ii S^{<}_{\alpha\beta}(x_{1},x_{2})=-\langle \bar{\Psi}_{\beta}(x_{2})\Psi_{\alpha}(x_{1})\rangle\,.
\label{SbackA}
\end{eqnarray}  
Here $\alpha$ and $\beta$ are spinor indices, which we suppress in the following; flavour indices can be included equivalently.
The $\langle\ldots\rangle$ is to be understood in the sense of the usual quantum statistical average $\langle\ldots\rangle={\rm Tr}(\varrho \ldots) $ of a system characterised by a density operator $\varrho$.
In the present context, we choose 
\begin{equation}\label{initialstate}
\varrho = \varrho_{\rm SM}^{\rm eq}\otimes\varrho_N^{\rm vac}\,,
\end{equation}
where $\varrho_{\rm SM}^{\rm eq}$ is an equilibrium density operator for all SM fields and $\varrho_N^{\rm vac}$ is the vacuum density operator for sterile neutrinos. Physically this represents a situation in which the $N_i$ are absent initially and all SM fields have reached thermal equilibrium before the $N_i$ have been produced in significant amounts, which is justified by the smallness of the Yukawa coupling $Y$.
The expressions~(\ref{SbackA}) apply to both, Majorana fields (such as $N_i$) and Dirac fields (such as $\ell_a$).
The linear combinations
\bse
\begin{align}
S^{\A}(x_{1},x_{2})&\equiv \frac{\ii}{2}\left(S^{>}(x_{1},x_{2})-S^{<}(x_{1},x_{2})\right)\label{SMinus}\, ,\\
S^{+}(x_{1},x_{2})&\equiv \frac{1}{2}\left(S^{>}(x_{1},x_{2})+S^{<}(x_{1},x_{2})\right)\label{Splus}\,,
\end{align}
\ese
have intuitive physical interpretations. 
The \textit{spectral function} $S^{\A}$ encodes the spectrum of quasiparticles in the plasma.
The \textit{statistical propagator} $S^{+}$ provides a measure for the occupation numbers. 
The correlators fulfil the symmetry relations
\bse
\label{S:Hermiticity}
\begin{align}
\ii\gamma_{0}S^\gtrless(x_{2},x_{1})&=\left(\ii\gamma_{0}S^\gtrless(x_{1},x_{2})
\right)^{\dagger}\, , \\
\ii\gamma_{0}S^{+}(x_{2},x_{1})&=\left(\ii\gamma_{0}S^{+}(x_{1},x_{2})\right)^{\dagger}\label{Splussym}\,,\\ 
\gamma_{0}S^{\A}(x_{2},x_{1})&=\left(\gamma_{0}S^{\A}(x_{1},x_{2})\right)^{\dagger}\label{Sminussym}\, ,\\
\gamma_{0}S^{H}(x_{2},x_{1})&=\left(\gamma_{0}S^{H}(x_{1},x_{2})\right)^{\dagger}
\,.
\end{align}
\ese
If $\Psi$ is a Majorana fermion, then there is an additional symmetry
\begin{equation}\label{MajoranaSymmetry}
S^\gtrless(x_1,x_2)=C S^\gtrless(x_2,x_1)^t C^\dagger\, ,
\end{equation}
where $C$ is the charge conjugation matrix and the transposition $t$ acts on spinor as
well as flavour indices.

It is often useful to introduce the retarded, advanced and Hermitian propagators,
\bse
\begin{align}
\ii S^{R}(x_{1},x_{2})&=2\theta(t_1-t_2)S^{\A}(x_{1},x_{2})\,,\label{Sretarded}
\\
\ii S^{A}(x_{1},x_{2})&=-2\theta(t_2-t_1)S^{\A}(x_{1},x_{2})\,,\label{Sadvanced}
\\
 S^{H}(x_{1},x_{2})&=\frac12\left(S^R(x_1,x_2)+S^A(x_1,x_2)\right)=-\ii\,\sign(t_1-t_2)S^{\A}(x_{1},x_{2})\,.\label{Shermitian}
\end{align}
\ese
From this it follows that 
\begin{equation}
 S^\A(x_{1},x_{2})=\frac{\ii}{2}\left(S^R(x_1,x_2)-S^A(x_1,x_2)\right)\,.\label{spectralAR}
\end{equation}
The usual Feynman propagator $S^F$ can be expressed as 
$S^F = S^R + S^< = S^A + S^>$.
Spectral, statistical, retarded, advanced and Hermitian self-energies $\slashed{\Sigma}^\A$, $\slashed{\Sigma}^+$, $\slashed{\Sigma}^R$, $\slashed{\Sigma}^A$ and $\slashed{\Sigma}^H$ are defined analogously, see \emph{e.g.} \cite{Beneke:2010wd,Beneke:2010dz} for a list of explicit definitions.

\paragraph{Equations of Motion}
The correlation functions for quantum fields out of thermal equilibrium can
be obtained from the Schwinger-Dyson equations
\bse
\begin{align}
(\ii \slashed{\partial}_{x_1}-M)S^\A(x_1,x_2)&=2\ii\int_{t_1}^{t_2} \!\dd  t' \int  \!\dd ^3\textbf{x}'\,\slashed{\Sigma}^\A(x_1,x')S^\A(x',x_2)\label{KBE1}\,,\\\nonumber
(\ii \slashed{\partial}_{x_1}-M)S^+(x_1,x_2)&=2\ii\int_{t_i}^{t_2} \!\dd t' \int  \!\dd^3\textbf{x}'\,\slashed{\Sigma}^+(x_1,x')S^\A(x',x_2)\\
&-2\ii\int_{t_i}^{t_1} \!\dd t' \int  \!\dd^3\textbf{x}'\,\slashed{\Sigma}^\A(x_1,x')S^+(x',x_2)\label{KBE2}\,,
\end{align}
\ese
which can be derived from two-particle irreducible effective action~\cite{Cornwall:1974vz} in the CTP framework~\cite{Calzetta:1986cq}.
An explicit derivation is given in Ref.~\cite{Berges:2004yj}.
If the initial state at time $t_i$ is Gaussian (\textit{i.e.}\ can entirely be specified by the initial conditions of the statistical propagators and one-point functions of all fields), then the above equations of motion are exact. Strictly speaking this is not true for~(\ref{initialstate}) because $\varrho_{\rm SM}^{\rm eq}$ is not Gaussian \cite{Garny:2009ni}. However, $\varrho_N^{\rm vac}$ is Gaussian, and we are primarily interested in the equation of motion for the heavy neutrinos.

The equations~(\ref{KBE1}) and~(\ref{KBE2}) can in principle be solved directly in position space \cite{Anisimov:2008dz,Anisimov:2010aq,Drewes:2010pf,Anisimov:2010dk,Garny:2011hg,Garbrecht:2011xw,Drewes:2012qw}, but it is often more practical to perform a Fourier transform in the relative coordinate $x_1-x_2$ to Wigner space \cite{Prokopec:2003pj,Prokopec:2004ic}.\footnote{See also \cite{Millington:2012pf,Millington:2013isa,Dev:2014laa} for an alternative approach.}
This is the approach we take here.
In order to perform the Wigner transformation, it is convenient to rewrite ~(\ref{KBE1}) and~(\ref{KBE2}) with integration limits $\pm\infty$. For this purpose, we send $t_i\rightarrow -\infty$\footnote{Boundary conditions at finite time can still be imposed by formally introducing singular external sources \cite{Drewes:2012qw}.} and note that it can be seen that causality is maintained when substituting the retarded and advanced propagators and self energies by virtue of the relations~(\ref{Sretarded}) and~(\ref{Sadvanced}). By using Eqs.~(\ref{Shermitian}) and~(\ref{spectralAR}) one finds $S^{A,R}=S^H\pm \ii S^\A$. Together with the definitions of $S^\A$ and $S^+$ this allows to rewrite~(\ref{KBE1}) and~(\ref{KBE2}) as 
\bse
\begin{align}
(\ii \slashed{\partial}_{x_1}-M)S^\A(x_1,x_2)&=
\int  \!\dd^4x'
\left(\slashed{\Sigma}^H(x_1,x')S^\A(x',x_2)
+\slashed{\Sigma}^\A(x_1,x')S^H(x',x_2)
\right)\label{KBE1bjoernstyle}\,,\\\nonumber
(\ii \slashed{\partial}_{x_1}-M)S^+(x_1,x_2)&=
\int  \!\dd^4x'
\left(
\slashed{\Sigma}^+(x_1,x')S^H(x',x_2)
+\slashed{\Sigma}^H(x_1,x')S^+(x',x_2)
\right)\label{KBE2bjoernstyle}\\
&+\frac{1}{2} \int  \!\dd^4x'\left(
\slashed{\Sigma}^>(x_1,x')S^<(x',x_2)
-\slashed{\Sigma}^<(x_1,x')S^>(x',x_2)
\right)\,,
\end{align}
\ese
which can easily be transformed to Wigner space by introducing new variables $x=(x_1+x_2)/2$ and $y=x_1-x_2$ and performing a Fourier transform with respect to $y$. 
In Wigner space, the symmetry relations~(\ref{S:Hermiticity}) 
of the propagators $S$ and, accordingly, of the self energies $\slashed{\Sigma}$ read
\bse
\begin{align}
\ii\gamma_{0}G^\gtrless(x;k)&=\left(\ii\gamma_{0}G^\gtrless(x;k)\right)^{\dagger}\,,\\
\ii\gamma_{0}G^{+}(x;k)&=\left(\ii\gamma_{0}G^{+}(x;k)\right)^{\dagger}\,,\\
\gamma_{0}G^{\A}(x;k)&=\left(\gamma_{0}G^{\A}(x;k)\right)^{\dagger}\,,\\
\gamma_{0}G^{H}(x;k)&=\left(\gamma_{0}G^{H}(x;k)\right)^{\dagger}
\label{SplussymWigner}
\,,
\end{align}
\ese
with $G$ being either $S$ or $\slashed{\Sigma}$.
Here $x$ denotes the real time and space coordinate and $k$ can be interpreted as the momentum
of a quasiparticle. In the following we mostly drop these arguments, and all correlation functions are to be understood as Wigner space functions.

Since the early Universe is homogeneous and isotropic, there is no dependence on the spatial part $\textbf{x}$ of $x=(t,\textbf{x})$.
During leptogenesis, all fields with gauge interactions are effectively kept in kinetic equilibrium. 
This means that we can describe the thermodynamic state of these degrees of freedom by a single temperature $T$ and chemical potentials $\mu_{\ell a}$ (for leptons) and $\mu_{\phi}$ (for the Higgs).
We can neglect the effect of the heavy neutrino production and decays on $T$ because of the large number of degrees of freedom $g_\star$ in the primordial plasma.
Compared to the typical time scale $1/T$ of microscopic processes, the temperature changes only slowly due to Hubble expansion, \textit{i.e.}\ $H\simeq \sqrt{8\pi^3g_\star/90}T^2/m_{\rm Pl} \ll T$, where $m_{\rm Pl}$ is the Planck mass.
Due to the smallness of the lepton flavour violating Yukawa couplings $Y$, also the chemical potentials only change at a small rate $||Y^t Y^*||T\ll T$. This separation of macroscopic and microscopic time scales justifies a gradient expansion in $t$ to leading order,\footnote{See \cite{Prokopec:2003pj,Prokopec:2004ic,Garbrecht:2011xw} for a more detailed discussion of this point.} such that in Wigner space, the Eqs.~(\ref{KBE1bjoernstyle}) and~(\ref{KBE2bjoernstyle}) read
\bse
\begin{align}
\Big(\slashed{p} +\frac{\ii}{2} \gamma_0\partial_t - M \Big)S^\A- 
\left(\slashed{\Sigma}^H S^\A + \slashed{\Sigma}^\A S^H\right)&=0\,,\label{kbe1}\\
\Big(
\slashed{p} +\frac{\ii}{2} \gamma_0\partial_t - M
\Big)S^+ - \slashed{\Sigma}^H S^+ - \slashed{\Sigma}^+ S^H &= \frac{1}{2}\left(\slashed{\Sigma}^> S^< - \slashed{\Sigma}^< S^>\right)\label{kbe2}\,.
\end{align}
\ese
By adding and subtracting the Kadanoff-Baym equation~(\ref{kbe2}) and its Hermitian conjugate, we obtain the constraint and kinetic equations
\bse
\begin{align}\label{constrainedSminu0}
\{ \H, \S^\A\}  - \{\G,\S^H\}&=0 \,, 
\\
\ii\partial_t \S^\A +[\H,\S^\A] 
-[\G,\S^H]
&=0 \,,\label{kineticSminus0}
\end{align}
\ese
and 
\bse
\begin{align}\label{constrainedSplus0}
\{ \H, \S^+\} 
-\{\mathcal{N},\S^H\}
&=\frac{1}{2}\left([\G^>,\S^<]-[\G^<,\S^>]\right)
\,,\\
\ii \partial_t \S^+ +[\H,\S^+]
-[\mathcal{N},\S^H]
&=\frac{1}{2}\left(\{\G^>,\S^<\}-\{\G^<,\S^>\}\right)\,,
 \label{kineticSplus0}
\end{align}
\ese
with 
\begin{eqnarray}
\nonumber
\S^+&\equiv&\ii\gamma^0 S^+ \ , \ \S^H\equiv \ii\gamma^0 S^H \ , \ \H\equiv(\slashed{p}-\slashed{\Sigma}^H-M)\gamma^0 
\,,\\ 
\G^>&\equiv&\slashed{\Sigma}^>\gamma^0 \  , \ \G^<\equiv\slashed{\Sigma}^<\gamma^0 , \ \G\equiv\frac{\ii}{2}(\G^>-\G^<) ,
\ \mathcal{N}\equiv \slashed{\Sigma}^+\gamma^0\,.
\end{eqnarray}
From the kinetic equation~(\ref{kineticSplus0}) it already becomes clear that $\H$ is the Hermitian
part of an effective Hamiltonian that leads to oscillations of the sterile neutrinos, and $\G^\gtrless$ are dissipative gain and loss terms. $\N$ can be interpreted as a noise term that owes its existence to the fluctuation-dissipation theorem.
It is convenient to express
\begin{equation}
\H=\bar{\H}+\delta\H \ , \ \G=\bar{\G} + \delta\G\,, 
\end{equation}
where $\bar{\H}$ and $\bar{\G}$ are $\H$ and $\G$ evaluated in local thermal equilibrium (with vanishing chemical potentials). The deviations   $\delta\H$ and $\delta\G$ arise due to finite chemical potentials of the SM fields.\footnote{In principle there are also contributions due to $\delta S_N$ in internal heavy neutrino propagators, but these are of order $\mathcal{O}[Y^4]$.}
We now define the static solutions $\bar{\S}^+=(\bar{\S}^>+\bar{\S}^<)/2$ and $\bar{\S}^\A=\ii(\bar{\S}^>-\bar{\S}^<)/2$ as the solutions to the algebraic equations 
\begin{eqnarray}
[\bar{\H},\bar{\S}^+]
-[\bar{\mathcal{N}},\bar{\S}^H]
=\frac12\left(\{\bar{\G}^>,\bar{\S}^< \}-\{\bar{\G}^<,\bar{\S}^> \}\right) \ \ , \ \ [ \bar{\H},\bar{\S}^\A ] = [ \bar{\G} , \S^H ]\,, 
\end{eqnarray}
and split
\begin{equation}
\S^+=\bar{\S}^+ + \delta\S\,.\label{splitSN}
\end{equation}
If the self energies $\slashed{\Sigma}$ are dominated by interactions with degrees of freedom that are in good approximation in equilibrium, then 
\begin{align}
\S^\A=\bar{\S}^\A  \,, \quad \S^H=\bar{\S}^H \,,  \quad 
\S^{\gtrless}=\bar{\S}^{\gtrless} + \delta\S \,,
\end{align}
to leading order in the small couplings and gradients \cite{Anisimov:2008dz,Garbrecht:2011xw}. 
This yields
\begin{eqnarray}
\nonumber
\partial_t \delta\S 
&=
-\partial_t \bar{\S}^+
+\ii[\bar{\H},\delta\S]
+\ii[\delta\H,\bar{\S}^+]
-\ii[\delta\mathcal{N},\bar{\S}^H]
-\{\bar{\G},\delta\S \}\\
&-\frac{\ii}{2}\left(\{\delta\G^>,\bar{\S}^< \}-\{\delta\G^<,\bar{\S}^> \}\right)
\,.\label{generalkinetic}
\end{eqnarray}
The term $\partial_t \bar{\S}^+$ is due to the fact that $\bar{\S}^+$ in the early Universe slowly changes due to Hubble expansion.

\subsection{Quantum Kinetic Equations for Heavy Neutrinos}
We now apply the general kinetic equation~(\ref{generalkinetic}) to the case of heavy neutrinos. The
associated correlators, self energies, effective Hamiltonians and rates will be attached with the subscript $N$.
It makes sense to split the self energies up into a part $\bar{\Sigma}$
computed in thermal equilibrium and for vanishing chemical potentials
and a deviation due to non-zero chemical potentials $\bar{\Sigma}$, such that
\begin{align}
\label{eq:equilibrium_decomposition}
\slashed\Sigma \equiv \gamma_\mu\Sigma^\mu=\gamma_\mu\left(\bar{\Sigma}^\mu+\delta \Sigma^\mu\right)\,.
\end{align}
In absence of chemical potentials, the  self energies of the sterile neutrinos $\slashed{\Sigma}_N$ and of 
the SM leptons $\slashed{\Sigma}_\ell$ can be factorised into a flavour dependent matrix of couplings and a reduced self energy $\hat{\Sigma}\!\!\!/$,

\begin{align}\label{hatSigmaNDef}
\bar{\Sigma}\!\!\!/_N&= g_w \hat{\Sigma}\!\!\!/ \left(
Y^* Y^t P_{\rm R}
+
Y Y^\dagger P_{\rm L}
\right) 
\,.
\end{align}
The absence of a superscript $^\A$, $^+$, $^R$, $^A$ or $^H$ indicates that the definition holds for either of these self energies. Here, $g_w=2$ accounts for the ${\rm SU}(2)$ multiplicity due to the SM doublets running in the loop. For non-zero chemical potentials, the reduced self energy also depends on the active flavour $a$. We can decompose
\begin{eqnarray}
(\slashed{\Sigma}_N)_{ij}&=& (\slashed{\Sigma}_{\rm R})_{ij} P_{\rm R} + (\slashed{\Sigma}_{\rm L})_{ij} P_{\rm L}\nonumber\\
&=&g_w\sum_{a=e,\mu,\tau}\left(
\hat{\Sigma}\!\!\!/_{{\rm R} a}Y^*_{i a} Y^t_{ a j} P_{\rm R}
+\hat{\Sigma}\!\!\!/_{{\rm L} a}Y_{i a} Y^\dagger_{ a j} P_{\rm L}
\right)\,.
\end{eqnarray}
Since gauge interactions keep all SM degrees of freedom in kinetic equilibrium, the deviations $\delta\H$, $\delta\G^\gtrless$ and $\delta\N$ can be parametrised in terms of the chemical potentials,
and the self energies thus fulfil the generalised Kubo-Martin-Schwinger (KMS) relations 
\begin{align}
\hat{\Sigma}\!\!\!/_{{\rm L},{\rm R} a}^>=-\ee^{(k^0\mp\mu_{\ell a} \mp \mu_\phi)/T}\hat{\Sigma}\!\!\!/_{{\rm L},{\rm R} a}^<\,,\label{generalKMS}
\end{align}
where $\mu_{\ell a}$ and $\mu_\phi$ are the chemical potentials of the SM leptons and the Higgs. 
These chemical potentials are small at all times of interest, and we expand in $\mu/T$ to linear order. This yields a linearised KMS relation
\begin{align}
\nonumber
\delta \hat{\Sigma}\!\!\!/^>_{{\rm L},{\rm R} a}&=
-\ee^{k^0/T}\left[\delta \hat{\Sigma}\!\!\!/^<_{{\rm L},{\rm R} a} \mp \frac{\mu_{\ell a}+\mu_\phi}{T} \left(\hat{\Sigma}\!\!\!/^< P_{{\rm L},{\rm R}}+\delta\hat{\Sigma}\!\!\!/^<_{{\rm L},{\rm R}a}\right)\right]\\
&\approx -\ee^{k^0/T}\left[\delta \hat{\Sigma}\!\!\!/^<_{{\rm L},{\rm R} a} \mp \frac{\mu_{\ell a}+\mu_\phi}{T}\hat{\Sigma}\!\!\!/^< P_{{\rm L},{\rm R}}\right]\,,
\end{align}
where in the second step we have suppressed the term that is quadratic in the chemical potentials.
Note that
\begin{align}
\bar{\Sigma}\!\!\!/_N^>=-\ee^{k^0/T}\bar{\Sigma}\!\!\!/_N^<\,.\label{KMSbar}
\end{align}
From the definition of $\slashed{\Sigma}_N^+$ and the KMS relation~(\ref{generalKMS}) it is clear that $\delta\N_N$ is quadratic in the chemical potentials, and we can neglect it.
We also neglect the term $\delta\H_N$. In principle, it is of the same order as the term with $\delta\G_N$, but  it only appears in a commutator, and $\delta\S_N$ is approximately proportional to a unit matrix for $T\gg M_i$, $Y\ll1$ ($\delta\S_N=-\bar{S}_N$ initially). 
This allows us to write 
\begin{align}
\label{eq:kinetic_eq_tensorial}
\partial_t \delta \mathcal{S}_N =2\frac{\partial_t f_F}{1-2f_F}\bar{\mathcal{S}}^+_N 
+\ii [\bar{\mathcal{H}}_N,\delta \mathcal{S}_N]-\{\bar{\mathcal{G}}_N,\delta \mathcal{S}_N\}-\frac{2}{1-2f_F}\sum_{a=e,\mu,\tau} \frac{\mu_{\ell a}+\mu_\phi}{T} \{\tilde{\mathcal{G}}^{a}_N, \bar{\mathcal{S}}^+_N\}\,,
\end{align}
with 
\begin{eqnarray}
\tilde{\mathcal{G}}^{a}_N=-g_w f_F[1-f_F]
\hat{\Sigma}\!\!\!/^\A_N
\left(
Y_{ia}^*Y_{aj}^tP_{\rm R} - Y_{ia}Y_{aj}^\dagger P_{\rm L}
\right)\gamma^0\,.
\end{eqnarray}
The Fermi-Dirac distribution $f_F=f_F(k^0)=1/(e^{k^0/T}+1)$ arises from the KMS relation~(\ref{KMSbar}).
We used the Leibniz rule for the term $\partial_t \bar{\mathcal{S}}^+_N$ that can be expressed in terms of the stationary quantity $\bar{\mathcal{S}}^\mathcal{A}_N$ and the distribution $f_F$, where the derivative only acts on the latter one.
$\bar{\H}_N$ and $\bar{\G}_N$ are the dispersive and dissipative part of an effective Hamiltonian.
The term $\tilde{\mathcal{G}}_N$ is responsible for the feedback or backreaction of the generated lepton asymmetry on the heavy neutrino dynamics. 

\paragraph{Lorentz Decomposition and Off-Shell Kinetic Equation}
We employ the decomposition of the non-equilibrium part $\delta \S_N=i\gamma^0\delta S_N$ of the heavy neutrino propagator $S_N$ in Wigner space in Lorentz components used in \cite{Garbrecht:2011aw,Fidler:2011yq},
\begin{eqnarray}
\label{eq:tensor_dec_propagator}
-\ii \gamma^0 \delta S_N = \sum_h \frac{1}{2} P_h \left(g_{0h} + \gamma^0 g_{1h} - \ii \gamma^0\gamma^5 g_{2h} -\gamma^5 g_{3h}\right)\,,
\end{eqnarray}
with the helicity projectors 
\begin{equation}
P_h \equiv \frac{1}{2}\left(1+h\hat{\bf k}\gamma^0\pmb{\gamma}\gamma^5\right)\,.
\end{equation}
Note that we are using the Weyl (chiral) representation of the Dirac matrices.
In the situation we consider here, they can all be expressed in terms of the  functions $ g_{0h}$, as shown explicitly in Refs.~\cite{Garbrecht:2011aw,Fidler:2011yq}. The relations can be found by taking traces over the products of the constraint equation~(\ref{constrainedSplus0}) with $P_h$ and different combinations of $\gamma$ matrices. To linear order in $M/k^0$ and  $Y$, they read\footnote{It remains to be seen if corrections to these relations of order $\mathcal{O}[Y^2]$ have a significant effect on the kinetic equation at $z=z_{\rm osc}$ when the vacuum masses are extremely degenerate.}
\bse
\begin{align}
\label{eq:constraint_propagator}
g_{1h}&=\frac{1}{2k^0}\left( \{\Re M,g_{0h}\} +
[\ii \Im M,g_{3h}] \right)\,, \\
g_{2h}&=\frac{1}{2\ii k^0}\left([\Re M,g_{3h}]+\{\ii \Im M,g_{0h}\} \right) \,,\\
g_{3h}&=h\sign(k^0)g_{0h}\,.
\end{align}
\ese
which is an approximation applicable in the regime $M_i\ll T$, $Y_{ia}\ll 1$.
The equilibrium function $\ii \bar{S}^+_N$ can be decomposed in analogy to $\ii \delta S_N$, where we replace $g_{bh}$ with the functions $\bar{g}_{bh}$.
In the self energies $\Sigma_N$ of heavy neutrinos and $\Sigma_\ell$ of SM leptons we consider terms up to second order in $Y$. 
The kinetic equation for $\delta g_{0h}$ can be obtained by taking the trace of Eq.~(\ref{eq:kinetic_eq_tensorial}) and inserting the relations~(\ref{eq:constraint_propagator}),
\begin{align}
\label{eq:off_shell}
\partial_t g_{0h} &= 2\frac{\partial_t f_F}{1-2f_F} \bar{g}_{0h} -\frac{\ii}{2} [ {\rm H}_N,g_{0h}] 
-
\frac12 \{ \Upgamma_N,g_{0h}
\} -\frac12 \frac{2}{1-2f_F}\sum_{a=e,\mu,\tau}
\frac{\mu_{\ell a}+\mu_\phi}{T}\{ \tilde{\Upgamma}_N^a,\bar{g}_{0h}
\}\,,
\end{align}
with
\bse
\begin{align}
{\rm H}_N&=2 g_w \left(
{\rm Re}[Y^*Y^t]\frac{k\cdot\hat{\Sigma}_N^{H}}{k^0}
-{\rm i}h {\rm sign}(k^0){\rm Im}[Y^*Y^t]\frac{k\cdot \hat \Sigma_N^{H}}{k^0}
\right)
\\&+\frac{1}{k^0}\left( \Re[M^\dagger M] +
\ii h \sign(k_0) \Im[M^\dagger M] \right)\,,
\\
\Upgamma_N
 &=2 g_w \left(
{\rm Re}[Y^*Y^t]\frac{k\cdot \hat \Sigma_N^{\A}}{k^0}
-{\rm i}h {\rm sign}(k^0){\rm Im}[Y^*Y^t]\frac{k\cdot \hat \Sigma_N^{\A}}{k^0}
\right) \, ,\\
(\tilde{\Upgamma}_N^a)_{ij} &=  2 h
g_w f_F(1-f_F) \left(
{\rm sign}(k^0){\rm Re}[Y^*_{ia}Y^t_{aj}]\frac{ k\cdot \hat \Sigma_N^{\A}}{k^0}-\ii h {\rm Im}[Y^*_{ia}Y^t_{aj}]\frac{k\cdot \hat \Sigma_N^{\A}}{k^0}
\right)\,.
\end{align}
\ese

\paragraph{On-Shell Kinetic Equation in an Expanding Universe}
The kinetic equation~(\ref{eq:off_shell}) for $g_{0h}$ very much resembles an equation for density matrices commonly used in neutrino physics~\cite{Sigl:1992fn}. However, it is still an equation of motion for correlation functions (rather than particle numbers) because all quantities are defined for general $k^0$ that may also be off shell. The feeble strength of the Yukawa interactions implies that the narrow width approximation holds for the sterile neutrino-quasiparticles, and all phase space integrals are strongly dominated by the quasiparticle poles $\Omega_i$, which are defined by the poles of ${\rm H}^{-1}$ in the flavour basis where ${\rm H}$ is diagonal. In that basis we can approximate
\begin{eqnarray}
\label{eq:quasiparticle_app}
\bar{g}_{0h}(k)_{ij}\approx
-\frac{1-2f_F}{2} \delta_{ij} 
2\pi\delta(k_0^2-\Omega_i^2)2k^0\sign(k^0)\,.
\end{eqnarray}
In the ultrarelativistic regime $T\gg M_i$, we can further approximate $\delta(k_0^2-\Omega_i^2)\simeq \delta(k^2)$ in Eq.~(\ref{eq:quasiparticle_app}) because kinematically, the (thermal and vacuum) masses are negligible. We do, however, have to include these as a part of ${\rm H}_N$ in the kinetic equation owing to their importance for flavour oscillations.
In Eq.~(\ref{eq:quasiparticle_app}) we have used that $\ii \bar{S}^+_N=\bar{S}^\A_N(1-2f_F)$ and $\ii \delta S_N=-2S^\A_N \delta f$. The above relations allow us to express the full equation~(\ref{eq:off_shell}) in terms of the equilibrium quantities $\bar{g}_{bh}$ and the perturbative part $\delta f_{bh}$ as
\begin{align}
\label{eq:quasiparticle_app:2}
g_{bh}=-\frac{2}{1-2f_F}\bar{g}_{bh}\delta f_{bh}\,.
\end{align}

It is convenient to define $\delta f_{0h}(\Omega_i,\k) + f_F(\Omega_i)$ as the number density for particles and 
$1 - \delta f_{0h}(-\Omega_i,\k) - f_F(-\Omega_i)$ as number density for antiparticles.
For the heavy neutrinos, the Majorana condition~(\ref{MajoranaSymmetry}) implies 
\begin{align}
\label{Majorana:dist}
\delta f_{0h}(-k^0)=\delta f_{0h}^*(k^0)\,,
\end{align}
 and there is no need to track particle and antiparticle numbers independently. For that reason we will focus on particles, hence we restrict to the case $\sign(k^0)=1$, and use Eq.~(\ref{Majorana:dist}) when needed.

Using
\begin{align}
\int  \!\frac{\dd k^0}{2\pi} g_{bh}=\delta f_{bh}\,,
\end{align}
we eventually obtain an equation for on-shell distribution functions by integrating over $k^0$,
\begin{align}
\partial_t \delta f_{0h} &= - \partial_t f_F -\frac{\ii}{2} [ {\rm H}_N,\delta f_{0h}] 
-
\frac12 \{ \Upgamma_N,\delta f_{0h}
\}+\sum_{a=e,\mu,\tau}
\frac{\mu_{\ell a}+\mu_\phi}{T}\tilde{\Upgamma}_N^a\,.
\end{align}
Note that we keep the same notation for $\Upgamma_N$, $\tilde{\Upgamma}_N$ and ${\rm H}_N$ while these
quantities are restricted to on-shell arguments $k^0=|\mathbf k|$ in above equation.

So far we have carried out our derivations in Minkowski spacetime.
During the radiation-dominated era, the expansion of the Universe can simply be included by using conformal time $\eta$ instead of physical time \cite{Beneke:2010wd},
\begin{align}
\delta f_{0hij}^\prime&
+\frac{\ii}{2} \left[{\rm H}_N,\delta f_{0h} \right]_{ij}
+(f^{\rm eq})^{\prime}_{ij}
= -\frac12\left\{\Upgamma_N, \delta f_{0h}\right\}_{ij} +\sum_{a=e,\mu,\tau}\frac{\mu_{\ell a}+\mu_\phi}{T}(\tilde{\Upgamma}_N^a)_{ij}\,.
\label{evolution:sterile}
\end{align}
A prime denotes a derivative with respect to the conformal time $\eta$, and we additionally have made the flavour content explicit.
During radiation domination, $a=a_{\rm R}\eta$, and when using the parametrisation
introduced in Section~\ref{sec:evo_eq}, $\eta=T/T_{\rm ref}$.
Note also that since $a_{\rm R}/a=T$ can be interpreted as a comoving temperature, the equilibrium distribution for massless sterile neutrinos that appears in Eq.~(\ref{evolution:sterile}) is given by
\begin{align}
f^{\rm eq}=\frac{1}{\ee^{|\k|/a_{\rm R}}+1}\,.
\end{align}
This distribution does not depend on conformal time, such that the term $f^{\rm eq '}=0$ in Eq.~(\ref{evolution:sterile}).

The effective Hamiltonian can be decomposed into a vacuum mass and a thermal mass term
such that $({\rm H}_N)_{ij}=({\rm H}_N^{\rm vac})_{ij}+({\rm H}^{\rm th}_N)_{ij}$ with
\bse
\label{Hamiltonian:effective}
\begin{align}
\label{H_vac}
({\rm H}_N^{\rm vac})_{ij}&=\frac{a^2}{|\k|} \left(\Re[M^\dagger M]_{ij} +\ii h  \Im[M^\dagger M]_{ij}\right)\,,\\
\label{H_therm}
({\rm H}_N^{\rm th})_{ij}&= 2g_w \left(
{\rm Re}[Y^*Y^t]_{ij}-{\rm i}h {\rm Im}[Y^*Y^t]_{ij}\right)
\frac{k\cdot \hat \Sigma_N^{H}}{|\k|}
\,,
\end{align}
\ese
where $M$ is the vacuum mass matrix [{\it cf.} Eq.~(\ref{eq:Lagrangian})] such that Eq.~(\ref{H_vac}) is written in the most general form allowing for a complex symmetric mass matrix. 
$\hat \Sigma_N^{H}$ is the Hermitian part of the reduced self-energy of the
sterile neutrino as defined in Eq.~(\ref{hatSigmaNDef})~\cite{Garbrecht:2011aw}. 
The anticommutator terms involve the reduced
spectral self-energy $\hat \Sigma_N^{\mathcal{A}}$,
and they are responsible for the decay of the deviations $\delta f_{0h}$ toward equilibrium.
Inserting the momentum and flavour structure one finds~\cite{Garbrecht:2011aw}
\begin{align}
\label{boltz:dec}
(\Upgamma_N)_{ij}=
2g_w \left(
{\rm Re}[Y^*Y^t]_{ij}
-{\rm i}h {\rm Im}[Y^*Y^t]_{ij}\right)\frac{k\cdot \hat \Sigma_N^{\cal A}}{|\k|}\,.
\end{align}
The oscillations of sterile neutrinos induce flavour asymmetries in the active sector. The produced SM charges, {\it i.e.} those within the doublet leptons $\ell_a$ of flavour $a$ and the Higgs field $\phi$, then lead to
a backreaction effect that is described by the term
\begin{align}
\nonumber
\label{boltz:back}
&\frac{\mu_{\ell a}+\mu_\phi}{T}(\tilde{\Upgamma}_N^a)_{ij}\to\\
&2h g_w
\left({\rm Re}[Y^*_{ia}Y^t_{aj}]-\ii h {\rm Im}[Y^*_{ia}Y^t_{aj}]\right)
\frac{k\cdot \hat \Sigma_N^{\cal A}}{|\k|}
\frac{\ee^{|\k|/a_{\rm R}}}{(\ee^{|\k|/a_{\rm R}}+1)^2}\left(\frac{\mu_{\ell a}}{a_{\rm R}}+\frac{\mu_{\phi}}{a_{\rm R}}\right)\,.
\end{align}
Here, $\mu_{\ell a}$ and $\mu_\phi$ are chemical potentials for the doublet leptons and the Higgs boson, 
where we assume kinetic equilibrium for these species.
We have linearised here in the chemical potentials, which is a valid approximation when
$\mu_{\ell a,\phi}\ll T$.

It is convenient to define helicity-even and helicity-odd parts of the distribution
functions 
\begin{subequations}
\label{f:odd:even}
\begin{align}
\delta\feven (k)&=\frac{\delta f_{0+}(k)+\delta f_{0-}(k)}{2}\,,\label{fevendef}\\
\delta\fodd (k)&=\frac{\delta f_{0+}(k)-\delta f_{0-}(k)}{2}\label{fodddef}\,.
\end{align}
\end{subequations}
In this work we assume that all lepton asymmetries remain small at all times. This allows to perform expansions to linear order in the $\mu_{\ell a}$ and $\delta\fodd$. 
We cannot expand in $\delta\feven$ because the initial state~(\ref{initialstate}) implies that 
\begin{equation}
\delta\feven_{ij}(k)=-f^{\rm eq}(|\k|) \delta_{ij} \simeq -\delta_{ij}
\end{equation} 
at initial time, where the second equality holds on shell and at $T\gg M_i, M_j$.

\paragraph{Rate Equations for Number Densities}
Though Eq.~(\ref{evolution:sterile}) has the same form as a density matrix equation for (quasi)particle occupation numbers, it is an equation of motion for the propagator $(S_N)_{ij}$ (which can be expressed in terms of the distribution functions $\delta f_{0h ij}$). 
In particular, it is valid for off-shell values of $k^0$ and holds for each momentum mode $\mathbf{k}$ individually. 
When accounting for backreaction effects, there will also be a coupling among the modes via $\tilde{\mathcal{G}}$. This is a considerable complication, and resolving
the full momentum dependence would be a road block toward the goal of
finding simple analytic approximations as well as fast numerical solutions.
We therefore follow the common procedure~\cite{Akhmedov:1998qx,Asaka:2005pn,Canetti:2012kh}
of reducing the problem to number densities in flavour of distribution functions
by averaging the rates over the momentum. As we discuss below,
this leads to order one uncertainties in the final result that should be resolved in future work. Some progress in this direction has been made in Ref.~\cite{Asaka:2011wq}.
The developments that we present here may be helpful in order to address this issue.

In order to cast Eq.~(\ref{evolution:sterile}) into a relation for the number densities of the
sterile neutrinos, we perform an integration over momentum space. We are lead to introduce
the equilibrium number density
\begin{align}
\label{n_eq}
n^{\rm eq}=\int \!\frac{\dd ^3k}{(2\pi)^3}f^{\rm eq}=\frac{3}{4\pi^2}a_{\rm R}^3\zeta(3)
\end{align}
and the deviations
\begin{align}
\delta n_{hij}=\int \!\frac{\dd ^3k}{(2\pi)^3}\,\delta  f_{0hij}(\textbf{k})\,.\label{nDeviationDef}
\end{align}
The number densities $\delta n^{\rm even}$ and $\delta n^{\rm odd}$ are then defined analogously
based on the distribution functions~(\ref{f:odd:even}).
We face the usual problem of approximating the momentum integral over products on the right-hand side of Eq.~(\ref{evolution:sterile})
by products of momentum integrals.
Under the integral, the distribution functions are in general multiplied
by different powers of the momentum. Inspection of the individual terms
in Eq.~(\ref{evolution:sterile}) (that we discuss explicitly below) reveals that there
are factors independent of $k$ as well as factors of $1/|\k|$. In order to account for the latter, we replace $1/|\k|$ by its average value
\begin{align}
\label{k:av}
\left\langle\frac{1}{|\mathbf{k}|}\right\rangle\equiv \frac{1}{{n^{\rm eq}}}\int \!\frac{\dd ^3k}{(2\pi)^3}\frac{1}{|\mathbf{k}|}f^{{\rm eq}}(\mathbf k)=\frac{\pi^2}{18 a_{\rm R}\zeta(3)}\,.
\end{align}

For $T\gg M_i$, the spectral self-energy of the sterile neutrinos, that appears in ${\cal G}$ and $\tilde{\cal G}$,
is dominated by the $t$-channel exchange of a doublet lepton in association with the radiation of a gauge boson~\cite{Besak:2012qm,Garbrecht:2013urw,Garbrecht:2014bfa}. 
We follow Ref.~\cite{Drewes:2012ma}, where the momentum averaging is applied through the replacement
\begin{align}
\label{col:av}
\frac{k\cdot\hat\Sigma^{\cal A}_N}{|\k|}\to \frac{\gamma_{\rm av} a_{\rm R}}{2 g_w}\,,
\end{align}
with the averaged relaxation rate $\gamma_{\rm av}\equiv\Gamma_{\rm av}/T$.
This rate has been computed in different regimes by various authors \cite{Anisimov:2010gy,Kiessig:2010pr,Garbrecht:2010sz,Laine:2011pq,Fidler:2011yq,Garbrecht:2011aw,Salvio:2011sf,Biondini:2013xua,Besak:2012qm,Garbrecht:2013bia,Bodeker:2014hqa,Laine:2013lka,Garbrecht:2013gd,Bodeker:2014hqa,Ghisoiu:2014ena,Ghiglieri:2016xye}.
Here we use $\gamma_{\rm av}=0.012$, corresponding to the value from Ref.~\cite{Garbrecht:2014bfa} based on Refs.~\cite{Besak:2012qm,Garbrecht:2013urw}.
In the backreaction term~(\ref{boltz:back}), it is useful to replace the
chemical potentials with charge densities according to Eq.~(\ref{ChargeDefs}).
In the effective Hamiltonian $H_N$, we substitute
the leading hard thermal loop contribution to the Hermitian self-energy given by
$k\cdot\hat\Sigma_N^H=T^2/8$~\cite{Weldon:1982bn} 
\footnote{
Note that in our definition, we account for the
fact that particles and antiparticles run in the loop correction to the Majorana propagator
while the gauge multiplicity enters through the explicit factors of $g_w$ in Eqs.~(\ref{Hamiltonian:effective}).
}
In total, we can decompose the Hermitian part as follows
\begin{align}
\nonumber
\frac{k\cdot\hat\Sigma^{H}_N}{|\k|}
=\frac{a_{\rm R}^2}{16 |\k|}
\to \frac{a_{\rm R}}{2 g_w} 
\mathfrak{h}_{\rm th}
\end{align}
where using Eq.~(\ref{k:av}), the coefficient $\mathfrak{h}_{\rm th}$ is given by
\begin{align}
\label{eq:higgs_thermal}
\mathfrak{h}_{\rm th}\approx 0.23.
\end{align}

In summary, when integrating Eq.~(\ref{evolution:sterile}) over the three momentum $\vect{k}$,
we obtain the momentum averaged evolution equation for the sterile number densities
\begin{align}
\label{diff:sterile}
\frac{\dd}{\dd z}\delta n_{h} = -\frac{\ii}{2}[H^{\rm th}_N+z^2 H^{\rm vac}_N,\delta n_{h}]-\frac{1}{2}\{\Gamma_N,\delta n_{h}\}+\sum_{a=e,\mu,\tau}\tilde{\Gamma}^a_N \left(q_{\ell a}+\frac12 q_{\phi}\right)\,,
\end{align}
and the rates given in equations (\ref{avg:hamiltonian})-(\ref{avg:backreaction})
\bse
\begin{align}
H^{\rm vac}_N &=
\frac{\pi^2 }{18 \zeta(3)}\frac{a_{\rm R}}{T_{\rm ref}^3}
\left(\Re[M^\dagger M] + \ii h  \Im[M^\dagger M]\right)\,,\nonumber\\
H^{\rm th}_N&=\mathfrak{h}_{\rm th}\frac{a_{\rm R}}{{T_{\rm ref}}}
\left(\Re[Y^* Y^t]-\ii h\Im [Y^* Y^t]\right)\,,\nonumber\\
\Gamma_N &=\gamma_{\rm av} \frac{a_{\rm R}}{{T_{\rm ref}}}
\left(\Re[Y^* Y^t]-\ii h \Im [Y^*Y^t]\right)\,,\nonumber\\
(\tilde{\Gamma}^a_N)_{ij}&= \frac{h}{2}\gamma_{\rm av} \frac{a_{\rm R}}{T_{\rm ref}}
\left(
\Re [Y^*_{ia}Y^t_{aj}] - \ii h  \Im [Y^*_{ia}Y^t_{aj}]
\right)\,.\nonumber
\end{align}
\ese
The result~(\ref{diff:sterile}) immediately leads to Eq.~(\ref{Diff:Sterile}) when including
spectator effects.
Note that if all distribution functions appearing under the momentum integral were of the form
$f^{\rm eq}(\mathbf k)$, the averaging procedure would not incur any inaccuracy. However, this form cannot
be assumed for $\delta f_{0h}(\mathbf k)$ and it neither holds for the statistical factor in Eq.~(\ref{boltz:back}).
Nonetheless, since all of these distributions take the form of a Boltzmann tail for $|\mathbf{k}|\gg a_{\rm R}$,
the error incurred is only of order one. For comparison, along the same lines, we can see that
momentum averaging for leptogenesis from non-relativistic sterile neutrinos does not lead to a
leading order inaccuracy because all distributions are well approximated by the
Maxwell-Boltzmann form.

\section{Evolution of SM Charges}
\label{AppendixSMCharges}

\subsection{Kinetic Equations}

The evolution equations for the charge densities $q_{\ell a}$ of doublet leptons are
\begin{align}
\label{evolution:active}
\frac{\dd q_{\ell a}}{\dd z}=-\frac{a_{\rm R}}{T_{\rm ref}}W_a \left(q_{\ell a}+\frac12 q_\phi-q_{Nii} \right)+
\frac{1}{T_{\rm ref}}S_{a}\,,
\end{align}
where  $S_a$ is the source for the asymmetry (which is defined as the part of the collision term that is non-vanishing even if all $\mu_{\ell a}=0$) and $W_a$ is the rate for washout (which is defined as the remainder of the collision terms).
Besides, we introduce the sterile charge as the helicity-odd part of the deviation
of the number densities from equilibrium,
\begin{align}\label{qNDef}
q_{Nij}=\delta n_{+ij}-\delta n_{-ij}
=2n_{ij}^{\rm odd}
\,.
\end{align}
This quantity is useful because
in the limit $M\to 0$, $q_N$ can be identified with a charge
density contributing to a conserved (modulo weak sphalerons) generalised lepton number along with
the doublet leptons and the charged right-handed leptons. In the present
context, where the sterile neutrinos are
relativistic, we can neglect the reactions that violate the generalised lepton
number. This is because for a typical momentum mode
the admixture of opposite chirality to a spinor of given helicity is of
order $M/T$, such that the processes mediated by the Yukawa couplings
$Y$ that violate the generalised lepton number
are suppressed by a relative factor of $M^2/T^2$ compared those that conserve
the generalised lepton number and that are accounted for in the present work.\footnote{In a situation where the eigenvalues of $YY^\dagger$ are very different in size, this argument might not hold because the $M^2/T^2$-suppression of lepton number violating processes involving the larger coupling may not be sufficient to suppress them relative to the lepton number conserving processes mediated by the weaker Yukawa coupling.
This issue, which should be addressed in future work, introduces an uncertainty in the results found in Section~\ref{sec:str}.
}
We also note that there are contributions from the off-diagonal correlations
in the sterile neutrinos that we attribute implicitly to the source term $S_a$.
In analogy with the terms proportional to $\tilde\Gamma^a$ in Eq.~(\ref{diff:sterile}), we
refer to the contribution involving $q_{N ii}$ in Eq.~(\ref{evolution:active}) as
a backreaction term.

The washout rate is complementary to the damping rates for sterile neutrinos,
{\it cf.} Eqs.~(\ref{RHN:rates}), and is given by
\begin{align}
W_a&=\frac{\gamma_{\rm av}}{g_w}\sum_i Y_{ia}Y_{ai}^\dagger\,.
\end{align}
The off-diagonal correlations of the sterile neutrinos give rise to the source
for charge asymmetries in the doublet leptons,
\begin{align}
\label{sourceterm}
S_{ab}&=-\sum\limits_{\overset{i,j}{i\not=j}}
Y^*_{ia}Y_{jb}\int \!\frac{\dd ^4 k}{(2\pi)^4}
{\rm tr}\left[P_{\rm R} {\rm i} \delta S_{N ij}(k) 2 P_{\rm L}\hat {\slashed\Sigma}_N^{\cal A}\right]\,,
\end{align}
where in Eq.~(\ref{evolution:active}) we use the shorthand
notation $S_a\equiv S_{aa}$. While in principle, there can also be off-diagonal
correlations in the doublet charges, we set these to zero in the present context because
at the temperatures we consider, processes mediated by the $\mu$ and $\tau$ Yukawa couplings
erase these by the mechanism described in Ref.~\cite{Beneke:2010dz,Blanchet:2011xq} corresponding
to leptogenesis in the fully flavoured regime~\cite{Abada:2006fw,Nardi:2006fx,Abada:2006ea}.

Because $T\sim |\mathbf k|\gg M_{ii}$ for the typical momentum scale,
we can focus on the limit of massless (ultrarelativistic) sterile neutrinos,
where $|\mathbf k|\approx {\rm sign}(k^0) k^0$.
Further, it is useful to decompose sterile propagator 
in the relativistic regime
\begin{align}
\label{deltaS:relativistic}
\notag
{\rm i}\delta S_N(k)&=2\pi\delta(k^2)2k^0 \frac{\delta f_{0+}(k)+\delta f_{0-}(k)}{2}
\left(
-\frac12\gamma^0+\frac12 \hat{\mathbf k} \cdot{\pmb{\gamma}}\,{\rm sign}(k^0)
\right)
\\
&+2\pi\delta(k^2)2k^0 \frac{\delta f_{0+}(k)-\delta f_{0-}(k)}{2}
\left(
\frac12\gamma^0\gamma^5\,{\rm sign}(k^0)-\frac12 \hat{\mathbf k} \cdot{\pmb{\gamma}}\gamma^5
\right)
\end{align}
in terms of helicity odd and even functions~(\ref{fevendef}) and~(\ref{fodddef}),
\begin{align}
{\rm i}\delta S_N(k)&=2\pi\delta(k^2)
\left[
-\slashed k \delta\feven (k)
+{\rm sign}(k^0)\slashed k \gamma^5 \delta\fodd (k)
\right]\,.
\end{align}
Here we have used Eqs.~(\ref{eq:tensor_dec_propagator}) and~(\ref{eq:constraint_propagator}), where the off-shell correlators $g_{1h}$ and $g_{2h}$ are suppressed by a factor $k^0/M$ and where $g_{3h}$ is related to $g_{0h}$. Additionally, the on-shell condition~(\ref{eq:quasiparticle_app:2}) has been used
in the form of
\begin{align}
g_{0h}=2\pi \delta(k^2)\sign(k^0)2k^0\delta f_{0h}\,.
\end{align}
Substitution into the source term~(\ref{sourceterm}) yields
\begin{align}
\label{source:relativistic}
S_{ab}&=\sum\limits_{\overset{i,j}{i\not=j}}Y^*_{ia}Y_{jb}
\int \!\frac{\dd ^3 k}{(2\pi)^3}\frac{1}{2|\mathbf k|}\sum\limits_{s_k=\pm}4k\cdot\hat\Sigma^{\cal A}_N
\left[
\delta\feven_{ij} (k)+s_k \delta\fodd_{ij} (k)
\right]\,.
\end{align}
This corresponds to the relativistic limit of the more general result derived in
Ref.~\cite{Garbrecht:2011aw}. 

Provided we can neglect the term $\delta f_{0hij}^\prime$ compared to the commutator term
in Eq.~(\ref{evolution:sterile}) and we can also drop the backreaction term as well as the
thermal masses, it follows
\begin{align}
\delta f_{0h ij}=-\ii g_w  \frac{4 k\cdot\hat\Sigma^{\cal A}_N}{a^2(M_{ii}^2-M_{jj}^2)}f^{\rm eq}\left({\rm Re}[Y^*Y^t]_{ij}-\ii h \sign(k^0){\rm Im}[Y^*Y^t]_{ij}\right)\,,
\end{align}
and we recover the result from Ref.~\cite{Drewes:2012ma} for the source term.
This gives rise to a first approximation for the asymmetry in the \emph{weak washout} regime,
which we improve upon in Section~\ref{sec:weak}.

In terms of momentum averaged expressions, the source term can be written as
\begin{align}
\label{source_av}
S_{ab}&=2\frac{\gamma_{\rm av}}{g_w}a_{\rm R}\sum\limits_{\overset{i,j}{i\not=j}}Y^*_{ia}Y_{jb}
\left[\ii{\rm Im}(\delta n^{\rm even}_{ij})+{\rm Re}(\delta n^{\rm odd}_{ij})\right]\,,
\end{align}
such that in total,
we obtain the differential equation for the evolution of the SM charges 
\begin{align}
\label{diff:active}
\nonumber
\frac{\dd q_{\ell a}}{\dd z}=
&-\frac{\gamma_{\rm av}}{g_w} \frac{a_{\rm R}}{T_{\rm ref}}\sum_i Y_{ia}Y_{ai}^\dagger
\,\left(q_{\ell a}+\frac12 q_\phi -q_{Ni}\right)\\
&+2\frac{\gamma_{\rm av}}{g_w} \frac{a_{\rm R}}{T_{\rm ref}}\sum\limits_{\overset{i,j}{i\not=j}}Y^*_{ia}Y_{ja}
\left[\ii{\rm Im}(\delta n^{\rm even}_{ij})+{\rm Re}(\delta n^{\rm odd}_{ij})\right]\,.
\end{align}
Eqs.~(\ref{diff:sterile}) and~(\ref{diff:active}) form a coupled system of differential equation for the active and sterile charges. In order to solve the whole system one can decompose Eq.~(\ref{diff:sterile}) into even and odd parts and seek for numerical solutions. However, we can identify different parameter regions, such as the oscillatory and overdamped regime, where approximate analytic solutions can be found,
as presented in Sections~\ref{sec:weak} and~\ref{sec:str}, respectively.

\subsection{Spectator Processes}\label{spectatorSec}
Standard Model processes redistributing charges during leptogenesis are called
spectator effects and affect the final result for the
baryon asymmetry~\cite{Barbieri:1999ma,Buchmuller:2001sr}.
In order to account for these it is useful to work with the asymmetries $\Delta_a=B/3-L_a$
defined in Eq.~(\ref{Delta:a}) which are conserved by all interactions other than those mediated by the Yukawa couplings $Y$ between the active and sterile sectors.
We then need to relate these asymmetries to the charge densities that appear on the right
hand sides of the evolution equations~(\ref{evolution:sterile}).
At temperatures below $T\lesssim 10^5\GeV$, when the electron as the SM particle
with the smallest Yukawa coupling finally reaches chemical equilibrium (see {\it e.g.} Ref.~\cite{Garbrecht:2014kda} for an overview of the equilibration rates of spectator processes),
the SM Yukawa-mediated processes lead to the constraints
\begin{subequations}
\label{equil:yukawa}
\begin{align}
\mu_{Qi}-\mu_{ui}+\mu_\phi&=0\,,\\
\mu_{Qi}-\mu_{di}-\mu_\phi&=0\,,\\
\mu_{\ell i}-\mu_{ei}-\mu_\phi&=0\,.
\end{align}
\end{subequations}
Besides, weak and strong sphaleron processes force the relations
\begin{subequations}
\label{equil:sphal}
\begin{align}
g_s(\mu_{Q1}+\mu_{Q2}+\mu_{Q3})+\mu_{\ell 1}+\mu_{\ell 2}+\mu_{\ell 3}&=0\,,\\
g_w(\mu_{Q1}+\mu_{Q2}+\mu_{Q3})-(\mu_{u1}+\mu_{u2}+\mu_{u3})-(\mu_{d1}+
\mu_{d2}+\mu_{d3})&=0\,,
\end{align}
\end{subequations}
where $Q_i$ denote left-handed quark doublets of flavour $i$, $u_i,\,d_i$ are the corresponding right-handed electroweak singlets and the factor $g_s=3$ accounts for the three colour states.
A common chemical potential for the weak doublets and colour triplets implies that the charge
densities associated with the diagonal generators for weak and strong interactions vanish.
Correspondingly a vanishing density of weak hypercharge leads to the condition
\begin{align}
\label{equil:hyper}
g_w Y_\phi q_\phi+\sum_{a=e,\mu,\tau}\left( g_w g_s Y_{Qa}q_{Qa}+g_w Y_{\ell a}q_{\ell a}+g_s Y_{ua}q_{ua}+g_s Y_{da}q_{da}+Y_{ea}q_{ea}\right)=0\,,
\end{align}
where we explicitly note the summation over the three active flavour indices.
We can now can solve Eqs.~(\ref{equil:yukawa}, \ref{equil:sphal}, \ref{equil:hyper}) in order
to obtain the desired relations
between the
charge densities of doublet leptons $q_{\ell 1,2,3}\equiv q_{\ell e,\mu,\tau}$ as well as of the Higgs bosons $q_\phi$ and the asymmetries $\Delta_{1,2,3}\equiv \Delta_{e,\mu,\tau}$. These are conveniently expressed as $q_\ell=A \Delta$ and $q_\phi=C \Delta$.
This way we obtain the matrices $A$ and $C$ given in equation (\ref{spectator}) as
\begin{align}
A=
\frac{1}{711}
\left(
\begin{array}{ccc}
-221 & 16 & 16\\
16 & -221  & 16\\
16 & 16 & -221
\end{array}
\right)
\,,
\quad
C=
-\frac{8}{79}
\left(
\begin{array}{ccc}
1 & 1 & 1
\end{array}
\right)\nonumber
\end{align}
and where $q_\ell=(q_{\ell 1},q_{\ell 2},q_{\ell 3})^t$ as well as $\Delta=(\Delta_1,\Delta_2,\Delta_3)^t$ are understood as column vectors in lepton flavour space. For completeness, we also define the column vector  $q_N=(q_{N_1},q_{N_2},\dots,q_{N_{ns}})^t$ for $n_s$ sterile neutrinos.
Besides, in terms of $\Delta$ we can express the baryon asymmetry as
\begin{align}
\label{baryon_charge_Delta}
B=D\Delta\,, \quad D=
\frac{28}{79}
\left(
\begin{array}{ccc}
1 & 1 & 1
\end{array}
\right)\,.
\end{align}
One may also relate the asymmetry in doublet leptons to the baryon asymmetry,
\begin{align}
\label{baryon_charge_q}
B=E q_\ell\,, \quad E=
-\frac{4}{3}
\left(
\begin{array}{ccc}
1 & 1 & 1
\end{array}
\right)\,.
\end{align}
Note that this calculation is consistent with the well-known
relation~\cite{Harvey:1990qw} $B=\frac{28}{79}(B-L)$. Because of the crossover nature of the
electroweak phase transition in the SM, there is another ${\cal O}(10\%)$ correction to this
relation~\cite{Khlebnikov:1996vj,Laine:1999wv}. In view of the sensitivity of the asymmetries from ${\rm GeV}$-scale leptogenesis
to spectator effects, it should be of interest to include this correction along with the
time dependence of the rate of weak sphaleron transitions prior to their quench. Both corrections
will lead to a temperature dependence in above conversion relations, a detailed study of which
we leave to future work.

\section{Oscillatory Regime}
\label{App:Oscillatory}
\subsection{Time Scales in the Oscillatory Regime}
\label{app:time_scales}

For the validity of the approximations used to calculate the initial asymmetry in the oscillatory regime, the equilibration time
\begin{align}
z_\text{eq}
\approx
\left(
2 g_w \lVert Y^* Y^t \rVert \frac{k\cdot \hat \Sigma_N^{\cal A}}{|\k|}
\right)^{-1} T_{\rm ref}
\approx
\frac{ T_{\rm ref}}{
\lVert Y^* Y^t \rVert \gamma_\text{av} a_{\rm R}
}
\,,
\end{align}
given here by the inverse of the smallest eigenvalue of the decay matrix~(\ref{boltz:dec}) needs to be much larger than the time by which the first oscillation is over. This oscillation time scale is determined by the difference of the squared masses
\begin{align}
z_\text{osc} \approx \left(a_{\rm R} |M_i^2 - M_j^2|\right)^{-1/3} T_{\rm ref}\,.
\end{align}
In the coordinates we have chosen, Eq.~(\ref{diff:sterile}) implies that the frequency of the oscillation $\omega_{\rm vac}$ induced by the vacuum term $H_N^{\rm vac}$ increases with $z^2$, whereas the thermal contribution $H_N^{\rm th}$ results in a constant oscillation frequency $\omega_{\rm th}$. For this reason the nonzero thermal oscillation may be of importance at early times when the vacuum oscillation has not started yet. However, one can show that $\omega_{\rm vac}$ is automatically larger than $\omega_{\rm th}$ at the time of the first oscillation $z_{\rm osc}$ when imposing $z_\text{eq}\gg z_\text{osc}$:
\begin{align}
\omega_{\rm vac}=a_{\rm R}|M_i^2-M_j^2|\eta_{\rm osc}^2=a_{\rm R}^{1/3}|M_i^2-M_j^2|^{1/3}\gg \lVert Y^* Y^t \rVert \mathfrak{h}_\text{th} a_{\rm R}= \omega_{\rm th}\,,
\end{align}
with 
$\mathfrak{h}_{\rm th}=0.23$.
This implies that in the oscillatory regime the thermal effects may only have lead to a small fraction of a full flavour oscillation by the time when the first oscillation due to the vacuum masses already has been completed. Since the main part of the active charge is generated during the first oscillation, one can consider the contribution from the thermal masses as a perturbation. 

It is easy to show that the perturbative corrections to $\delta n_{ij}$ arising due to the presence of $H_N^{\rm th}$ vanish at order $\mathcal{O}(\mathfrak{h}_{\rm th}/\gamma_{\rm av}|Y^*Y^t|)$
as the leading order term of the out-of-equilibrium distribution is $\delta n_{ij} = -  n^\text{eq} \delta_{ij}$ and hence
\begin{align}
\left[H_N^\text{th},\delta n \right] = 0\,.
\end{align}
The first non-vanishing contribution from the thermal masses is of order $\mathcal{O}(\mathfrak{h}_{\rm th}/\gamma_{\rm av}|Y^*Y^t|^2)$,
which can be neglected compared to the contributions coming from the vacuum masses
$\delta n_{ij}$ of order $\mathcal{O}(|Y^*Y^t|)$, {\it cf.} Eqs.~(\ref{sol:f0hav}).

\subsection{Momentum Dependence of the Source}
\label{app:act_charg}

In Section~\ref{sec:weak} we have calculated the active charge produced through the off-diagonal oscillations of the sterile neutrinos to order $|Y^*Y^t|^2$ with the simplification of fully momentum averaged expressions. We can go one step further and consider the momentum dependence of the vacuum term ${\rm H}_N^{\rm vac}$ as in Eq.~(\ref{H_vac}) but still keep the replacement~(\ref{col:av}) in order to able to solve the remaining momentum integral analytically. For this reason we solve
\begin{align}
\delta f_{0hij}^\prime&
+\frac{\ii}{2} \left[{\rm H}_N^{\rm vac},\delta f_{0h} \right]_{ij}
= -\frac12\left\{\Upgamma_N, \delta f_{0h}\right\}_{ij} \,,
\end{align}
by analogy with Eq.~(\ref{osc:weakwashoutav}) for the even and odd parts of the off-diagonal distributions $\delta f_{ij}$ whose solution  to order $|Y^*Y^t|$ can be obtained analogously

\bse
\begin{align}
\fodd_{ij}&=-\ii{\rm Im}[Y^*Y^t]_{ij}\tilde{G}\mathcal{F}_{ij}\,,\quad \quad 
\feven_{ij}={\rm Re}[Y^*Y^t]_{ij}\tilde{G}\mathcal{F}_{ij}\,,\\
\tilde{\Omega}_{ij}&=\frac{a_{\rm R}^2}{T_{\rm ref}^3 2 k^0}(M_{ii}^2-M_{jj}^2)\,, \quad \quad
\tilde{G}=2 g_w  \frac{k\cdot\hat\Sigma^{\cal A}_N}{|\mathbf k|T_{\rm ref}}f^{\rm eq}(\vect{k})\,.
\end{align}
\ese
with $\mathcal{F}_{ij}$ from Eq.~(\ref{F_tilde}) where $\Omega_{ij}$ is replaced by $\tilde{\Omega}_{ij}$. These can be plugged in into Eq.~(\ref{q:source}) with the source term~(\ref{source:relativistic}), where summation over positive and negative $k^0$ yields
\begin{align}
\exp\left(\frac{\ii\pi}{3}\sign(M_{ii}^2-M_{jj}^2)\right)-\exp\left(-\frac{\ii\pi}{3}\sign(M_{ii}^2-M_{jj}^2)\right)=\ii\sqrt{3}\,\sign(M_{ii}^2-M_{jj}^2)\,,
\end{align}
while the integration over $z$ remains unchanged, so that the active charge is given by
\begin{align}
\frac{\tilde{\Delta}_{a}^{\rm sat}}{s}&=\frac{20{\rm i}g_w}{g_\star}
\frac{3^{\frac23}\Gamma(\frac16)}{\pi^{\frac{3}{2}}a_{\rm R}^{13/3}}
\sum\limits_{\overset{i,j,c}{i\not=j}}
Y_{ai}^\dagger Y_{ic} Y_{cj}^\dagger Y_{jb}\frac{{\rm sign}(M_{ii}^2-M_{jj}^2)}{|M_{ii}^2-M_{jj}^2|^{\frac23}}\times \mathcal{I}
\end{align}
with a function that carries all momentum information
\begin{align}
\label{mom_int}
\mathcal{I}=\int \!\frac{\dd ^3 k}{(2\pi)^3}|\mathbf{k}|^{-\frac43}\frac{(k\cdot\hat\Sigma^{\cal A}_N)^2|_{k^0=|\vect{k}|}}{{\rm e}^{|\mathbf{k}|/a_{\rm R}}+1}
\,.
\end{align}
Solving this integral exactly is beyond the scope of this paper since $k\cdot\hat\Sigma^{\cal A}_N$ has a non-trivial momentum structure~\cite{Asaka:2011wq}. Nevertheless, we can use the momentum averaged replacement~(\ref{col:av}), which leaves us with a momentum integral that can easily be solved analytically:
\begin{align}
\int \!\frac{\dd ^3 k}{(2\pi)^3}|\mathbf{k}|^{\frac23}\frac{1}{{\rm e}^{|\mathbf{k}|/a_{\rm R}}+1}
=\frac{1}{2\pi^2}a_{\rm R}^{\frac{11}{3}}\left(1-2^{-\frac83}\right)\Gamma\left(\frac{11}{3}\right)\zeta\left(\frac{11}{3}\right)\,.
\end{align}
Thus, the total active charge produced in the \emph{weak washout} regime, before the washout kicks in, is given by
\begin{align}
\nonumber
\frac{\tilde{\Delta}_{a}^{\,\rm sat}}{s}&=\frac{\rm i}{g_\star^{\frac 53}}
\frac{3^{2} 5^{\frac53}(2-2^{-\frac53})\Gamma(\frac16)\Gamma(\frac{11}{3})\zeta(\frac{11}{3})}{2^{\frac{10}{3}}\pi^{\frac{11}{2}}}
\sum\limits_{\overset{i,j,c}{i\not=j}}
\frac{Y_{ai}^\dagger Y_{ic} Y_{cj}^\dagger Y_{jb}}{{\rm sign}(M_{ii}^2-M_{jj}^2)}\left(\frac{m_{\rm Pl}^2}{|M_{ii}^2-M_{jj}^2|}\right)^{\frac23}\frac{\gamma_{\rm av}^2}{g_w}
\\
&\approx
-\sum\limits_{\overset{i,j,c}{i\not=j}}
\frac{\Im[Y_{ai}^\dagger Y_{ic} Y_{cj}^\dagger Y_{ja}]}{{\rm sign}(M_{ii}^2-M_{jj}^2)}\left(\frac{m_{\rm Pl}^2}{|M_{ii}^2-M_{jj}^2|}\right)^{\frac23}
\times 4.18284\times 10^{-4}\frac{\gamma_{\rm av}^2}{g_w}\,.
\end{align}

Comparing with Eq.~(\ref{sol:act_av}), we see that momentum averaging the vacuum oscillation term yields an error of about $23 \%$:
\begin{align}
\tilde{\Delta}_{a}^{\,\rm sat}\approx 1.23\times  \Delta_{a}^{\rm sat}\,,
\end{align}
whereas we expect the error in Eq.~(\ref{mom_int}) of to be of order one~\cite{Asaka:2011wq} and hence sufficient our purposes.

\subsection{Sterile Charges in the Oscillatory Regime}
\label{app:2_3_flavours}
In Section~\ref{sec:weak} we have pointed out that up to order $|Y^*Y^t|^2$ no sterile charge $q_N$ is generated by the off-diagonal oscillations. We will show in the following that this is true to all orders for $n_s=2$ sterile flavours, whereas this is not true for $n_s\geq 3$ since a non-vanishing contribution appears at $\mathcal{O}(|Y^*Y^t|^3)$. In order to do so, we introduce a function
\begin{align}
T_{ij}={\rm Re}[Y^*Y^t]_{ij}\delta \nodd_{ji}-\ii {\rm Im}[Y^*Y^t]_{ij} \delta \neven_{ji}\,,
\end{align}
for $i\neq j$. Its derivative with respect to $z$ reads
\begin{align}
\frac{\dd}{\dd z}T_{ij}={\rm Re}[Y^*Y^t]_{ij}\frac{\dd}{\dd z}\delta \nodd_{ji}-\ii {\rm Im}[Y^*Y^t]_{ij} \frac{\dd}{\dd z} \delta \neven_{ji}\,.
\end{align}
The deviations $\delta \neven_{ji}$ and $\delta \nodd_{ji}$ are determined by solving  Eq.~(\ref{osc:weakwashoutav}) for non-diagonal components $(i\neq j)$. In case of $n_s=2$ flavours, one can express the anticommutators as
\begin{subequations}
\begin{align}
\{{\rm Re}[Y^*Y^t],\delta n\}_{ij}&=({\rm Re}[Y^*Y^t]_{ii}+{\rm Re}[Y^*Y^t]_{jj})\delta n_{ij}+{\rm Re}[Y^*Y^t]_{ij}(\delta n_{ii}+\delta n_{jj})\,,\\
\{{\rm Im}[Y^*Y^t],\delta n\}_{ij}&={\rm Im}[Y^*Y^t]_{ij}(\delta n_{ii}+\delta n_{jj})\,,
\end{align}
\end{subequations}
since the diagonal entries of $Y^*Y^t$ are purely real due to its Hermitian property. After some calculation we are left with
\begin{align}
\label{eq:T_1}
\nonumber
\frac{\dd}{\dd z}T_{ij}&=-\ii A_{ji}z^2 T_{ij}-\gamma_{\rm av}\frac{a_{\rm R}}{2T_{\rm ref}}({\rm Re}[Y^*Y^t]_{ii}+{\rm Re}[Y^*Y^t]_{jj})T_{ij}\\
&-\gamma_{\rm av}\frac{a_{\rm R}}{2T_{\rm ref}} (\delta \nodd_{ii}+\delta \nodd_{jj})\left({\rm Re}[Y^*Y^t]_{ij}{\rm Re}[Y^*Y^t]_{ji}-{\rm Im}[Y^*Y^t]_{ij}{\rm Im}[Y^*Y^t]_{ji}\right)\,.
\end{align}

It is easy to see that Eq.~(\ref{relaxation_force}) can be expressed in terms of ${\rm Re}[T_{ij}]$
\begin{align}
\label{eq:T_2}
\frac{\dd}{\dd z}\nodd_{ii}=-\gamma_i\nodd_{ii}-\gamma_{\rm av}\frac{a_{\rm R}}{T_{\rm ref}} \sum\limits_{\overset{j}{j\not=i}}{\rm Re}[T_{ij}]\,.
\end{align}

In order to require zero sterile charge, $\delta \nodd_{ij}$, $\delta \nodd_{ii}$ as well as $\delta \neven_{ij}$ have to vanish for $z\rightarrow 0$ and so does $T_{ij}$. Thus, Eqs.~(\ref{eq:T_1}) and~(\ref{eq:T_2}) can be solved to
\begin{align}
T_{ij}(z)=\delta \nodd_{ii}(z)=0\,,
\end{align} 
which is true for all $z$. Additionally, this even results in a condition between $\neven$ and $\nodd$:
\begin{subequations}
\begin{align}
{\rm Re}[Y^*Y^t]_{ij}{\rm Re}[\delta \nodd_{ij}]&={\rm Im}[Y^*Y^t]_{ij}{\rm Im}[\delta \neven_{ij}]\,,\\
{\rm Re}[Y^*Y^t]_{ij}{\rm Im}[\delta \nodd_{ij}]&=-{\rm Im}[Y^*Y^t]_{ij}{\rm Re}[\delta \neven_{ij}]\,.
\end{align}
\end{subequations}

Whereas this holds for $n_s=2$ sterile flavours one can show that for $n_s\geq 3$, already at $\mathcal{O}(|Y^*Y^t|^3)$, there appears a non-vanishing contribution to $F_i$. For that, we solve Eq.~(\ref{osc:weakwashoutav}) for off-diagonal $\delta n_{ij}$ recursively to $\mathcal{O}(|Y^*Y^t|^2)$ by using solutions for $\delta n$ at $\mathcal{O}(|Y^*Y^t|)$. This result can be used as an input for $F_i$ in Eq.~(\ref{relaxation_force}), such that for $n_s=3$, we have:
\begin{subequations}
\begin{align}
F_i(z)&=\frac{\gamma_{\rm av}^3a_{\rm R}^2}{2T_{\rm ref}^3}
\sum\limits_{j}|\epsilon_{ijk}|\mathcal{Y}_{ijk}{\rm Im}[\tilde{\mathcal{F}}_{jik}(z)]+\mathcal{O}(|Y^*Y^t|^4)\,,\\ 
\mathcal{Y}_{ijk}&={\rm Re}[Y^*Y^t]_{ij}{\rm Im}\left[(Y^*Y^t)_{jk}(Y^*Y^t)_{ki}\right]+{\rm Im}[Y^*Y^t]_{ij}{\rm Re}\left[(Y^*Y^t)_{jk}(Y^*Y^t)_{ki}\right]\,,\\
\tilde{\mathcal{F}}_{ijk}(z)&=\exp\left(-\frac{\ii}{3}\Omega_{ij}z^3\right)\times \int \limits_0^z  \!\dd t \,\exp\left(\frac{\ii}{3}\Omega_{ij}t^3\right)\big[\mathcal{F}_{kj}(t)+\mathcal{F}_{ik}(t)\big] \,,
\end{align}
\end{subequations}
with $|\epsilon_{ijk}|$ as the absolute value of the Levi-Civita-Symbol in order to account for $(i\neq j, k\neq i, k \neq j )$. Thus, as a perturbative expansion in the Yukawa coupling $Y$, it is justified to assume zero initial sterile charge $q_{Nii}$ after the first oscillations in the oscillatory regime.
\end{appendix}

\bibliographystyle{JHEP}
\bibliography{GeV_Seesaw}
\end{document}